\documentclass[11pt]{article}

\usepackage{amsmath, amsthm, amsfonts,amssymb}
\usepackage{graphicx,epsfig,rotating,algorithm,algpseudocode}
\usepackage{bm}
\usepackage[left=1in,right=1in,top=1.1in,bottom=1.2in]{geometry}
\usepackage{fancyhdr}
\usepackage{xurl}

\usepackage{authblk}
\usepackage{multirow}
\usepackage{graphicx, color,rotating}
\usepackage{subfigure}
\usepackage{epstopdf}
\usepackage{float}
\usepackage{hyperref}
\hypersetup{colorlinks,linkcolor={blue},citecolor={blue},urlcolor={red}}  
\usepackage{lipsum}% http://ctan.org/pkg/lipsum
\usepackage{ccaption}

\theoremstyle{plain}
\newtheorem{thm}{Theorem}[section]

\newtheorem{lem}[thm]{Lemma}

\theoremstyle{remark}

\newcommand{\beq}{\begin{equation}}
	\newcommand{\eeq}{\end{equation}}
\newcommand{\beas}{\begin{align*}}
	\newcommand{\eeas}{\end{align*}}
\newcommand{\bea}{\begin{align}}
	\newcommand{\eea}{\end{align}}
\newcommand{\bei}{\begin{itemize}}
	\newcommand{\eei}{\end{itemize}}
\newcommand{\ben}{\begin{enumerate}}
	\newcommand{\een}{\end{enumerate}}
\newcommand{\bet}{\begin{theorem}}
	\newcommand{\eet}{\end{theorem}}
\newcommand{\bel}{\begin{lemma}}
	\newcommand{\eel}{\end{lemma}}
\newcommand{\bep}{\begin{proposition}}
	\newcommand{\eep}{\end{proposition}}
\newcommand{\bed}{\begin{definition}}
	\newcommand{\eed}{\end{definition}}
\newcommand{\bec}{\begin{corollary}}
	\newcommand{\eec}{\end{corollary}}
\newcommand{\bex}{\begin{example}}
	\newcommand{\eex}{\end{example}}

\newcommand{\bu}{\bold{u}}

\newcommand{\bw}{\bold{w}}
\newcommand{\bv}{\bold{v}}
\newcommand{\bn}{\bold{n}}
\newcommand{\bfm}{\bold{m}}

\newcommand{\bs}{\bold{s}}

\newcommand{\bU}{\bold{U}}

\newcommand{\bW}{\bold{W}}

\newcommand{\bA}{\bold{A}}
\newcommand{\bR}{\bold{R}}

\newcommand{\bZ}{\bold{Z}}
\newcommand{\bL}{\bold{L}}
\newcommand{\bP}{\bold{P}}
\newcommand{\bY}{\bold{Y}}

\newcommand{\bX}{\bold{X}}
\newcommand{\bD}{\bold{D}}

\newcommand{\bGam}{\bold{\Gamma}}

\newcommand{\bTheta}{\bold{\Theta}}

\newcommand{\bSig}{\bold{\Sigma}}

\newcommand{\balpha}{\boldsymbol{\alpha}}

\newcommand{\bbeta}{\boldsymbol{\beta}}

\newcommand{\bgamma}{\boldsymbol{\gamma}}

\newcommand{\R}{\mathbb{R}}

\newcommand{\vertiii}[1]{{\left\vert\kern-0.25ex\left\vert\kern-0.25ex\left\vert #1 
		\right\vert\kern-0.25ex\right\vert\kern-0.25ex\right\vert}}

\newcommand{\xb}{\mathbf{x}}
\newcommand{\yb}{\mathbf{y}}

\usepackage{array}
\newcolumntype{L}{>{\centering\arraybackslash}m{3.5cm}}

\begin{document}

	\title{Is your data alignable? Principled and interpretable alignability testing and integration of single-cell data}

\author[1]{Rong Ma}
	\affil[1]{Department of Biostatistics, Harvard University}

		\author[2]{Eric D. Sun}
	\affil[2]{Department of Biomedical Data Science, Stanford University}	
	
	\author[3]{David Donoho}
		\affil[3]{Department of Statistics, Stanford University}	
	
		\author[2]{James Zou}

\date{}
	
	\maketitle

 \begin{abstract}
   Single-cell data integration can provide a comprehensive molecular view of cells, and many algorithms have been developed to remove unwanted technical or biological variations and integrate heterogeneous single-cell datasets. Despite their wide usage, existing methods suffer from several fundamental limitations. In particular, we lack a rigorous statistical test for whether two high-dimensional single-cell datasets are alignable (and therefore should even be aligned). Moreover, popular methods can substantially distort the data during alignment, making the aligned data and downstream analysis difficult to interpret. To overcome these limitations, we present a spectral manifold alignment and inference (SMAI) framework, which enables principled and interpretable alignability testing and structure-preserving integration of single-cell data with the same type of features. SMAI provides a statistical test to robustly assess the alignability between datasets to avoid misleading inference, and is justified by high-dimensional statistical theory. On a diverse range of real and simulated benchmark datasets, it outperforms commonly used alignment methods. Moreover, we show that SMAI improves various downstream analyses such as identification of differentially expressed genes and imputation of single-cell spatial transcriptomics, providing further biological insights. SMAI’s interpretability also enables quantification and a deeper understanding of the sources of technical confounders in single-cell data.
 \end{abstract}
 
 %\keywords{single-cell omics $|$ data alignment $|$ random matrix theory $|$ spectral method $|$ Procrustes analysis}

	\section*{Introduction}

The rapid development of single-cell technologies has enabled the characterization of complex biological systems at unprecedented scale and resolution. On the one hand, diverse and heterogeneous single-cell datasets have been generated, enabling opportunities for integrative profiling of cell types and deeper understandings of the associated biological processes \cite{trapnell2015defining,tanay2017scaling,elmentaite2022single}. On the other hand, the widely observed technical and biological variations across datasets also impose unique challenges to many downstream analyses \cite{lahnemann2020eleven,tran2020benchmark,luecken2022benchmarking}. The variations between datasets can originate from different experimental protocols, laboratory conditions, sequencing technologies, etc. It may also arise as biological variations when the samples come from distinct spatial locations, times, tissues, organs, individuals, or species. 

Several computational algorithms have been  developed to remove the unwanted variations and integrate heterogeneous single-cell datasets. 
To date, the most widely used data integration methods, such as Seurat \cite{stuart2019comprehensive}, LIGER \cite{welch2019single}, Harmony \cite{korsunsky2019fast}, fastMNN \cite{haghverdi2018batch}, and Scanorama \cite{hie2019efficient}, are built upon the key assumption that there is a shared latent low-dimensional structure between the datasets of interest. These methods attempt to obtain an alignment of the datasets by identifying and matching their respective low-dimensional structures. As a result, the methods would output some integrated cellular profiles, commonly represented as either a corrected feature matrix, or a joint low-dimensional embedding matrix, where the unwanted technical or biological variations between the datasets have been removed. These methods have  played an indispensable role in current single-cell studies such as generating large-scale reference atlases of human organs \cite{mereu2020benchmarking,tabula2022tabula},  inferring lineage differentiation trajectories \cite{ranek2022integrating,sugihara2022alignment}, and multiomic characterization of COVID-19 pathogenesis and immune response \cite{stephenson2021single,ahern2022blood}.

Despite their popularity, the existing integration methods also suffer from several fundamental limitations, which makes it difficult to statistically assess findings drawn from the aligned data. First, there is a lack of statistically rigorous methods to determine whether two or more datasets should be aligned. Without such a safeguard, existing methods are used to align and integrate single-cell datasets that do not have a meaningful shared structure, leading to problematic and misleading interpretations \cite{chari2021specious,cooley2019novel}.
Global assessment methods such as k-nearest neighbor batch effect test (kBET) \cite{buttner2019test}, guided PCA (gPCA) \cite{reese2013new}, probabilistic principal component and covariates analysis (PPCCA) \cite{nyamundanda2017novel}, and metrics such as local inverse Simpson’s index (LISI) \cite{korsunsky2019fast}, average silhouette width (ASW) \cite{batool2021clustering}, adjusted Rand index (ARI) \cite{wu2020accounting}, have been proposed to quantitatively characterize the quality of alignment or the extent of batch-effect removal, based on specific alignment procedures. However, these methods can only provide post-hoc evaluations of the mixing of batches, which may not necessarily reflect the actual alignability or structure-sharing between the original datasets. Moreover, these methods do not  account for the noisiness (ASW, ARI, LISI) or the effects of high dimensionality (kBET, gPCA, PPCCA) of the single-cell datasets, resulting in biased estimates and test results. Other methods such as limma \cite{ritchie2015limma} and MAST \cite{finak2015mast} consider linear batch correction, whose focus is restricted to  differential testing and does not account for possible covariance shifts.

\begin{figure}
	\centering
	\includegraphics[angle=0,width=13cm]{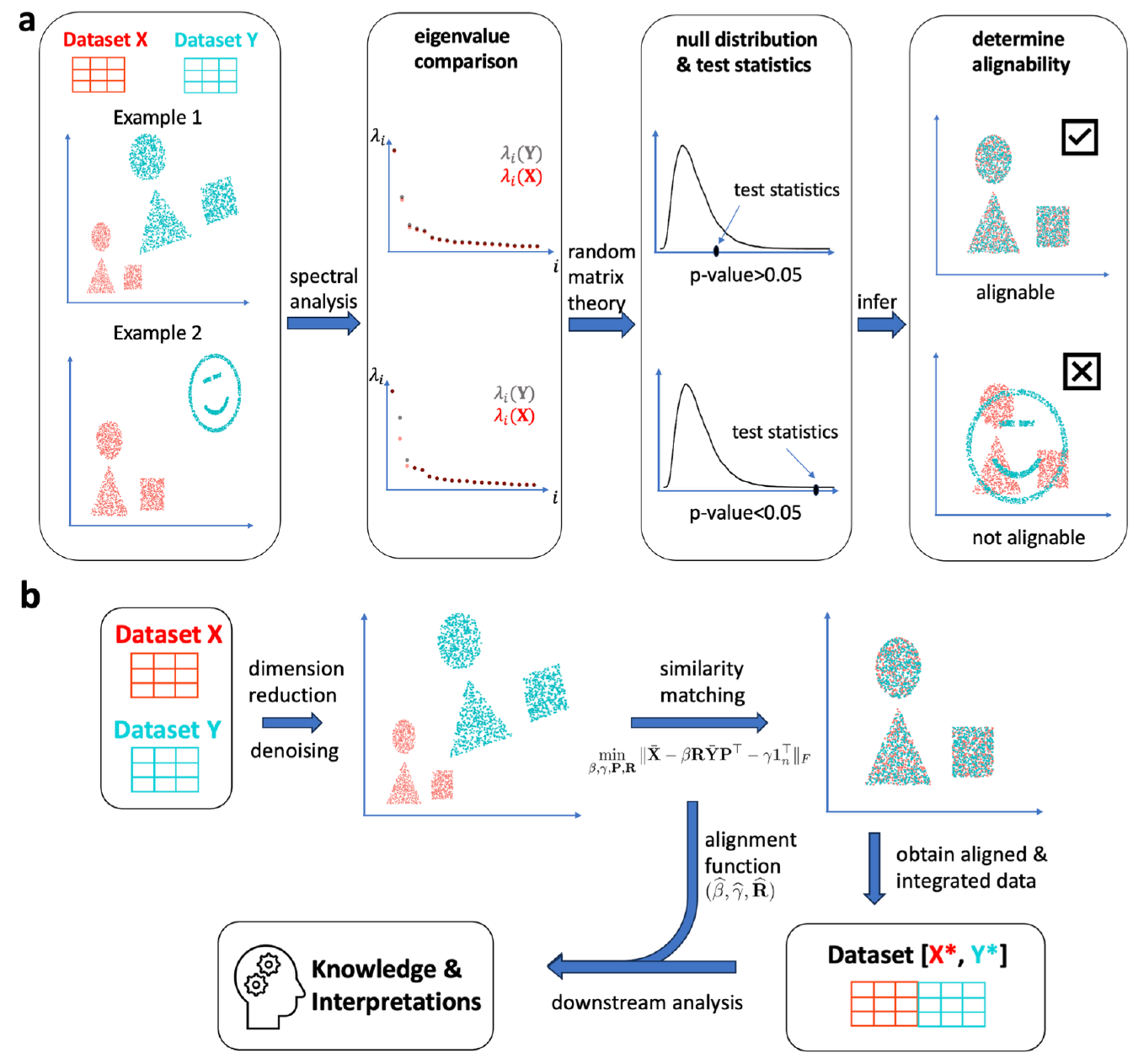}
	\caption{\emph{Overview and illustration of the SMAI algorithm.}  (a) SMAI-test imposes a low-rank spiked covariance matrix model where the low-dimensional signal structures of data matrices are encoded by a few largest eigenvalues of the population covariance matrices. 
		Under the null hypothesis that the underlying signal structures are alignable up to a similarity transformation, a test statistic based on comparing the leading eigenvalues of the empirical covariance matrices is computed,  whose theoretical null distribution as $(n,d)\to\infty$ is derived using random matrix theory. The final p-value returned by SMAI-test is used to infer the alignability of the two datasets. (b) SMAI-align aims to solve  the shuffled Procrustes optimization problem (\ref{align.opt}). To do so, SMAI-align starts with a denoising procedure, and then adopts an iterative spectral algorithm to achieve similarity matching between the two datasets using high-dimensional Procrustes analysis. The method returns an integrated dataset containing all the samples with the original features, along with a closed-form alignment function, which is interpretable and can be readily used for various downstream analyses. } 
	\label{fig1}
\end{figure}

Moreover, research suggests that  existing integration methods may introduce serious distortions to the individual datasets during their alignment process \cite{chari2021specious,cooley2019novel}. In this study, we systematically evaluate the severity and effects of such distortions for several popular integration methods. Our results confirm that these methods, while eliminating the possible differences between datasets, may also alter the original biological signals contained in individual datasets, causing misleading results such as power loss and false discoveries in downstream analyses. 
Finally, none of these popular integration methods admits a tractable closed-form expression of the final alignment function, with a clear geometric meaning of its constitutive components. As a result, these methods would suffer from a lack of interpretability, making it difficult to inspect and understand the nature of any removed variations, or to distinguish the unwanted variations from the potentially biologically meaningful variations.

To overcome the above limitations, we present a spectral manifold alignment and inference (SMAI) framework for accountable and interpretable integration of single-cell data {with the same type of features}. Our contribution is two-fold. First, we develop a rigorous statistical test (SMAI-test) that can robustly determine the alignability between two datasets. Secondly, motivated by this test, we propose an interpretable spectral manifold alignment algorithm (SMAI-align) that enables more reliable data integration without altering or corrupting the original biological signals. Our systematic experiments  demonstrate that SMAI improves various downstream analyses such as the identification of cell types and their associated marker genes, and the prediction of single-cell spatial transcriptomics. Moreover, we show that SMAI’s interpretability provides insights into the sources of technical confounders in single-cell data.

\section*{Results}

\subsection*{Overview of SMAI}

SMAI consists of two components: SMAI-test flexibly determines the global or partial alignability between the datasets, whereas SMAI-align searches for the best similarity transformation to achieve the alignment.

{SMAI-test evaluates the statistical significance against the null hypothesis that two single-cell datasets are alignable up to some similarity transformation, that is, combinations of scaling, translation, and rotation. In line with several previous works, it leverages random matrix theory to be robust to noisy and high-dimensional single-cell omic data \cite{bai2019statistical,aparicio2020random,nitzan2021revealing,leviyang2023random}. As a hypothesis testing framework that determines the presence and the nature of batch effects between two datasets, SMAI-test tells users if the datasets should or should not be aligned by SMAI-test or other integration methods, as enforcing alignment would introduce substantial distortions.}
To increase flexibility, SMAI-test also allows for testing against partial alignability between the datasets, where the users can specify a threshold $t\%$, so that the null hypothesis states that at least $t\%$ of the samples in both datasets are alignable. Recommended values for $t$ are between $50$ and $70$ depending on the context, to ensure both sufficient sample size (or power), and flexibility to local heterogeneity (Supplementary Notes); we used $t=60$ for the real datasets analyzed in this study. %\james{Explain what is the $t$ used in all of our experiments and how users should select $t$.} 
Importantly, the statistical validity of SMAI-test is theoretically guaranteed over a wide range of settings (Methods, Theorem \ref{test.thm}), suitable for modeling high-dimensional single-cell data. We support the empirical validity of the test with both simulated data and multiple real-world benchmark datasets, ranging from transcriptomics, chromatin accessibility, to spatial transcriptomics.

SMAI-align incorporates a high-dimensional shuffled Procrustes analysis, which iteratively searches for the sample correspondence and the best similarity transformation that minimizes the discrepancy between the intrinsic low-dimensional signal structures of the datasets. SMAI-align enjoys several advantages over the existing integration methods. First, SMAI-align returns an alignment function in terms of a similarity transformation, which has a closed-form expression (Supplementary Notes) equipped with a clear geometric meaning. The better interpretability enables quantitative characterization of the source and magnitude of any removed and remaining variations, and may bring insights into the  mechanisms underlying the batch effects. Second, due to the shape-invariance property of similarity transformations, SMAI-align preserves the relative distances between the samples within individual datasets throughout the alignment, making the final integrated data less susceptible to technical distortions and therefore more suitable and reliable for downstream analyses. Third, unlike many existing methods (such as Seurat, Harmony and fastMNN), which require specifying a target dataset for alignment and whose performance is asymmetric with respect to the order of datasets, SMAI-align obtains a symmetric invertible alignment function that is indifferent to such an order, making its output more consistent and robust to technical artifacts. %Finally, SMAI-align is very fast and can be easily scaled to datasets of \james{millions?} of cells. 
Below we sketch the main ideas of the SMAI algorithm and leave the details to Materials and Methods, and the Supplementary Notes.

\subsection*{SMAI-test}

Suppose  that $\bX\in\R^{d\times n_1}$ and $\bY\in\R^{d\times n_2}$ are the normalized count matrices generated from two single-cell experiments, with $d$ being the number of features (genes) and $n_1$ and $n_2$ being the respective numbers of cells. To test for the alignability between $\bX$ and $\bY$, SMAI-test %focuses on the centered  normalized matrices $\bar\bX=\bX({\bf I}-{\bf 1 1}^\top)$ and $\bar\bY=\bY({\bf I}-{\bf 1 1}^\top)$, and 
assumes a low-rank spiked covariance matrix model (Figure S1) where the low-dimensional signal structures of $\bX$ and $\bY$ are encoded by the leading eigenvalues and eigenvectors of their corresponding population covariance matrices $\bSig_1$ and $\bSig_2$. As a result, the null hypothesis that the signal structures underlying $\bX$ and $\bY$ are identical up to a similarity transformation implies that the leading eigenvalues of $\bSig_1$ and $\bSig_2$ are identical up to a global scaling factor. As such, a test statistic $T(\bX, \bY)$ based on comparing the leading eigenvalues of the empirical covariance matrices of $\bX$ and $\bY$ can be computed,  whose theoretical null distribution as $(n_1, n_2, d)\to\infty$ is derived  using random matrix theory. Thus, SMAI-test returns the p-value by comparing the test statistic $T(\bX, \bY)$ with its asymptotic null distribution (Figure \ref{fig1}a). 

For the test of partial alignability, a sample splitting procedure is adopted where the first part  is used to identify  subsets of the two datasets with maximal correspondence or structure-sharing (Methods), and the second part is used to compute the test statistic and the p-value concerning the alignability between such maximal correspondence subsets. As such, we avoid selection bias due to repeated use of the samples in both selection and test steps, ensuring a valid test.

\subsection*{SMAI-align} 

SMAI-align starts by filtering out the low-rank signal structures in $\bX$ and $\bY$ to obtain their denoised versions $\widehat\bX$ and $\widehat\bY$, %\james{explain how denoising works here} 
and then approximately solves the following shuffled Procrustes optimization problem
\beq \label{align.opt}
\min_{\beta,\bgamma,\bP,\bR}\|\widehat\bX-\beta\bR\widehat\bY\bP-\bgamma {\bf 1}_{n_1}^\top\|_F.
\eeq
Here, ${\bf 1}_n\in\R^n$ is an all-one vector, and the minimization is achieved for some global scaling factor $\beta\in \R$, some vector $\bgamma\in\R^d$ adjusting for the possible global mean shift between $\widehat\bY$ and $\widehat\bX$, %\james{Should we use $\bar\bX$ and $\bar\bY$ throughout here?},
some extended orthogonal matrix  (see Methods) $\bP\in\R^{n_2\times n_1}$ recovering the sample correspondence between $\widehat\bX$ and $\widehat\bY$, and some rotation matrix $\bR\in\R^{d\times d}$ adjusting for the possible covariance shift. Compared with the traditional Procrustes analysis \cite{goodall1991procrustes}, (\ref{align.opt}) contains an additional matrix $\bP$, allowing for a general unknown correspondence between the samples in $\bX$ and $\bY$, which is the case in most of our applications. To solve for (\ref{align.opt}), SMAI-align adopted an iterative spectral algorithm that alternatively solves for $\bP$ and $(\beta,\bgamma,\bR)$ using high-dimensional Procrustes analysis. The final solution $(\widehat\beta,\widehat\bgamma,\widehat\bR)$ then gives a good similarity transformation aligning the two datasets in the original feature space. In particular, to improve robustness and reduce the effects of potential outliers in the data on the final alignment function, in each iteration we remove some leading outliers  from both datasets, whose distances to the other dataset remain large.  Moreover, to allow for integration of datasets containing partially shared structures (up to a user-specified threshold, see Methods), users may also request SMAI-align to infer the final alignment function only based on the identified maximal correspondence subsets, rather than the whole datasets. This makes the alignment more robust to local structural heterogeneity. SMAI-align returns an integrated dataset  containing all the samples, along with the similarity transformation, which are interpretable and readily used for various downstream analyses (Figure \ref{fig1}b). The idea of SMAI-align is closely related to that of SMAI-test: a pass in SMAI-test essentially renders the goodness-of-fit of the model underlying SMAI-align, and therefore ensures its performance. 
In addition, since SMAI-align essentially learns some underlying similarity transformation, based on which all the samples are aligned, the algorithm is easily scalable to very large datasets. For example, one can first infer the alignment function by applying SMAI-align to some representative subsets of the datasets, and then use it to align all the samples (Methods).

To empirically evaluate the statistical validity of SMAI-test and the consistency of SMAI-align, we generate simulated data based on some signal-plus-noise matrix model with various signal structures, batch effects, and sample sizes (Supplementary Notes). Our simulation results indicate that SMAI-test has desirable type I errors  across all the simulation settings, that is, achieving the nominal  probability (0.05) of rejecting the null hypothesis when the datasets are truly alignable (Table S1, Methods). We also evaluate the performance of SMAI-align in recovering the true alignment function by measuring the estimation errors for each of the true parameters $(\beta^*,\bgamma^*,\bR^*)$ generating the data (Methods). We find that increasing the sample sizes leads to reduced estimation errors in general (Figure S2b), suggesting statistical consistency of SMAI-align. %\james{Explain out the simulation results more explicitly.}

\begin{figure}
\centering
	\includegraphics[angle=0,width=16cm]{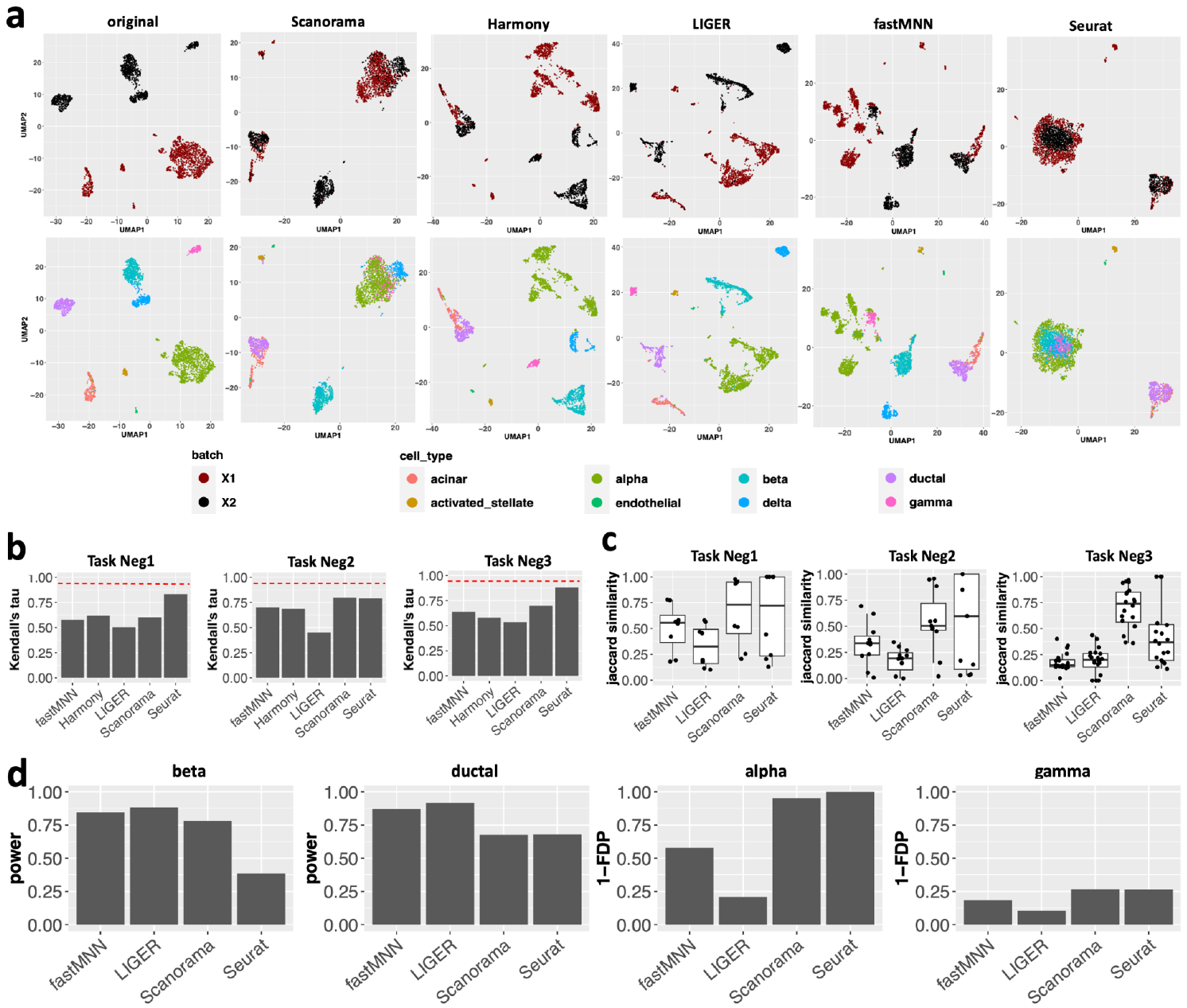}
	\caption{ \emph{Forcing uncertified data integration may cause false alignment, serious distortions and misleading inferences. } (a) UMAP visualizations of the original (pooled) data under negative control task Neg1, and the integrated data as obtained by five popular methods (Scanorama, Harmony, LIGER, fastMNN and Seurat). For each method, the top figure is colored to indicate the distinct datasets being aligned, whereas the bottom figure colored to indicate different cell types. {See Figures S3a and S3b for similar results about Pamona and SCOT, and the results about Neg2 and Neg3. }
		(b) Under the three negative control tasks, we show barplots of Kendall's tau correlations between relative distances among the cells before  integration and the distances after data integration, as achieved by each methods. The red dashed line benchmarks the average Kendall's tau correlation of 0.9 achieved by SMAI-align over the positive control tasks Pos1-Pos7. (c) Boxplots of Jaccard similarity between the set of differentially expressed (DE) genes associated with a distinct cell type detected based on the integrated data and the DE genes based on the original data. Each point represents a cell type. {See Figure S3c and S3d for similar results about Pamona and SCOT.} (d) For Task Neg1, we show some representative barplots of $(1-$false discovery proportion) (1$-$FDP) and the power of detecting DE genes for some cell types based on the integrated data. Harmony is not included in (c) and (d) as its integration is only achieved in the low-dimensional space. Notably, SMAI-test correctly detects that all the datasets in Tasks Neg1-Neg3 are not alignable.}% Continued caption
	\label{fig2}
\end{figure}

\subsection*{SMAI robustly determines alignability between diverse single-cell data} 

We apply SMAI-test to diverse single-cell data integration tasks and demonstrate its robust performance in determining the alignability between datasets. { The detailed information about each dataset and the corresponding test results are summarized in Table S1. In particular, our real and synthetic datasets involve diverse tissues including human livers, human pancreas, human blood (peripheral blood mononuclear cell, PBMC), human lung, human mesenteric lymph nodes (MLN), human lung-draining lymph nodes (LLN), mouse brain, mouse PBMC (either lipopolysaccharide (LPS) stimulated and control), mouse primary visual cortex (VISP), and mouse gastrulation, and contain multiple modalities measured by various sequencing technologies, such as single-cell transcriptomics (10X Genomics, Smart-seq, Smart-seq2, Drop-seq, and CEL-seq2), spatial transcriptomics (seqFISH, ISS and ExSeq), and chromatin accessibility (ATAC-seq). The 13 integration tasks cover 7 different scenarios that arise commonly in single-cell research, including (1) integration across different samples with the same cell types, (2) integration across different samples with partially overlapping cell types, (3) integration across samples with non-overlapping cell types, (4) integration across studies with different sequencing technologies, (5)  integration across studies with different tissues, (6) integration across studies with different experimental conditions, and (7) integration of single-cell RNA-seq and spatial-transcriptomic data. For each task, we test for partial alignability between each pair of datasets, determining whether at least 60\% of the cells are alignable in the sense of our null model ({Methods}). Among them, three out of 13 integration tasks, with zero or very low proportions ($\le 37\%$) of cells under the overlapping cell types, are taken as negative controls (Neg1-Neg3), whose alignability is doubtful in general; the rest of the tasks, including both non-spatial integration tasks (Pos1-Pos7) and spatial integration tasks (PosS1-PosS3), are taken as positive controls, whose alignability is expected due to the association with the same tissue or largely overlapping cell types.}

SMAI-test returns significant p-values ($<0.01$) for all the negative controls, correctly detecting their unalignability against our null model. The positive control tasks are assigned non-significant p-values, passing the (partial) alignability test as expected.
{For the 7 non-spatial integration tasks (Pos1-Pos7), SMAI-test confirms the alignability between datasets  with largely overlapping ($\ge84\%$) cell types but possibly different sequencing technologies, tissues or experimental conditions.} For the three spatial tasks (PosS1-PosS3), SMAI-test confirms the alignability of the paired single-cell RNA-seq and spatial-transcriptomic data from the same tissue, justifying its wide use for downstream analyses such as prediction of unmeasured spatial genes (Fig \ref{fig4}).

\subsection*{Necessity of certifying data alignability prior to integration}

 \begin{figure}
	\centering
	\includegraphics[angle=0,width=16cm]{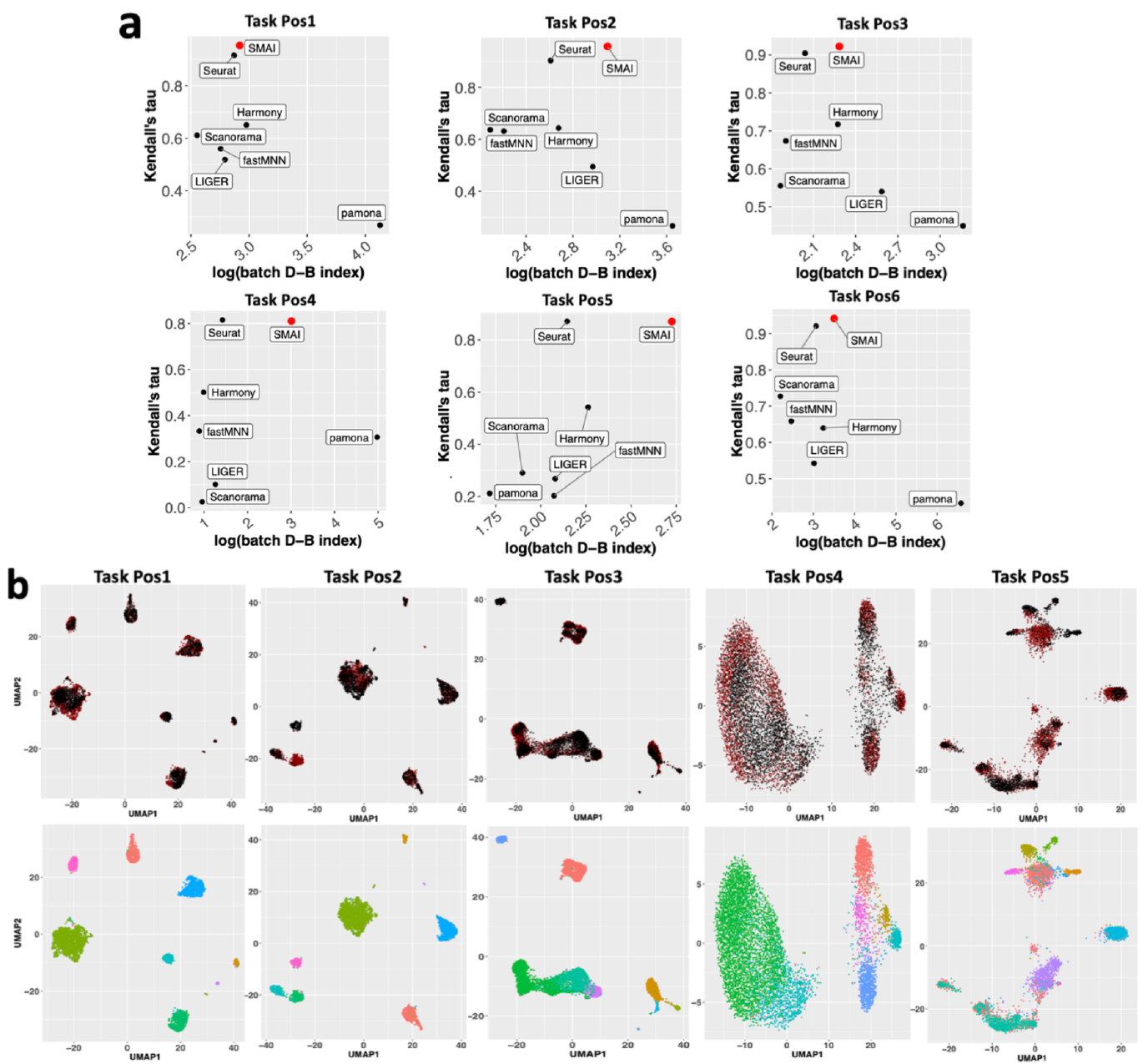}
	\caption{\emph{Performance of SMAI-align on the six positive control integration tasks.} { (a) Compared with the six existing algorithms (black), SMAI-align (red) has an overall best performance in preserving the within-data structures after integration while achieving a competitive performance in removing the unwanted variations. The former is characterized by the highest Kendall's tau correlations between the relative distances of the cells within a dataset before integration and the distances after integration (y-axis), whereas the latter is reflected by higher values of the batch-associated Davies-Bouldin (D-B) index (x-axis shown in log-scale). See Figure S5 for more  comparisons about additional datasets and metrics.} (b) UMAP visualizations of the integrated data as obtained by SMAI-align. For each integration task, the top figure is colored to indicate the two datasets being aligned, whereas the bottom figure is colored to indicate different cell types. See also Figure S7 for UMAP visualizations associated with other integration methods.} 
	\label{fig3}
\end{figure}

In the absence of a principled procedure for determining the alignability between datasets, the existing integration methods often end up forcing alignment between any datasets by significantly distorting and twisting each original dataset (Figure \ref{fig2}). Specifically, for each of the negative control tasks Neg1-Neg3, {we apply seven existing integration methods (Scanorama, Harmony, LIGER, fastMNN, Seurat, Pamona \cite{cao2022manifold}, and SCOT \cite{demetci2022scot})} to obtain the integrated datasets, and then evaluate how well the relative distances between the cells within each dataset before integration are preserved in the integrated datasets. As a result, {we find overall low correlations (Kendall's tau correlation on average 0.6 across the seven methods and three negative control tasks Neg1-Neg3, as compared with 0.9 achieved by SMAI-align on average across  the positive control tasks Pos1-Pos7) between the relative distances of the cells within each dataset before integration, and the distances after data integration (Methods, Figures \ref{fig2}b and S3c).} In addition, we observe many cases of false alignment of distinct cell types from different datasets, and sometimes serious distortion and creation of artificial cell clusters under the same cell type. {For example, in Task Neg1, we find the false alignment between ductal cells and acinar cells by Scanorama, Harmony, and fastMNN, %\james{instead of saying all methods except, better to list all the methods directly}, 
	between alpha cells and gamma cells by Scanorama, fastMNN and Seurat, between alpha cells and beta cells by Seurat, and between alpha cells and endothelial cells by Pamona and SCOT (Figures \ref{fig2}a and S3e);} %\eric{Revise preceding sentence for grammar?} 
we also observe significant distortion or dissolution of the alpha cell cluster and the beta cell cluster after integration by Harmony, LIGER, and fastMNN, as compared with the original datasets (Figure \ref{fig2}a). As such, the final integration results can be highly problematic and unfaithful to the original datasets, which may lead to erroneous conclusions from downstream analysis. SMAI-test is able to detect the lack of alignability (i.e., significant p-values) between these datasets, alerting users that the integrated data may not be reliable.

To evaluate the possible effects on downstream analysis, we focus on one important application following data integration, that is, the identification of differentially expressed (DE) genes for each cell type. {We consider the above five integration methods (Harmony and Pamona are not included as they only produce integrated data in the low-dimensional space). For the three negative control tasks, we find that for many cell types, the set of DE genes identified based on the integrated data have low overlap with those identified based on the original datasets (Figure \ref{fig2}c and Figures S3d, and S4)}. These discrepancies are likely artifacts created by the respective integration methods. For instance, in Task Neg1, we find that, compared with other methods, the integrated data based on Seurat has lower power in detecting DE genes for beta cells and ductal cells, which may be a result of collapsing beta and ductal cells with other cell types during Seurat integration (Figure \ref{fig2}d). Similarly, we observe a higher false discovery proportions (FDPs) in detecting DE genes of alpha cells based on the integrated datasets by fastMNN and LIGER than other methods, which is likely a consequence of the artificial split of the alpha cell cluster during fastMNN and LIGER integration (Figure \ref{fig2}d). 
%In this regard, SMAI-test may be used prior to any integration task, to recognize any intrinsic discrepancy between the datasets, and avoid potentially misleading inferences and conclusions.

\subsection*{SMAI enables principled structure-preserving integration of single-cell data}

 \begin{figure}
 	\centering
 	\includegraphics[angle=0,width=16cm]{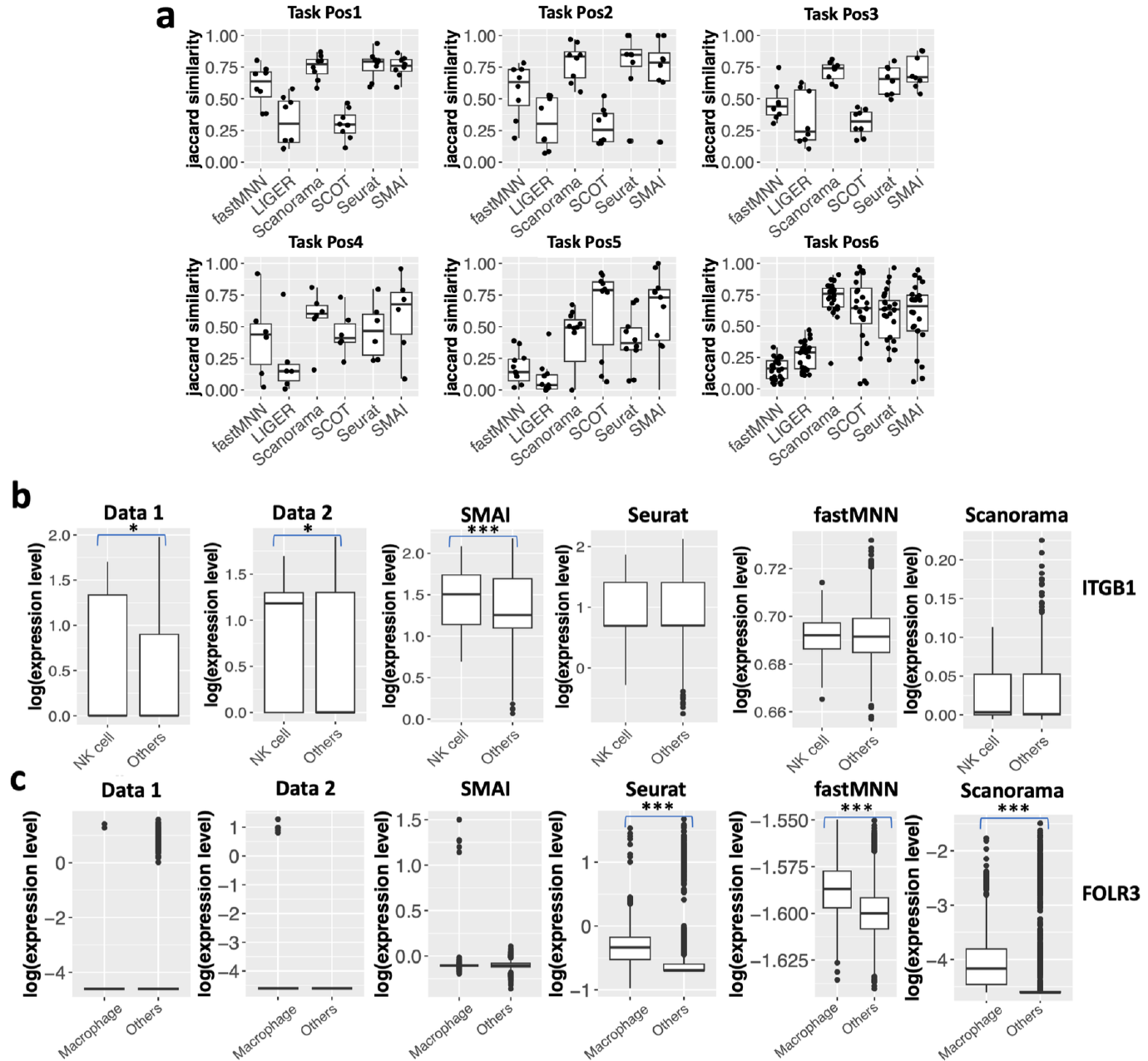}
 	\caption{\footnotesize\emph{SMAI improves reliability and power of DE analysis.} (a) Boxplots of Jaccard similarity between the DE genes for each cell type identified based on the integrated data, obtained by one of the six integration methods, and the genes identified based on the individual datasets before integration. Each point represents a distinct cell type. The results indicate that SMAI-align oftentimes leads to more consistent and more reliable characterization of DE genes, as compared with other methods. (b) Boxplots of log-expression levels of ITGB1 as grouped by cell types in the two datasets about human PBMCs (Task Pos3: Data 1 contains 219 natural killer (NK) cells and 3143 other cells, and Data 2 contains 194 NK cells and 3028 other cells), and in the integrated datasets (413 NK cells and 6171 other cells) as produced by SMAI-align, Seurat, fastMNN, and Scanorama. The DE pattern of ITGB1 is only preserved by SMAI after integration. (c) Boxplots of log-expression levels of FOLR3 as grouped by cell types in the two datasets about human lung tissues (Task Pos5: Data 1 contains 68 macrophages and 2285 other cells, and Data 2 contains 911 macrophages and 1000 other cells), and in the integrated datasets (979 macrophages and 3285 other cells) as produced by SMAI-align, Seurat, fastMNN, and Scanorama. Artificial DE patterns are created by existing integration methods. The stars above the boxplots indicate statistical significance of DE test. Specifically, * means adjusted p-value $<0.05$;  ** means adjusted p-value $<0.01$; *** means adjusted p-value $<0.001$.  Harmony and LIGER are not included in (b) and (c) as they do not produce gene-specific integrated data. %\james{Explain why other methods not shown here.}
 	} 
 	\label{fig3.5}
 \end{figure}

For the first six non-spatial positive control tasks (Tasks Pos1-Pos6) with annotated cell types,  we further apply SMAI-align to obtain the integrated datasets (Figures \ref{fig3}a and S5) and compare the quality of alignment with the above seven existing methods. {We find SMAI-align has overall better performance in preserving the within-data structures after integration (i.e., highest correlations between the relative distances of the cells before integration and the distances after integration), while achieving comparable if not better performance in removing the unwanted variations between the datasets (i.e., higher similarity in expression profiles for the same cell types across batches) (Figure \ref{fig3}a).} In particular, these features are consistent across multiple evaluation metrics, including Kendall's tau correlation and Spearman's rho correlation for structure preservation, and the Davies-Bouldin (D-B) index \cite{davies1979cluster} %,chen2023aster,yu2023topological,hawkins2023icat,ma2023single} 
and the inverse Calinski-Harabasz (C-H) index \cite{calinski1974dendrite} %,lin2017cidr,kimmel2021semisupervised,chidester2023spicemix,karin2023scprisma} 
for batch effect removal (Figure S5). The advantages of these indices over other metrics such as LISI or ARI are explained in Methods section. The desirable alignment achieved by SMAI and its advantages over the existing methods is further supported by visualizing low-dimensional embeddings of the integrated data. In Figure \ref{fig3}b and Figure S6, we observe that across all six tasks, SMAI-align in general achieves good alignment of cells from different datasets under the same cell types. In contrast, distortions and misalignment of certain cell types are found for some existing methods. For instance, in Task Pos2, we observe false integration of gamma cells and beta cells by Harmony and Seurat, and significant distortion, that is, stretching and creation of multiple artificial subclusters, of the alpha cell cluster by LIGER and fastMNN (Figure S7a). As another example, for Task Pos4, we observe strong distortion and artificial split of the excitatory neurons and the inhibitory neurons, by Harmony, LIGER, and fastMNN (Figure S7b). In general, compared with the existing methods, the integrated datasets obtained by SMAI-align are overall of higher integration quality, and less susceptible to technical artifacts, structural distortions, and information loss, making them more reliable for downstream analyses.

{Preserving and characterizing rare cell types is a desirable property of data integration methods. In this respect, SMAI is by design less likely to merge rare cell clusters with other cell types during alignment. To evaluate SMAI's performance in preserving and aligning rare cell types compared with the existing methods, we focus on two positive tasks with more cell types (Pos5 and Pos6), and assess for each rare cell type (containing less than 5\% of the total cells) the performance of different methods in both structure preservation and batch-effect removal. Our results confirm the advantages of using SMAI in dealing with rare cell types (Figure S8).}

\subsection*{SMAI improves reliability and power of differential expression analysis}

A common and important downstream analysis following data integration is to identify the marker genes associated with individual cell types based on the integrated data \cite{welch2019single,hie2019efficient,stuart2019comprehensive}. To demonstrate the advantage of SMAI-align in improving the reliability of downstream differential expression analysis, we focus on the first six positive control tasks and evaluate how many DE genes for each cell type are preserved after integration, and how many new DE genes are introduced after integration. %\james{People might argue that not all new DE genes are false discoveries; if integration is done correctly, then it can increase power and discover new true DE genes.} 
Specifically, for each integrated dataset produced by fastMNN, LIGER, Scanorama, Seurat, or SMAI, we identify the DE genes for each cell type based on the Benjamini-Hochberg adjusted p-values, and compare their agreement with the DE genes identified from the individual datasets before integration using the Jaccard similarity index, which accounts for both power and false positive rate in signal detection (Supplementary Notes). As a result, we find that compared with other methods, SMAI-align oftentimes leads to more consistent and more reliable characterization of DE genes based on the integrated data (Figure \ref{fig3.5}a).

% \james{Suggest condensing this paragraph.} 
Biological insights can be obtained from the improved DE analysis with SMAI-align. For instance, under Task Pos3 concerning human PBMCs, an important protein coding gene ITGB1 (CD29), involved in cell adhesion and recognition \cite{hynes1992integrins}, has been found differentially expressed in natural killer (NK) cells compared with other cell types in the SMAI-integrated data (adjusted p-value $< 10^{-11}$), but not in the Seurat-, fastMNN-, or Scanorama-integrated data. In both original datasets before integration, we also find statistical evidence supporting ITGB1 as a DE gene for NK cells (Figure \ref{fig3.5}b). %\james{is this true for dataset 1? If the ARPC1B case is less clear, fine to move it to Appendix and just focus on the false positive example below.}. In particular, while 
The functional relevance of ITGB1 to NK cells has been reported previously \cite{bezman2012molecular}. In this case, the biological signal is blurred and compromised during data alignment by existing methods. See Figure S9 for similar examples. %As another example, we observe the preservation of the DE gene PRDX1 of CD14+ monocytes after integration, again only achieved by SMAI-align, as shown in Figure \ref{sup.fig3-3}a. The association between PRDX1 (i.e., peroxiredoxin 1), an important antioxidant protein playing a pivotal role in regulating inflammation \cite{ding2017peroxiredoxin}, and CD14+ monocytes, has been reported previously in the experimental settings \cite{szabo2017monocyte}. 
On the other hand, we also observe that SMAI-align is less likely to introduce artificial signals or false discoveries as compared with the existing methods. For example, under Task Pos5 concerning human lung tissues, we find almost no expression of the gene FOLR3 in macrophages in both datasets before integration. However, after integration, this gene is detected as DE gene based on the Seurat-, fastMNN-, and Scanorama-integrated datasets, but not based on the SMAI-integrated dataset (Figure \ref{fig3.5}c). Similar examples are shown in Figure S10. %Such artificial signals are likely consequences of overly distorted expression profiles of these cell types created by existing integration methods that locally search for the best alignment. Such a limitation can be overcome by SMAI-align, which is a global alignment method.

\subsection*{SMAI improves integration of scRNA-seq with spatial transcriptomics}

\begin{figure}%[tbhp]
	\centering
	\includegraphics[width=13cm]{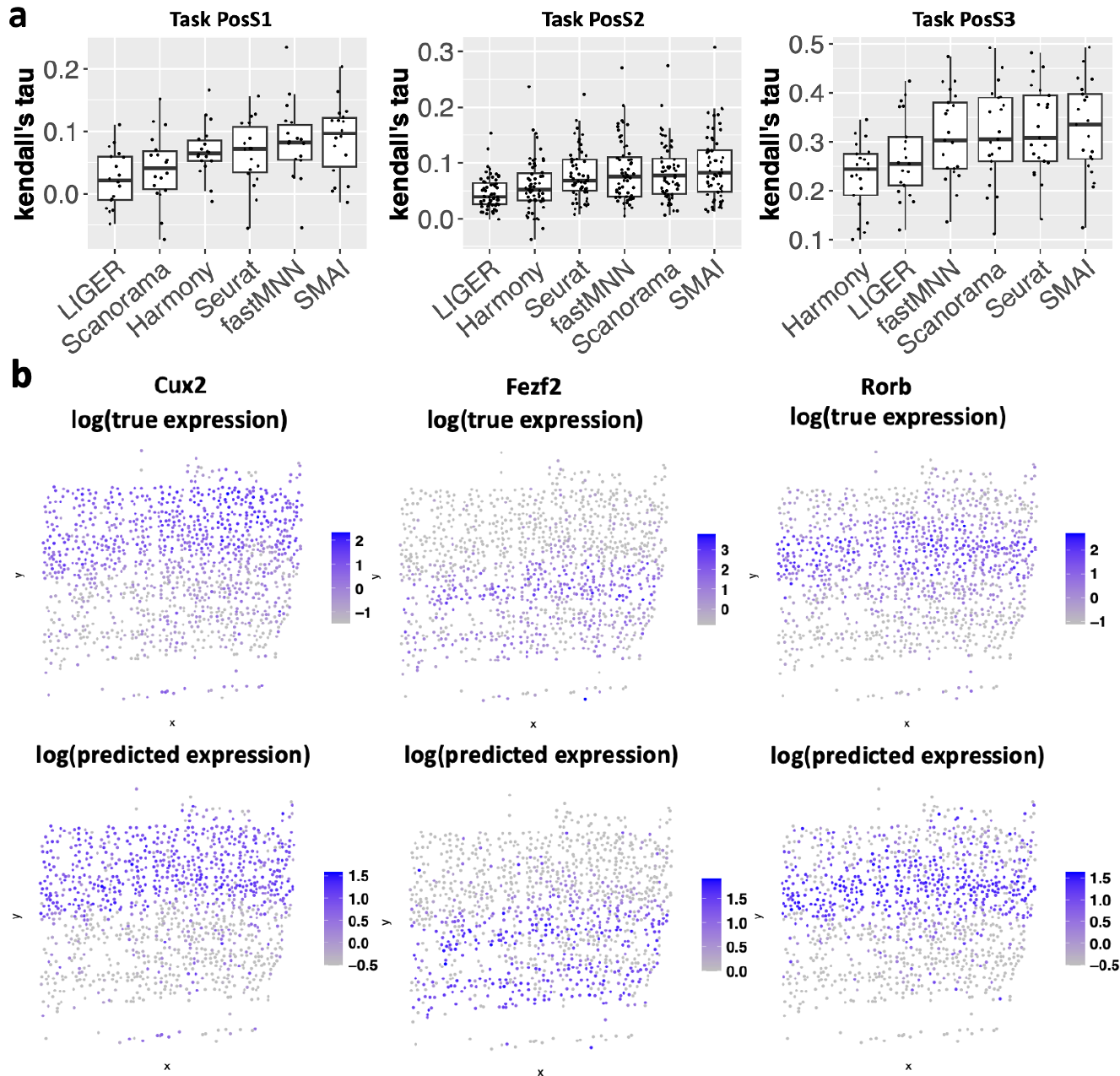}
	\caption{\emph{SMAI improves prediction of single-cell spatial transcriptomic data.} (a) Boxplots of Kendall's tau correlation between the actual expression levels of the spatial genes and their predicted values based on the two-step procedure (alignment followed by $k$ nearest neighbor regression) where the data alignment is achieved by LIGER, Scanorama, Harmony, Seurat, fastMNN, or SMAI-align. Each point represents a distinct spatial gene. The methods are ordered according to their median predictive performance, showing the overall best performance of SMAI. (b) Examples of true expression levels of some spatial genes from Task PosS3, presented according to the cells' spatial layout, and their predicted values based on SMAI-align. The colors are in log scale.} 
\label{fig4}
\end{figure}

Another important application of data integration techniques is the imputation of the spatial expression levels of unmeasured transcripts in single-cell spatial transcriptomic data. Spatial transcriptomics technologies extend high-throughput characterization of gene expression to the spatial dimension and have been used to characterize the spatial distribution of cell types and transcripts across multiple tissues and organisms \cite{asp2019spatiotemporal,%moncada2020integrating,ji2020multimodal,maynard2021transcriptome,biancalani_deep_2021,
	moses2022museum,wei2022spatial}. A major trade-off across all spatial transcriptomics technologies is between the number of genes profiled and the spatial resolution such that most spatial transcriptomics technologies with single-cell resolution are limited to the measurement of a few hundred genes rather than the whole transcriptome \cite{li_benchmarking_2022}. Given the resource-intensive nature of single-cell spatial transcriptomics data acquisition, computational methods for upscaling the number of genes and/or predicting the expression of additional genes of interest have been developed, which oftentimes make use of some paired single-cell RNA-seq data. Among the existing prediction methods, an important class of methods \cite{abdelaal_spage_2020,shengquan_stplus_2021,allen_molecular_2023,welch2019single} are based on first aligning the spatial and RNA-seq datasets and then predicting the expression of new spatial genes by aggregating the nearest neighboring cells in the RNA-seq data. Applications of these methods have been found, for example, in the characterization of spatial differences in the aging of mouse neural and glial cell populations \cite{allen_molecular_2023}, and recovery of immune signatures in primary tumor samples \cite{vahid2023high}. As a key step within these prediction methods, we show that data alignment achieved by SMAI-align may lead to improved performance in predicting unmeasured spatial genes. To ensure fairness, we compare various prediction workflows which only differ in the data integration step (Methods). For the three spatial positive control tasks (PosS1-PosS3), we withhold each gene from the spatial transcriptomic data, and compare its actual expression levels with the predicted values based on the aforementioned two-step procedure where the data alignment is achieved by one of the six methods (LIGER, Scanorama, Harmony, Seurat, fastMNN, and SMAI-align). Due to the intrinsic difficulty of predicting some spatial genes that are nearly independent of any other genes, we only focus on predicting the first half of spatial genes that have overall higher correlations with  other genes. Our analysis of the three pairs of datasets yields the best predictive performance of the SMAI-based prediction method (Figure \ref{fig4}).

%\blue{ Fig 4. imputation results.}

\begin{figure}
	\centering
	\includegraphics[angle=0,width=16cm]{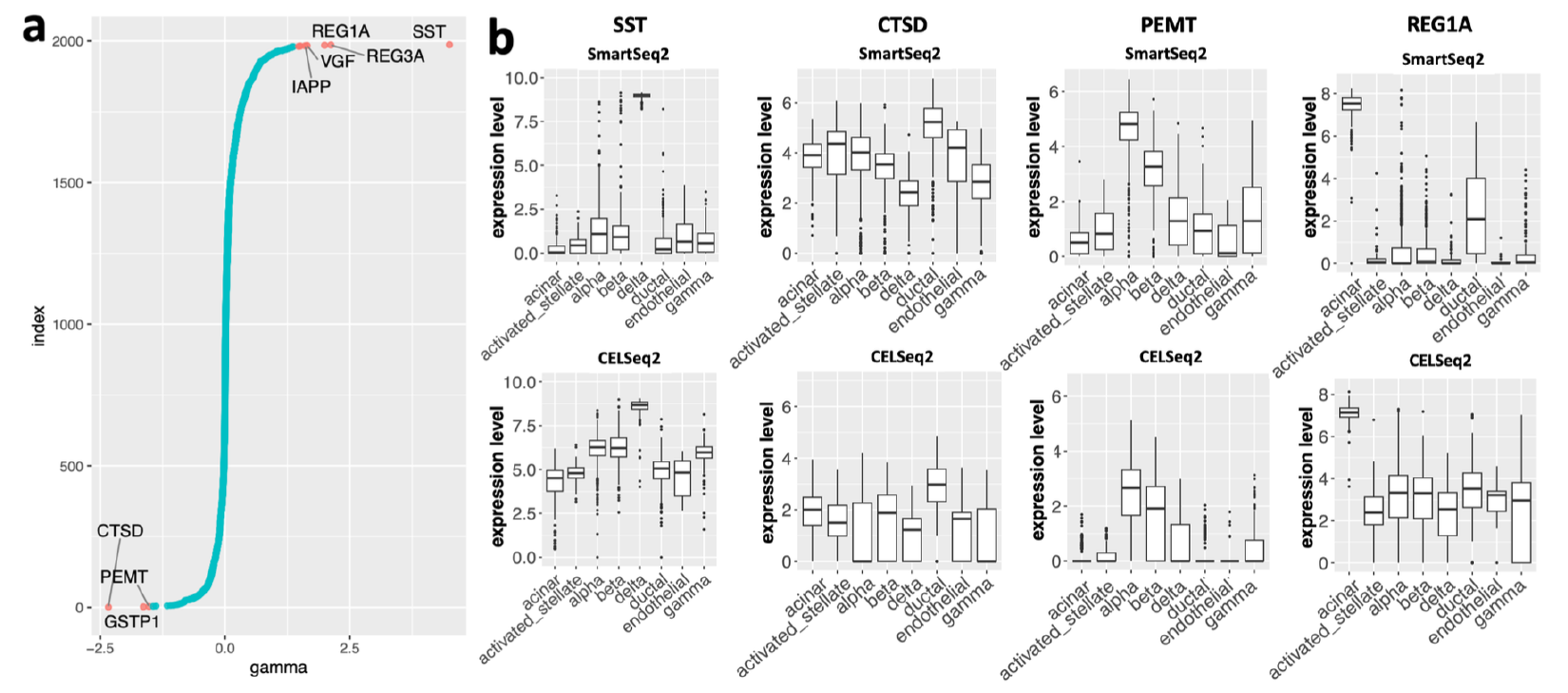}
	\caption{\emph{SMAI's interpretability brings insights into the batch effects.} (a) Visualization of the estimated mean-shift vector $\widehat\bgamma$ from integrating the pancreatic data (Task Pos1) by SMAI-align, whose components are ordered from the smallest (bottom) to the largest (top). In $\widehat\bgamma$ only a sparse set of genes such as SST, CTSD, PEMT, and REG1A are substantially affected by the batch effects. (b) %ROC curves between the DE genes associated with one of the cell types in the pancreatic data (Task Pos1), and the batch-effect-related genes as captured by $\widehat\gamma$ (Left), or by $\widehat\bR$ (Right). The high AUCs suggest most of the genes affected by the batch effects are simultaneously DE genes. (c) 
		Boxplots of expression levels of some genes highlighted in panel (a), grouped by different cell types. Top: Smart-Seq2 dataset with $n=2364$. Bottom: CEL-Seq2 dataset with $n=2244$. Note that SST, CTSD, PEMT, and REG1A are all DE genes. The batch effects on these genes are relatively uniform across cell types, and therefore SMAI-align doesn't affect DE results after integration.} 
	\label{fig5}
\end{figure}

\subsection*{SMAI's interpretability reveals  insights into the sources of batch effects} 

Unlike the existing methods, SMAI-align not only returns the aligned datasets, but also outputs explicitly the underlying alignment function achieving such alignment, which enables further inspection and a deeper understanding of possible sources of batch effects. Specifically, recall that the final alignment function obtained by SMAI-align consists of a scaling factor ($\widehat\beta$), a global mean-shift vector ($\widehat\bgamma$), and a rotation matrix ($\widehat\bR$); each of them may contain important information as to the nature of the corrected batch effects. For example, applying SMAI-align to the human pancreas data (Task Pos1) leads to an alignment function in terms of $(\widehat\beta, \widehat\bgamma, \widehat\bR)$, linking the CEL-Seq2 dataset to the Smart-Seq2 dataset. The scaling factor $\widehat\beta=1.01$ suggests little scaling difference between the two datasets. However, the obtained global mean-shift vector $\widehat\bgamma$ highlighted a sparse set of genes such as SST, CTSD, PEMT, and REG1A, affected by the batch effects (Figure \ref{fig5}a). In particular, for both datasets, we observe similar patterns in the relative abundances of the transcript across different cell types (Figure \ref{fig5}b), suggesting the batch effects on these genes are relatively uniform across cell types. Moreover, while SST, CTSD, PEMT, and REG1A are all DE genes associated with some cell types,  SMAI-align doesn't affect DE patterns after integration due to its ability to identify and remove such global discrepancy. Similar observations can be made on %the rotation matrix (Figure \ref{sup.fig5b}) as well as 
other integration tasks such as human PBMCs (Task Pos3, Figure S12). As for the rotation matrix $\widehat\bR$, it  captures and removes the batch effects altering the gene correlation structures in each dataset (Figure S13).  The obtained SMAI-align parameters $(\widehat\beta, \widehat\bgamma, \widehat\bR)$ can also be converted into distance metrics to quantify and compare the geometrically constitutive features  of the batch effects (Figure S14a) and their overall magnitudes (Figure S14b).

\section*{Discussion}

We develop a spectral manifold alignment and inference algorithm that can determine general alignability between datasets and achieve robust and reliable alignment of single-cell data. The method explores the best similarity transformation that minimizes the discrepancy between the datasets, and thus is interpretable and able to reveal the source of batch effects. { Despite SMAI being limited to modeling  batch effects as linear transformations, our analyses of the ten positive control integration tasks suggest its practical suitability and competence in removing batch effects between datasets across diverse scenarios, oftentimes  achieving superior performance compared with the existing nonlinear data integration methods such as Seurat or Harmony.}
In terms of computational time, standard implementation of SMAI requires a similar running time as existing methods such as Seurat (Figure S15a). The hyperparameters of SMAI, such as the number of informative eigenvalues for each data matrix, have been carefully determined (Supplementary Notes), and shown to work robustly in our benchmark datasets.  %A key hyperparameter of SMAI is the number of informative eigenvalues $r_{\max}$ for each data matrix, which can be determined consistently using statistical procedures (Methods). Other hyperparameters such as the number of iterations and the number of outliers to be removed per iteration, have recommended default values (Methods), shown to work robustly in our benchmark datasets. %Moreover, the availability of a closed-form alignment function allows for fast alignment of very large datasets, by first learning  the alignment function from some representative subsets and then applying the alignment function to the whole dataset (Methods, Figure S15b\&c).

SMAI has a few limitations that deserve further development. {First, although our restriction to the similarity class already yields promising performance over diverse applications, extending to and allowing for more flexible nonlinear transformations may lead to further improvement, especially in tracking and addressing local discrepancies associated with particular cell types. We describe two possible nonlinear extensions of SMAI in the Supplementary Notes. 
Second, the current method only makes use of the overlapping features in both datasets. However, in many other applications, such as integrative analysis of single-cell DNA copy number variation data and single-cell RNA-seq data \cite{zachariadis2020highly,yu2023scone}, the features are related but not shared in general. To achieve integration across different modalities, one could first embed multimodal data into a common lower-dimensional embedding space using, for example, MOFA+\cite{argelaguet2020mofa+}, and then apply SMAI to the embedded data with common features.} Third, although the current framework mainly concerns testing and aligning two datasets, direct extension to multiple datasets is available, which is achieved by applying SMAI in a sequential manner, based on some pre-specified order for integrating the datasets. Moreover, we point out that given the nature of our alignability test and the spectral alignment algorithm, extension to the simultaneous (non-sequential) testing and alignment of multiple datasets may be achieved by replacing the Procrustes analysis objective in (\ref{align.opt}) with a generalized Procrustes analysis objective involving all the available datasets and multiple alignment functions \cite{goodall1991procrustes,dryden2016statistical}. Expanding SMAI's scope of applications along these lines are interesting direction for future work.

\section*{Materials and Methods}

For clarity, here we introduce SMAI-test and SMAI-align along with their theoretical properties, focusing on the global alignment of two datasets with the same sample size. Extensions to aligning datasets with unequal sample sizes and SMAI for partial alignment, can be achieved by slightly modifying these basic algorithms, whose details are provided in the Supplementary Notes. %In later sections, we show that the complete version of SMAI, which flexibly allows for partial alignment, as used in our data analysis, can be obtained with slight modifications of these basic algorithms.

\subsection*{SMAI-test algorithm}

The basic SMAI-test algorithm requires as input the data matrices $\bX\in\R^{d\times n}$ and $\bY\in\R^{d\times n}$, each containing $n$ cells and $d$ features (genes), a pre-determined parameter $r_{\max}$ corresponding to the number of leading eigenvalues to be used in the test, and a global rescaling factor $b>0$. In principle, the number $r_{\max}$ should reflect the dimension of the underlying true signal structure of interest, whereas the global scaling factor $b>0$ adjusts for the potential scaling difference between the two matrices. Estimators of $r_{\max}$ and $b$ are discussed in the Supplementary Notes. The formal procedure of the test is summarized as follows:
\begin{enumerate}
\item {\bf Compute and normalize eigenvalues:} Define the centered matrices $\bar\bX=\bX({\bf I-11}^\top)$ and $\bar\bY=\bY({\bf I-11}^\top)$. Let $\lambda^{(1)}_1\ge \lambda^{(1)}_2\ge ...\ge \lambda^{(1)}_{n}$ and $\lambda^{(2)}_1\ge \lambda^{(2)}_2\ge ...\ge \lambda^{(2)}_{n}$ be the ordered eigenvalues of matrices $d^{-1}\bar\bX^\top\bar\bX$ and $d^{-1}\bar\bY^\top\bar\bY$, respectively. Update $\lambda_i^{(1)}\leftarrow b\lambda_i^{(1)}$ for all $i=1,2,..., n$.
\item {\bf Construct test statistic:}  (i) define $\alpha_i^{(\ell)}=\big(\frac{1-n/d}{\lambda_i^{(\ell)}}+\frac{1}{d}\sum_{j=r_{\max}+1}^{d}\frac{1}{\lambda_j^{(\ell)}-\lambda_i^{(\ell)}} \big)^{-2},$ and $\phi_i^{(\ell)}=\frac{1}{\sqrt{\alpha_i^{(\ell)}}}\big( -\frac{1-n/d}{(\lambda_i^{(\ell)})^2}+\frac{1}{d}\sum_{j=r_{\max}+1}^d\frac{1}{(\lambda_j^{(\ell)}-\lambda_i^{(\ell)})^2}\big)^{-1}$
for $\ell=1,2$ and $i=1, 2, ..., r_{\max}$.

(ii)  define the test statistic
$
T_n = \sum_{i=1}^{r_{\max}} \frac{{d}(\lambda^{(1)}_i-\lambda^{(2)}_i)^2}{{2\alpha_i^{(1)}\phi_i^{(1)}+2\alpha_i^{(2)}\phi_i^{(2)}}}.
$
\item {\bf Obtain p-value:} the p-value is defined as $1-F^{-1}(T_n; r_{\max})$, where $F(\cdot; s)$ is the cumulative distribution function of the $\chi^2$ random variable with degree of freedom $s$.
\end{enumerate}

\subsection*{Theoretical guarantee of SMAI-test} 

We first introduce our assumption on the centered matrices $\bar\bX$ and $\bar\bY$, as well as the formal null hypothesis, based on which our testing procedure is developed. Suppose $\bar\bX = \bZ_1\bSig_1^{1/2}$ and $\bar\bY= \bZ_2\bSig_2^{1/2}$, where $\bSig_1, \bSig_2\in\R^{n\times n}$ are positive definite matrices, and $\bZ_1,\bZ_2$ are independent copies of random matrix $\bZ=(z_{ij})\in\R^{d\times n}$ with entries $z_{ij}=d^{-1/2}q_{ij}$ where the double array $\{q_{ij}: i=1,2,...,d, j=1,2,...,n\}$ consists of independent and identically distributed random variables whose first four moments match those of a standard normal random variable. We assume each of $\bSig_1$ and $\bSig_2$ has $r$ spiked/outlier eigenvalues, and the remaining bulk eigenvalues have some limiting spectral distribution. In other words, for each  $\bSig_\ell, \ell=1,2,$ there are  exactly $r$ eigenvalues $\theta^{(\ell)}_1\ge \theta^{(\ell)}_2\ge ...\ge \theta^{(\ell)}_r$ larger than a certain threshold, characterizing the dominant global signal structures in the data, whereas for the rest of the eigenvalues $\theta^{(\ell)}_{r+1}\ge \theta^{(\ell)}_2\ge ...\ge \theta^{(\ell)}_n$ characterizing the remaining signal structures of much smaller magnitude, their empirical distribution (i.e., histogram) has the same deterministic limit as $n\to\infty$. The precise statements of these conditions are given as assumptions (A1)-(A3) in the Supplementary Notes. Moreover, to account for the high dimensionality of the datasets, we assume that the number of genes is comparable to the number of cells, in the sense that $n/d\to c\in(0,\infty)$ as $n\to\infty$. The above model is commonly referred to as the high-dimensional generalized spiked population model \cite{bai2012sample,li2020asymptotic,cai2020limiting,zhang2022asymptotic}, which is widely used for modeling high-dimensional noisy datasets with certain  low-dimensional signal structures. In particular, the key assumption about the spiked eigenvalue structure can be empirically verified in our real single-cell datasets (Figure S1 and references \cite{landa2022biwhitening,landa2023dyson}).  Essentially, such a model ensures the existence and statistical regularity of some underlying low-dimensional signal structure. For instance, it implies that when the signal strength of the low-dimensional structure is strong enough, that is, when top eigenvalues $\{\theta_i^{(\ell)}\}_{1\le i\le r}$ are large enough, the underlying low-dimensional signal structure would be roughly captured by the leading $r$ eigenvalues and eigenvectors of $d^{-1}\bar\bX^\top\bar\bX$ and $d^{-1}\bar\bY^\top\bar\bY$. Unlike many earlier works \cite{johnstone2001distribution,baik2006eigenvalues,paul2007asymptotics,bao2022statistical}, where the bulk eigenvalues $\{\theta^{(\ell)}_i\}_{i>r}$ are assumed to be identical or all ones, the current framework allows for more flexibility as to the possible heterogeneity in the signal and/or noise structures. 

Suppose the eigendecompositions of $\bSig_1$ and $\bSig_2$ are expressed as 
$
\bSig_{\ell}= \bU_\ell\bTheta_\ell\bU_\ell^{\top}, \ell=1,2,
$
where $\bTheta=\text{diag}(\theta_1^{(\ell)}, \theta_2^{(\ell)}, ..., \theta_n^{(\ell)}\}$ is a diagonal matrix containing the ordered eigenvalues, and the columns of $\bU_{\ell}=[\bu^{(\ell)}_{1}, \bu^{(\ell)}_{2}, ..., \bu^{(\ell)}_{n}]$ are the corresponding eigenvectors. By definition, the (centered) low-dimensional structure associated with $\bar\bX$ or $\bar\bY$ can be represented by the leading $r$ eigenvectors of $\bSig_\ell$ weighted by the square root of their corresponding eigenvalues, that is, 
$$
\bL_\ell=\begin{bmatrix}
	\sqrt{\theta_1^{(\ell)}}\cdot\bu^{(\ell)}_{1} & \sqrt{\theta_2^{(\ell)}}\cdot\bu^{(\ell)}_{2}& ...& \sqrt{\theta_r^{(\ell)}}\cdot\bu^{(\ell)}_{r}
\end{bmatrix}\in\R^{n\times r},
$$
where $\ell=1,2.$
With these, the null hypothesis about the general alignability between $\bX$ and $\bY$ can be formulated as the alignability between their low-dimensional structures $\bL_1$ and $\bL_2$ up to a possible rescaling and a rotation (note that translation is not needed as $\bL_{\ell}$'s are already centered). Formally, the null hypothesis $H_0$ under the above statistical model states that ``there exists a rotation matrix $\bR\in\R^{n\times n}$ and a scalar $\beta>0$ satisfying $\bL_1=\beta\bR\bL_2.$"
To develop a statistical test against such a null hypothesis, we notice that under $H_0$, it necessarily follows that $\theta^{(1)}_i=\beta\theta^{(2)}_i$ for all $1\le i\le r$. As a result, it suffices to develop a statistical test that can  evaluate if the spiked eigenvalues of $\bSig_1$ and $\bSig_2$ are identical up to a global scaling factor. This leads us to the proposed test. The next theorem, proved in the Supplementary Notes, theoretically  justifies the proposed test and ensures its statistical validity in terms of type I errors.

\begin{thm} \label{test.thm}
	Suppose $\bar\bX$ and $\bar\bY$ are independent and satisfy the above high-dimensional generalized spiked population model. Let $p(\bar\bX,\bar\bY,r_{\max}, \beta)$ be the p-value returned by SMAI-test with $(r_{\max},b)=(r,1/\beta)$. Under the null hypothesis $H_0$, for any $0<\alpha<1/2$, it holds that
$$
	\lim_{(d,n)\to\infty} P_{H_0}(p(\bar\bX,\bar\bY,r_{\max},b)<\alpha)<\alpha.
$$
\end{thm}

%\red{Explain Procrustes analysis and the interpretation. Cite Theorem on procrustes analysis.}

\subsection*{SMAI-align algorithm}

The basic SMAI-align algorithm requires as input the normalized data matrices $\bX$ and $\bY$, the eigenvalue threshold $r_{\max}$, the maximum number of iterations $T$, and the outlier control parameter $k$. The algorithm starts with a denoising procedure during which the centered, best rank $r_{\max}$ approximations ($\bar\bX$ and $\bar\bY$) of the data matrices are obtained. Multiple ways are available to denoise a high-dimensional data matrix with low-rank signal structures \cite{nadakuditi2014optshrink,gavish2017optimal,leeb2021matrix}. To ensure both computational efficiency and theoretical guarantee, we adopt the hard-thresholding denoiser (Supplementary Notes). The second step is a robust iterative manifold matching and correspondence algorithm, motivated by the shuffled Procrustes optimization problem (\ref{align.opt}). Spectral methods that generalize the classical ideas of  shape matching and correspondence analysis in computer vision \cite{shapiro1992feature,belongie2002shape} and the theory of Procrustes analysis \cite{goodall1991procrustes,dryden2016statistical}, are used to account for high dimensions. Specifically, the algorithm alternatively searches for the best basis transformation over the sample space and the feature space. In each iteration, the following ordinary Procrustes analyses are considered in order:
\beq
\min_{\bP\in \mathbb{O}(n), \alpha\in\R}\|\bar\bX-\alpha\bR\bar\bY\bP\|_F^2,\quad\text{with $\bR$ given,}
\eeq
\beq
\min_{\bR\in \mathbb{O}(d), \beta\in\R, \bgamma\in\R^d}\|\bar\bX-\beta\bR\bar\bY\bP-\bgamma{\bf 1}_n^\top\|_F^2,\quad \text{with $\bP$ given.}
\eeq 
The first optimization looks for an orthogonal matrix $\bP\in\mathbb{O}(n)$ and a scaling factor $\alpha\in\R$ so that the data matrix $\bar\bX$ is close to the transformed data matrix $\alpha\bR\bar\bY$, subject to recombination of its samples. In the second optimization, a similarity transformation $(\beta,\bgamma,\bR)$ is obtained to minimize the discrepancy between the data matrix $\bar\bX$ and the sample-matched data matrix $\bar\bY\bP$. The matrix $\bP$ relaxes the original permutation class, allowing for more flexible sample matching between the two data matrices; $\bR$ represents the rotation needed to align the features between the two matrices.   To improve robustness against potential outliers in the data, in each iteration we remove the top $k$ outliers from both datasets, whose distances to the other dataset are the largest after alignment. $T$ and $k$ are tunable parameters, about which we find  $3\le T\le 5$ and $5\le k\le20$ work robustly for our benchmark datasets. In the last step, we determine the direction of alignment, that is, whether aligning $\bY$ to $\bX$, or aligning $\bX$ to $\bY$. In general, such directionality is less important as our similarity transformation is invertible and symmetric with respect to both datasets. As a default setting, our algorithm automatically determines the directionality  by pursuing whichever direction that leads to a smaller between-data distance. Users can also specify the preferred directionality, which can be useful in some applications such as in the prediction of unmeasured spatial genes using single-cell RNA-seq data. Additional technical details of SMAI-align are provided in the Supplementary Notes.

\subsection*{Evaluation of within-data structure preservation and batch effect removal} 

For the negative control tasks Neg1-Neg3, and the positive control tasks Pos1-Pos7, we compare the overall correlation between the pairwise distances of cells before and after alignment. Specifically, suppose $\bX\in\R^{d\times n_1}$ and $\bY\in\R^{d\times n_2}$ are the original normalized datasets, and $\bX^*\in\R^{r\times n_1}$ and $\bY^*\in\R^{r\times n_2}$ are the aligned datasets obtained by one of the integration methods, where $r$ could be different from $d$. We first obtain $\mathcal{D}(\bX)\in\R^{n_1\times n_1}$, whose entries contain pairwise Euclidean distances between the columns of $\bX$. Similarly, we obtain $\mathcal{D}(\bY), \mathcal{D}(\bX^*),$ and $\mathcal{D}(\bY^*)$. Let $[\mathcal{D}(\bX)]_{i.}$ be the $i$-th row of $\mathcal{D}(\bX)$. { We calculate Kendall's tau correlations and Spearman's rho correlations between $[\mathcal{D}(\bX)]_{i.}$ and $[\mathcal{D}(\bX^*)]_{i.}$, for all $i$'s, and then use the average correlation $\bar r_X$ across all $i$'s to quantify how well the structure within $\bX$ is preserved after integration, in $\bX^*$. Similarly, we also calculate the average correlation $\bar r_Y$ between the rows of $\mathcal{D}(\bY)$ and the rows of $\mathcal{D}(\bY^*)$, to quantify the preservation of structure within $\bY$ after integration. Then we take the mean of $\bar r_X$ and $\bar r_Y$ as the final metric for the within-data structure preservation of the method, reported as the y-axis of the scatter plots in Figures \ref{fig2}b, \ref{fig3}a  and S5.}  

Once the integrated data $(\bX^*,\bY^*)$ are obtained, we calculate the Davies-Bouldin (D-B) index %the mean silhouette index, 
	and the inverse Calinski-Harabasz (C-H) index, to quantify how well the two integrated datasets are mixed. For a given integrated dataset consisting of $K$ batches $C_1, ..., C_K$, we define 
$$
\text{D-B index} =  \frac{1}{K}\sum_{k=1}^K \max_{j\ne k}\frac{S_k+S_j}{M_{ij}},
$$
where $S_i=\big[\frac{1}{|C_k|}\sum_{i\in C_k}\|X_i-A_k\|_2^2\big]^{-1/2}$, $M_{kj}=\|A_k-A_j\|_2$,
with $A_k$ being the centroid of batch $k$ of size $|C_k|$, and $X_i$ being the $i$-th data point. 
We define 
$$
\text{inverse C-H index}=\big[ \frac{\sum_{k=1}^K\sum_{i\in C_k}\|X_i-A_k\|_2^2}{N-K} \big]/\big[\frac{\sum_{k=1}^K |C_k|\cdot\|A_k-A\|_2^2}{K-1}\big],
$$
where $A$ is the global centroid, and $N$ is the total sample size. {These metrics essentially quantify the ratio between within-batch variations and between-batch variations. }
{As a result, a method achieving better alignment quality will have a higher D-B index, %higher values of ($1-$ mean silhouette index), 
	and a higher inverse C-H index. These metrics are shown as the x-axis of the scatter plots in Figures \ref{fig3}a  and S5, {where we evaluate, for a given pair of datasets (i.e., fixing the within-batch variations), which alignment method leads to smaller between-batch variations}.} Compared with other metrics such as LISI and ARI, the D-B and inverse C-H indices are calculated from the pairwise distances,  and do not rely on pre-specified nearest neighbor graphs (LISI), or the predicted cluster labels (ARI), which may be sensitive to specific methods.

\subsection*{Differential expression analysis} 

For each of the positive control tasks Pos1-Pos6, we first obtain an integrated dataset containing all the samples, by using one of the alignment methods. Then for each cell type $k$, we identify the set of marker genes $S_k$ based on the aforementioned procedure. We use a threshold of 0.01 on the BH-adjusted p-values to select the differentially expressed genes. For each cell type $k$, we also obtain the benchmark set $S^b_k$ of marker genes, which contains all the differentially expressed genes identified based on the individual datasets before alignment. Finally, we compute the Jaccard similarity index between the set $S_k$ and the set $S_k^b$. The results are reported in Figure \ref{fig3}c.

\subsection*{Prediction of spatial genes} 

For the three spatial positive control tasks PosS1-PosS3, we withhold each gene from the spatial transcriptomic data, and predict its values based on the following procedure. In particular, we denote $\bX\in\R^{d\times n_1}$ as the spatial transcriptomic data, and $\bY\in\R^{d\times n_2}$  as the paired RNA-seq data with the same features. We also denote $\bX_{-i.}\in\R^{(d-1)\times n_1}$ as the submatrix of $\bX$ after removing the $i$-th row, and denote $\bX_{i.}\in\R^{d}$ as the $i$-th row of $\bX$. For each $1\le i\le d$, we
\begin{enumerate}
	\item Apply one of the alignment methods (LIGER, Scanorama, Seurat, fastMNN, or SMAI) to $\bX_{-i.}$ and $\bY_{-i.}$, and obtain the aligned datasets $\bX^*_{-i.}$ and $\bY^*_{-i.}$;
	\item Fit a $k$-nearest neighbor regression (with $k=5$ in all our analysis) between the predictor matrix $\bY^*_{-i.}\in\R^{(d-1)\times n_2}$ and the outcome vector $\bY_{i.}\in\R^{n_2}$;
	\item Predict the outcome vector $\widehat\bX_{i.}$ associated with the predictor matrix $\bX_{-i.}^*$ based on the above regression model.
\end{enumerate}
To evaluate the prediction accuracy, we calculate  Kendall's $\tau$ correlation between $\bX_{i.}$ and $\widehat\bX_{i.}$.

\subsection*{Cell type specific alignment using SMAI} 
	
{	To characterize possible intra-cell type variability, one can apply SMAI to each cell type separately, to  learn the cell-type specific alignment. In this way, each cell type will have its own alignment function, characterizing the intra-cell type variability across different studies or conditions. In particular, the availability of a closed-form expression of the obtained alignment function allows for quantitatively comparing the similarity between the obtained alignment functions for different cell types. We show a scatter plot, where each point represents a cell-type specific alignment function, whose coordinates demonstrate the magnitude of the associated rotation and translation, obtained by converting the obtained rotation and translation parameters into normalized metrics between 0 and 1.
	Focusing on two positive integration tasks (Tasks Pos4 and Pos5), we applied SMAI to each cell type to learn the cell-type specific alignment functions. Interestingly, for each of the tasks, we found a remarkable similarity in the obtained cell-type specific alignment functions (Figure S14c), which supports the default design of SMAI using a common global alignment across cell types. }

\subsection*{Data preprocessing} 

The raw counts data listed in Table S1 were filtered, normalized, and scaled by following the standard procedure (R functions \texttt{CreateSeuratObject}, \texttt{NormalizeData} and \texttt{ScaleData} under default settings) as incorporated in the R package \texttt{Seurat}. For  datasets with more than $2000$ genes, we also applied the R function \texttt{FindVariableFeatures} in \texttt{Seurat} to identify the top $2000$ most variable genes for subsequent analysis. 
For the human pancreatic data associated with Task Pos1, we remove the cell types containing less than 20 cells. {For the cross-tissue integration tasks (Neg3 and Pos6), a subset of 4000 cells were randomly sampled from each of the tissues for our analysis.}

\subsection*{Data Availabiliy}

{The human pancreatic data can be accessed in the R package \texttt{SeuratData} [\url{https://github.com/satijalab/seurat-data}] under the dataset name \texttt{panc8}.  The PBMC data can be accessed in the R package \texttt{SeuratData} [\url{https://github.com/satijalab/seurat-data}] under the dataset name \texttt{pbmcsca}. {The mouse brain chromatin accessibility data were downloaded from Figshare [\url{https://figshare.com/ndownloader/files/25721789}], containing a dataset from Fang et al. \cite{fang2021comprehensive} (single-nucleus ATAC-seq protocol), and a 10X Genomics dataset for fresh adult mouse brain cortex  (sample retrieved from [\url{https://support.10xgenomics.com/single-cell-atac/datasets/1.2.0/atac_v1_adult_brain_fresh_5k}]). Both ATAC-seq datasets have been preprocessed by Luecken et al \cite{luecken2022benchmarking} to characterize gene activities.} The human lung data were downloaded as Anndata objects (samples 1, A3, B3 and B4) on Figshare [\url{https://figshare.com/ndownloader/files/24539942}]. {The human liver, MLN, and LLN immune cell data were downloaded from \url{https://www.tissueimmunecellatlas.org}. The mouse PBMC datasets (samples `Control 1h' and `LPS 1h') were downloaded from Gene Expression Omnibus (GSE178431) \url{https://www.ncbi.nlm.nih.gov/geo/query/acc.cgi?acc=GSE178429}.}
	The mouse gastrulation  seqFISH data were downloaded from  \url{https://content.cruk.cam.ac.uk/jmlab/SpatialMouseAtlas2020/}, and the RNA-seq (10X Chromium) data can be accessed as ‘Sample 21’ in \texttt{MouseGastrulationData} within the R package \texttt{MouseGastrulationData}. For the mouse VISP data, the ISS spatial transcriptomic data can be downloaded from \url{https://github.com/spacetx-spacejam/data}, the ExSeq spatial transcriptomic data can be downloaded from \url{https://github.com/spacetx-spacejam/data}, and the Smart-seq data can be downloaded from \url{https://portal.brain-map.org/atlases-and-data/rnaseq/mouse-v1-and-alm-smart-seq}.}
%Source data are provided with this paper.

\subsection*{Code Availabiliy}

The R and Python packages of SMAI, and the R codes for reproducing our simulations and data analyses, are available at our GitHub repository \url{https://github.com/rongstat/SMAI}.

\section*{Acknowledgment} 

We would like to thank the editors and two referees for their helpful comments and suggestions, which significantly improved the quality and presentation of the paper.

	%\copyrightnotice
% Bibliography
%-----------------------------------------------------------------
\bibliographystyle{abbrv}
\bibliography{reference}

\appendix

	\section{Supplementary Notes}

			\subsection{Advantages of SMAI-test as suggested by statistical theory} 
		
		A few remarks are in order about the advantages of SMAI-test implied by our theory. First, unlike many classical statistical tests for shape conformity, in which $d$ is assumed to be fixed (effectively a small integer) as $n\to\infty$, the statistical validity of the proposed test is guaranteed (Theorem \ref{test.thm}) under the high-dimensional asymptotic regime, where $d$ is comparable to $n$ as $n\to\infty$. This makes our testing framework more reliable and coherent with the high-dimensional single-cell omic data. Second, SMAI-test is developed based on the spiked population model, which essentially assumes a few large (``spiked") eigenvalues encoding the dominating low-dimensional signal structure, followed by a large number of smaller bulk eigenvalues. As mentioned earlier, such a spectral property has been widely observed in single-cell omic data, and allows for straightforward empirical verification (Figure \ref{eigs.fig}). Third, in contrast with many existing tests whose validity relies on rather rigid signal/noise regularity (e.g., identical bulk eigenvalues), the proposed test incorporates a self-calibration scheme in constructing the test statistic $T_n$, allowing for general bulk eigenvalues and is thus applicable to a wider range of real-world settings. Finally, the proposed alignability test is data-driven and does not require side information such as cell type, tissues, or species identities. This ensures an unbiased characterization of the similarity between datasets according to their actual geometric measures rather than potentially imprecise annotations.

	\subsection{SMAI-align algorithm: technical details} 
	
	The basic SMAI-align method is summarized in Algorithm \ref{alg.smai}. In particular, we denote by $\bA_{\mathcal{M}}$ the submatrix of $\bA$ containing the columns indexed by the elements in set $\mathcal{M}$, and denote by ${\bf 1}_{|\mathcal{M}|}$ the vector containing all ones whose dimension is the same as the cardinality of set $\mathcal{M}$. The algorithm requires as input the normalized data matrices $\bX$ and $\bY$, the eigenvalue threshold $r_{\max}$, the maximum number of iterations $T$, and the outlier control parameter $k$. The algorithm starts with a denoising procedure during which the best rank $r_{\max}$ approximations ($\widehat\bX$ and $\widehat\bY$) of the data matrices are obtained. Like the determination of $r_{\max}$, multiple ways are available to denoise a high-dimensional data matrix with low-rank signal structures \cite{nadakuditi2014optshrink,gavish2017optimal,leeb2021matrix}. To ensure both computational efficiency and theoretical guarantee, we adopt the hard-thresholding denoiser, summarized in Algorithm \ref{alg.denoise}. The second step is a robust iterative manifold matching and correspondence algorithm, motivated by the shuffled Procrustes optimization problem (1) in the main text. The use of spectral methods in Steps 2(ii) and 2(iii) generalizes the classical ideas of  shape matching and correspondence analysis in computer vision \cite{shapiro1992feature,belongie2002shape} and the theory of Procrustes analysis (Theorem \ref{opa.thm}) to account for high dimensions.   To improve robustness against potential outliers in the data, in each iteration we remove the top $k$ outliers from both datasets, whose distances to the other dataset are the largest after alignment. $T$ and $k$ are tunable parameters, about which we find  $3\le T\le 5$ and $5\le k\le20$ work robustly for our benchmark datasets. In the last step, we determine the direction of alignment, that is, whether aligning $\bY$ to $\bX$, or aligning $\bX$ to $\bY$. In general, such directionality is less important as our similarity transformation is invertible and symmetric with respect to both datasets. As a default setting, our algorithm automatically determines the directionality in Step 3, by pursuing whichever direction that leads to a smaller between-data distance. Users can also specify the preferred directionality, which can be useful in some applications such as in the prediction of unmeasured spatial genes using single-cell RNA-seq data.

	\begin{algorithm} 
		\caption{Basic SMAI-align algorithm} \label{alg.smai}
		\begin{algorithmic}
			\State {\bf Input:} Data matrices $\bX = [\xb_1\quad \xb_2\quad ... \quad \xb_{n}]\in\R^{d\times n}$ and $\bY = [\yb_1\quad \yb_2 \quad ... \quad \yb_{n}]\in\R^{d\times n}$, eigenvalue threshold $r_{\max}>0$, maximum iteration number $T$, outlier control parameter $k$. %\james{Is $n_1$ $n$? Clarify that the data here are not matched.}
			\State 1. {\bf Denoise:} Obtain the denoised data matrices
			$
			\widehat{\bX} = \mathcal{D}(\bX; r_{\max}), \widehat{\bY} = \mathcal{D}(\bY; r_{\max}),
			$
			where $\mathcal{D}(\cdot; r)$ is the denoising function defined in Algorithm \ref{alg.denoise}, or any other suitable denoiser. 
			\State 2. {\bf Robust iterative Procrustes analysis:} Initialize  $\mathcal{M}_1=\mathcal{M}_2=\{1,2,...,n\}$ and $\bR={\bf I}_d$. For $t=1, 2, ..., T$, do
			\State \hspace{4mm} (i) Centering: obtain centered matrices $\bar\bX_{\mathcal{M}_1} \leftarrow \widehat\bX_{\mathcal{M}_1}({\bf I}-{\bf 1 1}^\top)$ and $\bar\bY_{\mathcal{M}_2}\leftarrow \widehat\bY_{\mathcal{M}_2}({\bf I}-{\bf 11 }^\top)$, where ${\bf 1}$ is the vector of all ones.
			\State \hspace{4mm} (ii) Sample space transform: %update $\bar\bX \leftarrow \bar\bX\bQ$ where
			set $
			\bP \leftarrow \sum_{i=1}^{r_{\max}}\bfm_i\bn_i^\top\in\R^{|\mathcal{M}_1|\times |\mathcal{M}_2|}
			$
			where $\bfm_i$ and $\bn_i$ are the $i$-th left and right singular vectors of
			$\bar\bX_{\mathcal{M}_1}^\top\bR\bar\bY_{\mathcal{M}_2}$.
			\State \hspace{4mm} (iii) Feature space transform: update $\bR\leftarrow\sum_{i=1}^{r_{\max}}\balpha_i\bbeta^\top_i\in\R^{d\times d}$ where $\balpha_i$ and $\bbeta_i$ are the $i$-th left and right singular vectors of $\bar\bX_{\mathcal{M}_1}\bP^\top\bar\bY_{\mathcal{M}_2}^\top$.
			\State \hspace{4mm} (iv) Outlier removal: identify $
			S_1\subset  \mathcal{M}_1$ such that the residuals $\{\| [\bar\bX_{\mathcal{M}_1}]_{i.}-[\beta\bR\bar\bY_{\mathcal{M}_2}\bP]_{i.}\|_2\}_{i\in S_1}$ are the $k$ largest ones among all the samples in $\mathcal{M}_1$. Similarly, identify $S_2\subset  \mathcal{M}_2$ such that the residuals $\{\| [\bR^\top\bar\bX_{\mathcal{M}_1}\bP^\top]_{i.}-[\beta\bar\bY_{\mathcal{M}_2}]_{i.}\|_2\}_{i\in S_2}$ are the $k$ largest ones among all the samples in $\mathcal{M}_2$.
			Then update $\mathcal{M}_1\leftarrow \mathcal{M}_1\setminus S_1$ and $\mathcal{M}_2\leftarrow \mathcal{M}_2\setminus S_2$. 
			\State 3. {\bf Obtain best similarity transformation and aligned datasets:}  Define {$\beta=\frac{\text{tr}(\bar\bX_{\mathcal{M}_1}\bP^\top\bar\bY_{\mathcal{M}_1}^\top\bR)}{\|\bar\bY_{\mathcal{M}_2}\bP\|_F^2}$}. If 
			$$
			\sum_{i\in\mathcal{M}_1}\| [\bar\bX_{\mathcal{M}_1}]_{i.}-[\beta\bR\bar\bY_{\mathcal{M}_2}\bP]_{i.}\|_2
			<\sum_{i\in\mathcal{M}_2}\| [\bR^\top\bar\bX_{\mathcal{M}_1}\bP^\top]_{i.}-[\beta\bar\bY_{\mathcal{M}_2}]_{i.}\|_2,
			$$
			then define $\bX^*=\bX$, $\bY^*=\beta\bR\bY({\bf I}-{\bf 11}^\top)+\bar\bX_{\mathcal{M}_1}{\bf 1}_{|\mathcal{M}_1|}{\bf 1}_{n}^\top$, and $(\widehat\beta,\widehat\bgamma,\widehat\bR)=(\beta, \bar\bX_{\mathcal{M}_1}{\bf 1}_{|\mathcal{M}_1|},\bR)$; otherwise, define $\bX^*=\beta^{-1}\bR^\top\bX({\bf I}-{\bf 11}^\top)+\bar\bY_{\mathcal{M}_2}{\bf 1}_{|\mathcal{M}_2|}{\bf 1}_{n}^\top$, $\bY^*=\bY$, and $(\widehat\beta,\widehat\bgamma,\widehat\bR)=(\beta^{-1}, \bar\bY_{\mathcal{M}_2}{\bf 1}_{|\mathcal{M}_2|}, \bR^\top)$.
			\State {\bf Output:} The best similarity transformation $(\widehat\beta,\widehat\bgamma, \widehat\bR)$ and the aligned data $\{\bX^*,\bY^*\}$.
		\end{algorithmic}
	\end{algorithm}	
	
	\begin{algorithm} 
		\caption{Hard-thresholding denoiser} \label{alg.denoise}
		\begin{algorithmic}
			\State {\bf Input:} Data matrix $\bX\in\R^{d\times n}$, eigenvalue threshold $r_{\max}>0$.
			\State 1. {\bf Singular value decomposition:} Obtain the singular value decomposition of $\bX$ where $\lambda_1\ge \lambda_2\ge ...\lambda_r$, $r=\min\{n,d\}$ are the singular values, and $\{\bu_i\}_{1\le i\le r}$ and $\{\bv_i\}_{1\le i\le r}$ are the associated left and right singular vectors. 
			\State 2. {\bf Signal reconstruction:} define the denoising map $\mathcal{D}: \R^{d\times n}\times \mathbb{N}^+ \mapsto \R^{d\times n}$ as
			$
			\mathcal{D}(\bX; r_{\max}) = \sum_{i=1}^{r_{\max}} \lambda_i\bu_i\bv_i^\top.
			$
			\State {\bf Output:} the denoised matrix $\widehat\bX=\mathcal{D}(\bX; r_{\max}) $.
		\end{algorithmic}
	\end{algorithm}

	\subsection{Rationale and geometric interpretation of SMAI-align} 
	
	The key step of SMAI-align is an iterative manifold matching and correspondence algorithm. The algorithm alternatively searches for the best basis transformation over the sample space and the feature space. Specifically, in each iteration, the following ordinary Procrustes analyses are considered in order:
	\beq
	\min_{\bP\in \mathbb{O}(n), \alpha\in\R}\|\bar\bX_{\mathcal{M}_1}-\alpha\bR\bar\bY_{\mathcal{M}_2}\bP\|_F^2,\quad\text{with $\bR$ given,}
	\eeq
	\beq
	\min_{\bR\in \mathbb{O}(d), \beta\in\R, \bgamma\in\R^d}\|\bar\bX_{\mathcal{M}_1}-\beta\bR\bar\bY_{\mathcal{M}_2}\bP-\bgamma{\bf 1}_n^\top\|_F^2,\quad \text{with $\bP$ given.}
	\eeq 
	The first optimization problem looks for an orthogonal matrix $\bP\in\mathbb{O}(n)$ and a scaling factor $\alpha\in\R$ so that the data matrix $\bar\bX_{\mathcal{M}_1}$ is close to the rescaled data matrix $\alpha\bR\bar\bY_{\mathcal{M}_2}$, subject to recombination of its samples. In the second optimization problem, a similarity transformation $(\beta,\bgamma,\bR)$ is obtained to minimize the discrepancy between the data matrix $\bar\bX_{\mathcal{M}_1}$ and the sample-matched data matrix $\bar\bY_{\mathcal{M}_2}\bP$. Intuitively, $\bP$ can be considered as a relaxation of permutation matrices allowing for general linear combinations for sample matching between the two data matrices, which may account for differences in sample distributions, such as proportions of different cell types, between the two datasets; $\bR$ represents the rotation needed to align the features between the two matrices. 
	The above optimization problems admit closed-form solutions, which only require a singular value decomposition of some product matrix. The following theorem summarizes some mathematical facts from the classical Procrustes analysis \cite{dryden2016statistical}.
	
	\begin{thm} \label{opa.thm}
		Let $\bX_1,  \bX_2\in\R^{p\times q}$ be any matrices, whose column-centered counterparts $\bar\bX_1,\bar\bX_2\in\R^{p\times q}$ are such that ${\bf 1}_p^\top\bar\bX_\ell=0$ for $\ell=1,2$. 
		\begin{enumerate}
			\item Let $(\widehat\alpha, \widehat\bGam)$ be the solution to the optimization problem 
			\beq
			\min_{\bGam\in \mathbb{O}(q), \alpha\in\R}\|\bX_{2}-\alpha\bX_1\bGam\|_F^2.
			\eeq
			Then it holds that $\widehat\bGam=\sum_{i=1}^q\bu_i\bv_i^\top$, where $\{\bu_i\}_{1\le i\le q}$ and $\{\bv_i\}_{1\le i\le q}$ are the ordered left and right singular vectors of $\bX_2^\top\bX_1$, respectively.
			\item Let $(\widehat\beta,\widehat\bgamma,\widehat\bR)$ be the solution to the optimization problem
			\beq
			\min_{\bR\in \mathbb{O}(q), \beta\in\R, \gamma\in\R^d}\|\bar\bX_{2}-\beta\bar\bX_1\bR-{\bf 1}_p\bgamma^\top\|_F^2.
			\eeq 
			Then it holds that
			\beq
			\widehat\bgamma=0, \qquad \widehat\bR=\sum_{i=1}^q\bw_i\bs_i^\top,\qquad \widehat\beta=\frac{\textup{tr}(\bar\bX_2^\top\bar\bX_1\widehat\bR)}{\|\bar\bX_1\|_F^2},
			\eeq
			where $\{\bw_i\}_{1\le i\le q}$ and $\{\bs_i\}_{1\le i\le q}$ are the ordered left and right singular vectors of $\bar\bX_2^\top\bar\bX_1$, respectively.
		\end{enumerate}
	\end{thm}
	
	In particular, Part 1 of Theorem \ref{opa.thm} explains the rationale behind Steps 2(ii) of Algorithm \ref{alg.smai}, whereas Part 2 of Theorem \ref{opa.thm} justifies Step 2(iii), and the form of the estimator for $\beta$ in Step 3 of Algorithm \ref{alg.smai}. Note that in Algorithm \ref{alg.smai} we slightly abuse the notation so that $\widehat\bgamma$ is defined to account for the global mean shift between $\bX$ and $\bY$, rather than between $\bar\bX$ and $\bar\bY$ as in Theorem \ref{opa.thm}. In both Steps 2(ii) and 2(iii), to adjust for high dimensionality and reduce noise, only the leading singular vectors are used to reconstruct the transformation matrices $\bP$ and $\bR$.

	\subsection{Simulation I: empirical consistency and statistical validity of SMAI}
	
	To verify the statistical validity and consistency of SMAI, we generate two families of noisy datasets, each containing a distinct low-dimensional structure as its underlying true signal. Under each setting, a pair of (alignable) datasets are generated whose underlying low-dimensional structures are identical up to a similarity transformation. More specifically, for a given sample size $n$, we generate $d$-dimensional noisy data matrix $\bY\in\R^{d\times n}$ from the signal-plus-noise model $\bY=\bY^*+\bZ$, where the columns of $\bY^*$ are the underlying noiseless samples (signals), and $\bZ$ is the random noise matrix. The signal matrix $\bY^*$ are generated such that its columns $\{\bY^*_i\}_{1\le i\le n}$ are sampled from some low-dimensional structure isometrically embedded in the $d$-dimensional Euclidean space. Each low-dimensional structure lies in some $r$-dimensional linear subspace, and is subject to an arbitrary rotation in $\R^d$, so that $\{\bY^*_i\}_{1\le i\le n}$ are generally $d$-dimensional vectors with dense (nonzero) coordinates. The entries of $\bZ$ are generated independently from a standard Gaussian distribution $N(0,1)$. Similarly, we generate another noisy data matrix $\bX\in\R^{d\times n}$ from $\bX=\beta^*\bR^*\bY^*(\bP^*)^\top+(\bgamma^*){\bf 1}_n^\top+\bW$, for some fixed scaling factor $\beta^*\in\R$, mean shift vector $\bgamma^*\in\R^d$, permutation matrix $\bP^*\in\R^{n\times n}$, rotation matrix $\bR^*\in\R^{d\times d}$, and random noise matrix $\bW\in\R^{d\times n}$, containing independent standard Gaussian entries. Observe that both datasets $\bX$ and $\bY$ are based on the same low-dimensional noiseless signal matrix $\bY^*$. In this way, we have simulated noisy datasets $\bX$ and $\bY$, each having  an intrinsically $r$-dimensional structure, that are identical up to a similarity transformation encoded by the parameters $(\beta^*,\bgamma^*,\bR^*)$. Note that there is an one-to-one correspondence between $(\beta^*,\bgamma^*,\bR^*)$ and the parameters in the optimization (\ref{align.opt}). We consider a smiley face with $r=2$ (Figure \ref{sup.fig2}a) and a Swiss roll manifold with $r=3$ as the true signal structures (Figure \ref{sup.fig2}b), where the signals $\{\bY_i^*\}_{1\le i\le n}$ are sampled uniformly from each object. %isometrically embedded in $\R^d$ and subject to an arbitrary rotation.

For SMAI-align, we assess the quantitative errors ({Methods}) of the SMAI-inferred similarity transformation parameters $(\widehat\beta,\widehat\bgamma,\widehat\bR)$, obtained from Algorithm \ref{alg.smai}, as compared with the underlying true transformation parameters $(\beta^*,\bgamma^*,\bR^*)$, respectively, for various $n$. For SMAI-test, we evaluate the empirical type I errors against the above null model at the $\alpha=0.05$ level. Each setting is repeated for 300 times to obtain the empirical evaluation. We observe that for both signal structures the estimation errors decrease as $n$ increases (Figure \ref{sup.fig2}c), suggesting the consistency of the SMAI estimators $(\widehat\beta,\widehat\bgamma,\widehat\bR)$ in the large-sample limit. In addition, the empirical type I errors (Table \ref{table:t1e}) are close to the nominal level $\alpha=0.05$ across all the settings, demonstrating the statistical validity of the test.
% [Convergence of rotation/translation/scaling estimators; type I error] In particular, we are not aware of any existing methods directly applicable the current setting; for example, the existing methods based on shape analysis have not take into account the disordered/unmatched samples, and are susceptible to high dimensionality \cite{dryden2016statistical}.

\subsection{Simulation II: consistency and type I errors} 

We generate data matrices $\bX$ and $\bY$ following the simulation setup described in the Supplementary Notes. Under each round of the simulation, we apply SMAI-align to obtain estimates of the underlying similarity transformation, as parametrized by $(\beta^*,\bgamma^*, \bR^*)$. The estimated parameters $(\widehat\beta,\widehat\bgamma,\widehat\bR)$ as returned from Algorithm \ref{alg.smai}, up to possible inversion of the similarity transformation, are evaluated based on the following loss functions
\beq \label{loss.dist}
\text{Error}_{\beta}=\big|1-{\beta^*}/{\widehat\beta}\big|,
\qquad \text{Error}_{\bgamma}=\frac{\|\bgamma^*-\widehat\bgamma\|_2}{\|\bgamma^*\|_2},\qquad \text{Error}_{\bR}=1-\frac{1}{d} \text{tr} (\widehat\bR^\top\bR^*).
\eeq
The estimation errors under various sample sizes and signal structures are shown in Figure \ref{sup.fig2}c. Moreover, these functions are also adapted to characterize the magnitude of the estimated batch effects, by comparing with the identify map (Figure \ref{sup.fig5d}).

To evaluate the empirical type I error of SMAI-test, we apply SMAI-test to the simulated datasets sharing the same low-dimensional structure but subject to some global similarity transformation. We reject the null hypothesis whenever the p-value is less than 0.05. The empirical type I error is defined as the proportion of false rejections under each setting (Table \ref{table:t1e}).

\subsection{Determine the rank parameter $r_{\max}$} 

In both SMAI-test and SMAI-align, an important parameter to be specified is $r_{\max}$, the minimal number of leading eigenvalues and eigenvectors expected to capture all of the signal structures. In the statistical literature, under the spiked population model,  various methods have been developed to consistently estimate this value in a data-driven manner \cite{kritchman2009non,ke2023estimation,donoho2023screenot}. In our software implementation, we provide two options for estimating $r_{\max}$. The simpler and computationally more effective approach is to consider
\beq\label{rmax}
r_{\max} = \max\big\{ r: 1\le r\le n,  {\lambda^{(\ell)}_{r}}/{\lambda^{(\ell)}_{r+1}}>1+{\sf c} \text{ for } \ell=1,2 \big\},
\eeq
for some small constant ${\sf c}>0$ such as $0.01$. Theoretical justifications of such a method have been established based on standard concentration argument \cite{yao2015sample}.  A more recent approach, ScreeNOT \cite{donoho2023screenot}, is also implemented as an alternative; it is theoretically more appealing but requires additional computational efforts.

\subsection{Unequal sample sizes} 

The basic versions of SMAI-test and SMAI-align can be easily adjusted to allow for unequal sample sizes. For example, if $\bX\in\R^{d\times n_1}$ and $\bY\in\R^{d\times n_2}$ where, say, $n_1> n_2$, then one can simply randomly subsample the columns of $\bX$ to obtain $\tilde\bX\in \R^{d\times n_2}$ as an input in place of $\bX$. One advantage of the current framework is the generalizability of the obtained similarity transformation. Specifically, in Algorithm \ref{alg.smai}, the similarity transform $(\widehat\beta,\widehat\bgamma,\widehat\bR)$, obtained based on subsamples $\tilde\bX$ and $\bY$, can still be applied to all the samples, leading to the aligned data $\bX^*\in\R^{d\times n_1}$ and $\bY^*\in\R^{d\times n_2}$ containing all the samples. In other words, the subsampling step does not reduce the total sample size of the final output.

	\subsection{Partial alignability testing and partial alignment}  \label{part.sec}

A sample splitting procedure is incorporated into SMAI-test, so that one can determine if there exist subsets containing $t\%$ of the samples in each dataset, that are alignable up to some similarity transformation.   The main idea is to first identify subsets of both datasets, referred to as the maximal correspondence subsets, each containing about $t\%$ of the samples, expected to contain more similar structures, and then perform SMAI-test on these identified subsets, to determine their alignability. Specifically, the samples in $\bX$ and $\bY$ are first split randomly into two parts $\{\bX_1,\bX_2\}\subset\R^{d\times (n_1/2)}$ and $\{\bY_1,\bY_2\}\subset\R^{d\times (n_2/2)}$, each containing half of the samples. Next, we identify subsets of $\bX_1$ and $\bY_1$ containing $t\%$ of their samples, that are structurally more similar. This is achieved based on the independent datasets $\bX_2$ and $\bY_2$ with the following three steps: 
\begin{enumerate}
	\item Obtain an initial alignment between $\bX_2$ and $\bY_2$ using SMAI-align or any existing methods;
	\item Identify submatrices $\bX^S_2$ of $\bX_2$ and $\bY^S_2$ of $\bY_2$, each containing $t\%$ of their samples, so that $\bX^S_2$ and $\bY^S_2$ have the highest structural similarity; this can be achieved, for example, by searching for the mutual nearest neighbors \cite{haghverdi2018batch} between the aligned datasets;  
	\item Identify submatrices $\bX_1^S$ of $\bX_1$ and $\bY_1^S$ of $\bY_1$, each containing $t\%$ of their samples, which are  nearest neighbors of the samples in $\bX_2^S$ and $\bY_2^S$, respectively. 
\end{enumerate}
With these, the partial alignability between $\bX$ and $\bY$ can be determined by applying SMAI-test to the submatrices $\bX_1^S$ and $\bY_1^S$. Importantly, by introducing the sample-splitting, the subsequent hypothesis testing is based on the samples independent from those used in the prior subset selection step, therefore avoiding selection bias. As a result, by first conditioning on the subsets $\{\bX_2, \bY_2\}$ and then integrating out its uncertainty, it can be shown that the statistical validity (controlled type I error) of the final test results still hold as in Theorem \ref{test.thm}.  For practical use, we recommend setting $t$ between 50 and 70 to simultaneously retain robustness against local heterogeneity and ensure statistical power with sufficient sample size. Moreover, one should be aware that there is a tradeoff between the threshold $t$ and the associated p-value:  when $t$ is below the true proportion of the alignable samples, the p-values are usually not significant as the null hypothesis about partial alignability remains true; when $t$ becomes larger than that, the null hypothesis will be mostly rejected. This relationship can be seen empirically in Figure S15.

In line with the partial alignability test, in our software implementation, we also allow restricting SMAI-align to $\bX_2^S$ and $\bY_2^S$. In this way, the final alignment is essentially achieved on the maximal correspondence subsets rather than the whole set, making the alignment function less sensitive to local structural heterogeneity.

%\red{Comments on Figure \ref{sup.fig5b}.}

\subsection{Simulation III: DE gene detection} 

To strengthen our argument on the advantage of SMAI in improving downstream DE analysis, we also carry out simulation studies by generating pairs of datasets, each containing 2000 cells of 12 different cell types (clusters) and expression levels of 1000 genes, with different cell type proportions. We create batch effects between the two datasets mimicking those observed in real datasets. More specifically, we generate datasets $\bX\in\R^{d\times n}$ and $\bY\in\R^{d\times n}$, with $n=2100$ and $p=1000$, based on a Gaussian mixture model with variance $\sigma^2=1$ containing 12 clusters. Specifically, we generate dataset $\bX$, so that each of its clusters has a mean vector, each supported on some 20 features with the same value $\rho$. In other words, for each of the 12 clusters (cell types), there are 20 unique marker genes with nonzero mean expression levels. Moreover, we consider unbalanced class proportions for $\bX$, by setting the cluster proportion as $(0.18, 0.16, 0.15, 0.13, 0.11, 0.10, 0.03, 0.03,$ $ 0.03, 0.03, 0.03, 0.02).$ To generate $\bY$, we first generate a dataset $\bY_0\in\R^{d\times n}$ based on the above Gaussian mixture model, but with a different cluster proportion $(0.05, 0.05, 0.06, 0.07, 0.07, 0.08, 0.09,$ $ 0.09, 0.10, 0.11, 0.11, 0.12)$. Then we obtain $\bY$ by applying a similarity transform of $\bY_0$. In particular, we set $\bY=\beta\bR\bY_0+\bgamma{\bf 1}_n^\top$, where $\beta=2$, $\bgamma\in\R^d$ has random entries uniformly sampled from the interval $[0,\rho/7]$, and $\bR\in\R^{d\times d}$ is a rotation matrix that slightly rotate the marker genes associated with Cluster 1.

To evaluate the performance of various alignment methods for assisting detection of marker genes, we apply each of them to obtain an integrated dataset $\bD\in\R^{d\times 2n}$, concatenating the two datasets after alignment. Then, for each cluster, we perform a one-sided two-sample t-test on each gene, and determine the set of marker genes by selecting those genes whose p-values adjusted by the Benjamini-Hochberg procedure are below 0.01.  Finally, for each integration method, for each cluster, we evaluate the agreement between the set of identified marker genes and the set of true marker genes. Specifically, we use the Jaccard similarity  index
$$
\text{Jaccard index}(S_1, S_2)=\frac{|S_1\cap S_2|}{|S_1\cup S_2|},
$$
where $S_1$ and $S_2$ are two non-empty sets. In addition, we also assess the  false positive rate by
$$
\text{FDP}=\frac{\text{Number of false rejections}}{\text{Number of total rejections}},
$$
and the power by
$$
\text{Power}=\frac{\text{Number of true rejections}}{\text{Number of true positives}}.
$$
The simulation results under  two different settings of $\{\rho,\bgamma,\bR\}$ are summarized in Figure \ref{sup.fig4}. This analysis further confirms the advantage of SMAI-align over the alternative methods in  characterizing the true marker genes, as measured by  three different metrics (Figure S11). In particular, our simulation study demonstrates the tendency of creating more false positives by the existing methods as compared with SMAI-align.

\subsection{Computing time} 

We evaluate the computing time of SMAI-align as compared with two existing alignment methods, Seurat and LIGER. Specifically, we consider a sequence of integration tasks involving  datasets of various sample sizes. Seurat and LIGER are applied with the default parameters, and SMAI-align is applied with default parameters $t=60$ and $T=3$. As a result, we find SMAI-align has a similar running time as Seurat, and is in general much faster than LIGER (Figure S15a). In particular, to align two datasets with sample size $n_1=n_2=7000$ and $d=2000$, SMAI took at most 7 mins on a standard PC (MacBook Pro with 2.2 GHz 6-Core Intel Core i7) to obtain the alignment function and the aligned datasets. On the other hand, we find that even after including the SMAI-test procedure, the complete SMAI algorithm still has a reasonable running time, about twice the time used by running SMAI-align alone. For example, it takes less than 13 mins to test and align datasets containing in total 14K samples. 

{For aligning very large datasets, SMAI has the advantage of allowing for first learning the alignment function based on much smaller numbers of samples, and then applying the alignment function to the whole dataset. We evaluated SMAI and SMAI-align's running time for integrating two large datasets of immune cells associated with different tissues \cite{dominguez2022cross}, collectively containing over 134K cells. As a result, under different subsampling rates (10\%, 20\% or 40\%), SMAI takes about 44 to 220 minutes to complete both alignability testing and alignment of all the samples, whereas SMAI-align alone takes about 20 to 104 minutes to finish alignment ({Figure S15b}). As a comparison, neither Seurat nor LIGER was able to finish its running within 250 minutes. Moreover, we also observed a remarkable similarity in the final alignment functions learned based on different subsampling rates ({Figure S15c}), demonstrating the reliability of SMAI in leveraging only a small portion of samples (e.g., 10\%) for fast alignment of very large datasets.}

\subsection{Nonlinear extensions of SMAI}

We point out two ways to extend our current framework to account for possible nonlinear relationships. On the one hand, one could account for nonlinearity by combining locally linear transforms. Specifically, one could consider partitioning the datasets into different parts, for example according to cell types, and then apply SMAI to each part, to allow for locally distinct alignment maps. Interestingly, our analysis of Tasks Pos4 and Pos5 reveals a remarkable similarity in the obtained cell-type specific translations and rotations that achieve alignment (Figure S14c).
On the other hand, one could leverage kernel or graph-based methods to obtain nonlinear representations of each dataset before applying SMAI. For example, instead of using the  denoised data matrices $\widehat\bX$ and $\widehat\bY$ in Equation (1) of the main text, one could replace them with certain nonlinear (low-dimensional) embedding matrices of the two datasets, such as the outputs from  Isomap \cite{balasubramanian2002isomap}, local linear embedding \cite{roweis2000nonlinear}, Laplacian eigenmap \cite{belkin2003laplacian}, or kernel eigenfunction embedding \cite{ding2022learning}.  {In other words, the flexibility of the SMAI framework enables simple combination with nonlinear (kernel) embedding methods for aligning nonlinear structures of high-dimensional datasets.} In this way, the final alignment functions between the two datasets are necessarily nonlinear, likely capturing more complex relationships.

\subsection{Synthetic data} 

Synthetic datasets are generated to evaluate the performance of SMAI. Specifically, the datasets associated with the positive control Task Pos2 and the negative control Task Neg1 are generated by selecting subsets of cells from the datasets associated with the positive control Task Pos1. For Task Pos2, we generate the first dataset by removing the \emph{beta} and \emph{endothelial} cells from the Smart-seq2 dataset in Task Pos1, and generate the second dataset by removing the \emph{gamma} cells from the CEL-Seq2 dataset in Task Pos1. As a result, the datasets in Task Pos2 have partially overlapping cell types. Similarly, for Task Neg1, the first dataset only contains the \emph{acinar}, \emph{activated sttellate}, \emph{alpha} and \emph{endothelial} cells of the Smart-seq2 dataset  in Task Pos1, whereas the second dataset only contains the \emph{beta}, \emph{delta}, \emph{ductal} and \emph{gamma} cells of the CEL-Seq2 dataset  in Task Pos1. The datasets in Task Neg1 thus do not share any common cell types.

\subsection{Implementation details} 

For all the integration tasks analyzed in this study, we use the R implementations of the relevant alignment algorithms. For LIGER and Harmony, we set the dimension number for the dimension reduction step as 50. The other tuning parameters of LIGER and Harmony, as well as those of Seurat, Scanorama, and fastMNN, are set as their default values. We have implemented our SMAI-test and SMAI-align algorithms into a unified R/Python function, where under the default setting, we have $T=3$, $t=60$, and $k=20$. To obtain the initial alignment for the partial alignability test, we use Seurat, fastMNN, or SMAI-align to align $\bX_2$ and $\bY_2$. We find the results to be not sensitive to the choice of these methods. In our analysis, we used the above default values for $T$, $t$ and $k$, and used (\ref{rmax}) for determining the rank parameter, with ${\sf c}=0.001$ for SMAI-align and ${\sf c}=0.5$ for SMAI-test. The R package \texttt{SMAI}, along with its Python version, and the R codes for reproducing the analyses presented in this study, can be retrieved and downloaded from our online GitHub repository \url{https://github.com/rongstat/SMAI}.

\subsection{Proof of Theorem \ref{test.thm}}

We begin by providing more details of the model assumptions. Throughout, we assume $r$ is fixed and does not grow with $n$ or $d$, indicating the low-dimensionality of the underlying structure. Without loss of generality, we also assume $\beta=1$, as the discussion below can be easily adapted to allow for a global rescaling of all the eigenvalues of $\bSig_1$, or $\bSig_2$. In this case, under the null hypothesis, we are essentially interested in determining if $\theta_i^{(1)}=\theta_i^{(2)}$ for all $1\le i\le r$. Specifically, for the bulk eigenvalues $\{\theta_i^{(\ell)}\}_{i>r}$, we assume that 

{\bf (A1)} The empirical distribution $H_n^{(\ell)}=\frac{1}{n-r}\sum_{i=r+1}^n\delta({\lambda_i^{(\ell)}})$ converges to a limiting distribution $H^{(\ell)}$ as $n\to \infty$, where $\delta(x)$ denotes the Dirac mass at a point $x$.

For the assumption on the spiked eigenvalues $\{\theta_i^{(\ell)}\}_{i\le r}$, we  define for any $\alpha\ne 0$ lying outside the support of $H^{(\ell)}$ the following function
\beq \label{psi}
\psi^{(\ell)}(\alpha)=\alpha+c\alpha\int\frac{t}{\alpha-t}dH^{(\ell)}(t),
\eeq
and its derivative
\beq \label{psi'}
(\psi^{(\ell)})'(\alpha)=1-c\int\frac{t^2}{(\alpha-t)^2}dH^{(\ell)}(t).
\eeq
Similarly, we can define $\psi^{(\ell)}_n(\alpha)$ and $(\psi_n^{(\ell)})'(\alpha)$, with $H^{(\ell)}$ in (\ref{psi}) and (\ref{psi'}) replaced by $H^{(\ell)}_n$, and $c$ replaced by $n/d$.
Then we further assume that

{\bf (A2)} For each $\ell=1,2$, the spiked eigenvalues $\{\theta_i^{(\ell)}\}_{i\le r}$ lie outside the support of $H^{(\ell)}$, and satisfy $(\psi^{(\ell)})'(\theta_i^{(\ell)})>0$ for all $1\le i\le r$.

Finally, we also assume that the empirical bulk eigenvalues $\{\lambda_i^{(1)}\}_{i>r}$ and $\{\lambda_i^{(2)}\}_{i>r}$ have sufficiently similar empirical distributions, which can be obtained if a certain convergence rate requirement is imposed on assumption (A1) above. Specifically, we assume that

{\bf (A3)}  For each $i=1,2,...,r$, it holds that 
\beq
%\sqrt{d}\cdot\bigg|\sum_{i=r+1}^n\frac{\lambda_i^{(1)}}{\alpha-\lambda_i^{(1)}}-\sum_{i=r+1}^n\frac{\lambda_i^{(2)}}{\alpha-\lambda_i^{(2)}}\bigg|\to 0,\qquad \text{in probability},
\sqrt{d}\cdot\bigg|\int\frac{t}{\alpha-t}dH_n^{(1)}(t)-\int\frac{t}{\alpha-t}dH_n^{(2)}(t)\bigg|\to 0,\qquad \text{in probability},
\eeq
for all $\alpha\in \{\theta_i^{(\ell)}\}_{1\le i\le r}$.

With these preparations, we proceed to prove our theorem. Firstly, we recall the following result obtained by Zhang et al. \cite{zhang2022asymptotic}. 

\begin{lem} \label{clt.lem}
for each $\ell=1,2,$ we define $a^{(\ell)}_i=\psi_n^{(\ell)}(\theta_i^{(\ell)})$. Under the above high-dimensional generalized spiked covariance model and assumptions (A1) and (A2), it holds that
\beq
\bigg( \sqrt{d}\frac{\lambda_1^{(\ell)}-a^{(\ell)}_1}{a^{(\ell)}_1}, ...,  \sqrt{d}\frac{\lambda_r^{(\ell)}-a^{(\ell)}_r}{a^{(\ell)}_r} \bigg) \to_d N(0, \textup{diag}([\sigma^{(\ell)}_1]^2,...,[\sigma^{(\ell)}_r]^2)),
\eeq
where $\sigma^{(\ell)}_i=\frac{\theta_i^{(\ell)}}{\psi^{(\ell)}(\theta_i^{(\ell)})}\sqrt{2(\psi^{(\ell)})'(\theta_i^{(\ell)})}$, for $i=1, 2, ..., r.$
\end{lem}

Now for any $i=1,2,...,r$, we have
\begin{align*}
\frac{\sqrt{d}(\lambda_i^{(1)}-\lambda_i^{(2)})}{\sqrt{(a_i^{(1)}\sigma_i^{(1)})^2+(a_i^{(2)}\sigma_i^{(2)})^2}} &=\frac{\sqrt{d}(\lambda_i^{(1)}-a_i^{(1)})}{a_i^{(1)}\sigma_i^{(1)}}\frac{a_i^{(1)}\sigma_i^{(1)}}{\sqrt{(a_i^{(1)}\sigma_i^{(1)})^2+(a_i^{(2)}\sigma_i^{(2)})^2}}\\
&\quad+\frac{\sqrt{d}(\lambda_i^{(2)}-a_i^{(2)})}{a_i^{(2)}\sigma_i^{(2)}}\frac{a_i^{(2)}\sigma_i^{(2)}}{\sqrt{(a_i^{(1)}\sigma_i^{(1)})^2+(a_i^{(2)}\sigma_i^{(2)})^2}}\\
&\quad +\frac{\sqrt{d}(a_i^{(1)}-a_i^{(2)})}{\sqrt{(a_i^{(1)}\sigma_i^{(1)})^2+(a_i^{(2)}\sigma_i^{(2)})^2}}.
\end{align*}
By Lemma \ref{clt.lem}, it follows that
\beq
\frac{\sqrt{d}(\lambda_i^{(1)}-a_i^{(1)})}{a_i^{(1)}\sigma_i^{(1)}}\frac{a_i^{(1)}\sigma_i^{(1)}}{\sqrt{(a_i^{(1)}\sigma_i^{(1)})^2+(a_i^{(2)}\sigma_i^{(2)})^2}}+\frac{\sqrt{d}(\lambda_i^{(2)}-a_i^{(2)})}{a_i^{(2)}\sigma_i^{(2)}}\frac{a_i^{(2)}\sigma_i^{(2)}}{\sqrt{(a_i^{(1)}\sigma_i^{(1)})^2+(a_i^{(2)}\sigma_i^{(2)})^2}}\to_d N(0,1),
\eeq
whereas by Assumptions (A1) and (A3), it follows that 
\beq
\frac{\sqrt{d}(a_i^{(1)}-a_i^{(2)})}{\sqrt{(a_i^{(1)}\sigma_i^{(1)})^2+(a_i^{(2)}\sigma_i^{(2)})^2}}\to 0,\qquad \text{in probability}.
\eeq
As a result, by Slutsky's theorem, we have
\beq
\bigg(\frac{\sqrt{d}(\lambda_1^{(1)}-\lambda_1^{(2)})}{\sqrt{(a_1^{(1)}\sigma_1^{(1)})^2+(a_1^{(2)}\sigma_1^{(2)})^2}},...,\frac{\sqrt{d}(\lambda_r^{(1)}-\lambda_r^{(2)})}{\sqrt{(a_r^{(1)}\sigma_r^{(1)})^2+(a_r^{(2)}\sigma_r^{(2)})^2}}\bigg)\to_d N(0,{\bf I}_r),
\eeq
or
\beq
\sum_{i=1}^r\frac{{d}(\lambda_i^{(1)}-\lambda_i^{(2)})^2}{{(a_i^{(1)}\sigma_i^{(1)})^2+(a_i^{(2)}\sigma_i^{(2)})^2}}\to_d \chi^2(r),
\eeq
where $\chi^2(r)$ is a $\chi^2$ random variable with degree of freedom $r$. Finally, in light of the SMAI-test algorithm, it suffices to show that
\beq \label{var.eq}
2\alpha_i^{(\ell)}\phi_i^{(\ell)} - (a_i^{(\ell)}\sigma_i^{(\ell)})^2 \to 0,\qquad \text{ in probability}.
\eeq
Note that by definition, we have
\beq \label{4.15}
[a_i^{(\ell)}\sigma_i^{(\ell)}]^2=\bigg[\frac{\psi_n^{(\ell)}(\theta_i^{(\ell)})\cdot\theta_i^{(\ell)}}{\psi^{(\ell)}(\theta_i^{(\ell)})}\bigg]^2{2(\psi^{(\ell)})'(\theta_i^{(\ell)})},
\eeq
where
\beq
\psi_n^{(\ell)}(\theta_i^{(\ell)})\to \psi^{(\ell)}(\theta_i^{(\ell)}),\qquad \text{in probability}.
\eeq
By \cite{bai2012estimation}, we have
\beq
\alpha_i^{(\ell)}=\bigg(\frac{1-n/d}{\lambda_i^{(\ell)}}+\frac{1}{d}\sum_{j=r_{\max}+1}^{d}\frac{1}{\lambda_j^{(\ell)}-\lambda_i^{(\ell)}}\bigg)^{-2}\to [\theta_i^{(\ell)}]^2,\qquad \text{ in probability},
\eeq
whereas by Section 2.4 of \cite{zhang2022asymptotic}, we have
\beq \label{4.18}
\phi_i^{(\ell)}=\frac{1}{\sqrt{\alpha_i^{(\ell)}}}\bigg( -\frac{1-n/d}{(\lambda_i^{(\ell)})^2}+\frac{1}{d}\sum_{j=r_{\max}+1}^d\frac{1}{(\lambda_j^{(\ell)}-\lambda_i^{(\ell)})^2}\bigg)^{-1}\to (\psi^{(\ell)})'(\theta_i^{(\ell)}),\qquad \text{ in probability}.
\eeq
Combining Equations (\ref{4.15}) to (\ref{4.18}), we have shown Equation (\ref{var.eq}). This completes the proof of the theorem.

\newpage
\section{Supplementary Figures and Tables}

	\setcounter{table}{0}
\renewcommand{\thetable}{S\arabic{table}}	
\begin{table}
	\caption{Summary of single-cell datasets analyzed in this paper. The p-values indicate statistical significance against the null hypothesis that the two datasets are at least partially alignable (Section \ref{part.sec}). {Under each characteristic, whenever there are two entries, the former refers to dataset $\bX$ whereas the latter refers to $\bY$.}%\james{Explain what is the p value here testing. Add refs for each dataset.}
	} 
	\raggedright
	\begin{tabular}{|c|c|c|l|c|}
		\hline
		task id	&  {\# of cells} & {\# of features}  & same technology (Y-yes; N-no)  & reference\\  
		\hline
		Neg1& (1272, 1015) & (34363, 34363)   & synthetic&  \cite{panc8}\\
		Neg2& (904, 1226) & (15148, 15148)  & Y(10X Genomics) & \cite{benchmark} \\
		Neg3& (4000, 4000) & (36601, 36601)  & Y(10X Genomics) &  \cite{dominguez2022cross} \\
		Pos1& (2364, 2244)  & (34363, 34363)  & N(Smart-seq2, CEL-Seq2) & \cite{panc8}\\
		Pos2& (2035, 2134) & (34363, 34363) & synthetic & \cite{panc8} \\
		
		Pos3& (3362, 3222) & (33694, 33694) & N(10X Genomics v2, v3) & \cite{ding2019systematic} \\
		Pos4& (3618, 3750) & (3429, 3429)  & Y({ATAC-seq gene activity})& \cite{benchmark} \\
		Pos5& (2353, 1911) & (15148, 15148)  & Y(10X Genomics)& \cite{benchmark} \\
		Pos6& (4000, 4000) & (15148, 15148)  & Y(10X Genomics)& \cite{dominguez2022cross}\\
		Pos7& (3508, 2461) & (19222, 19222)  & Y(10X Genomics)& \cite{kartha2022functional}\\
		PosS1& (14891, 4651) & (36, 29452) & N(seqFISH, 10X Genomics) & \cite{lohoff2022integration,mousegas} \\
		PosS2& (6000, 34041) & (119, 34041) & N(ISS, Smart-seq) & \cite{gyllborg2020hybridization,tasic2018shared,long2023spacetx}\\
		PosS3& (1154, 34041) & (42, 34041) & N(ExSeq, Smart-seq) & \cite{alon2021expansion,tasic2018shared,long2023spacetx}\\
		\hline
	\end{tabular}
	
	\vspace{0.5cm}
	
	\begin{tabular}{|c|l|L|c|}
		\hline
		task id & same tissue/condition  (Y-yes; N-no)  & {\small \% of samples under overlapping cell types} &  p-values \\  
		\hline
		Neg1& synthetic & 0\% & $<10^{-6}$\\
		Neg2& Y(human lung) &  26\% & $<10^{-3}$\\
		Neg3& N(human liver, human MLN) & 37\% & $<10^{-9}$\\
		Pos1 & Y(human pancreas) & 100\% %(0.96)
		& 0.44\\
		Pos2& synthetic & 84\% %(0.90)
		& 0.50\\
		
		Pos3&  Y(human PBMC) &  99\% %(0.92)
		& 0.78 \\
		Pos4&  Y(mouse brain) &  100\% %(0.99) 
		& 0.88 \\
		Pos5& Y(human lung) & 100\% %(0.69) 
		& 0.14  \\
		Pos6& N(human LLN, human MLN) & 95\% %(0.69) 
		& 0.88  \\
		Pos7 & N(mouse PBMC, LPS stimulated/control) & 100\% %(0.69) 
		& 0.47  \\
		PosS1&  Y(mouse gastrulation) & -- & 0.84 \\
		PosS2& Y(mouse VISP) &  -- & 0.18 \\
		PosS3&  Y(mouse VISP) & -- & 0.38 \\
		\hline
	\end{tabular}
	\label{table:data}
\end{table}

\begin{table}
	\centering
	\caption{Empirical type I errors of SMAI-test at the nominal level $\alpha=0.05$ under various signal structures and sample size $n$, with $d=1000$.} 	
	\begin{tabular}{c|cccc}
		\hline
		Low-Dimensional Structure	&$n=1200$ & $n=1600$ & $n=2000$ & $n=2400$  \\  
		\hline
		Smiley Face	& 0.062&  0.063& 0.040 & 0.043\\  
		\hline
		Swiss Roll & 0.043 & 0.046 & 0.040 & 0.043\\
		\hline
	\end{tabular}
	\label{table:t1e}
\end{table}

\setcounter{figure}{0}
\renewcommand{\thefigure}{S\arabic{figure}}

	\begin{figure}[H]
		\centering
		\includegraphics[angle=0,width=15cm]{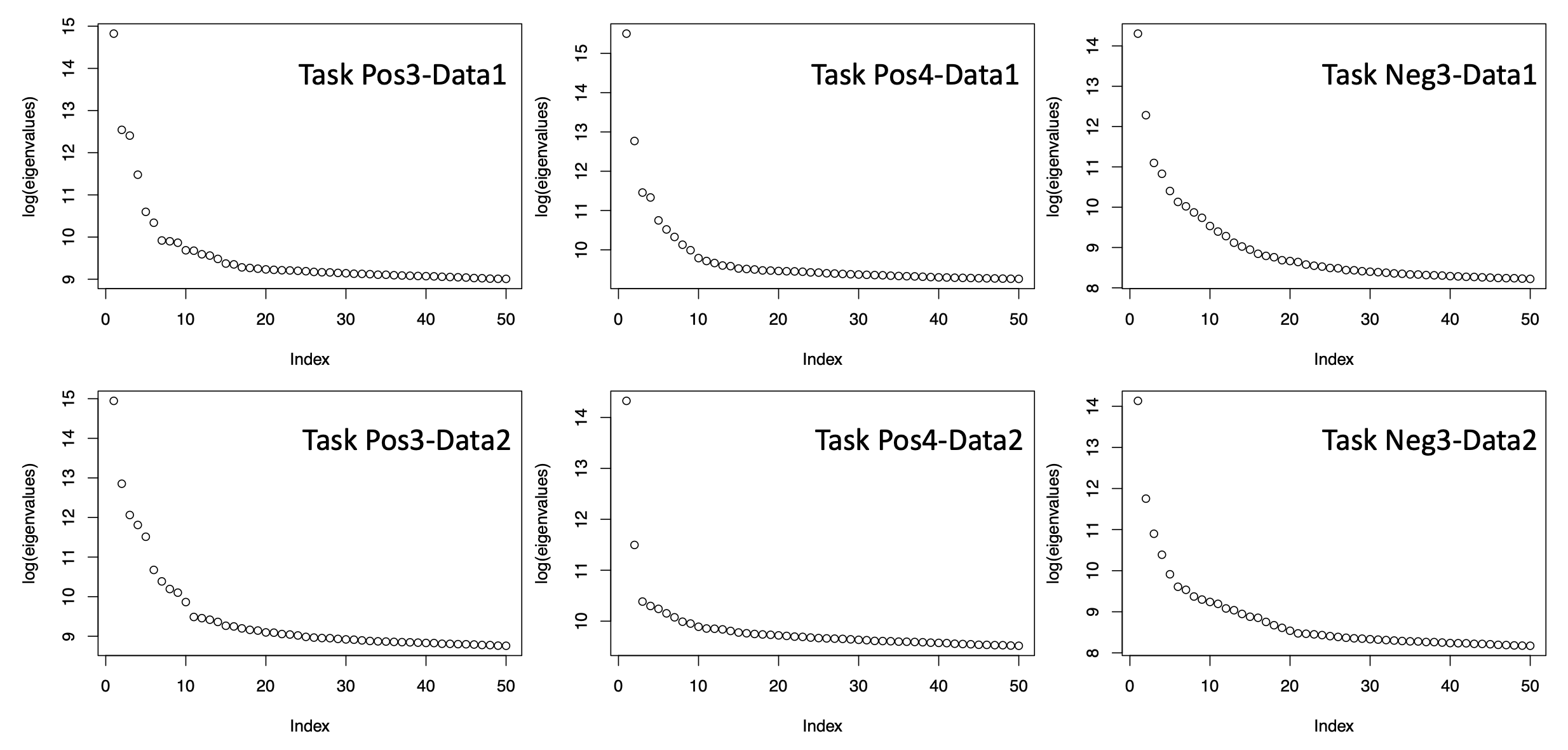}
			\caption{Empirical evidence on spiked eigenvalues. The top 50 eigenvalues of the normalized count data involved in some integration tasks are shown in order, at the log-scale. The presence of a few leading eigenvalues that are much larger than the rest of the eigenvalues indicates suitability of the spiked population models adopted by SMAI.} 
	\label{eigs.fig}
		\end{figure}
		
	\begin{figure}
		\centering
		\includegraphics[angle=0,width=16cm]{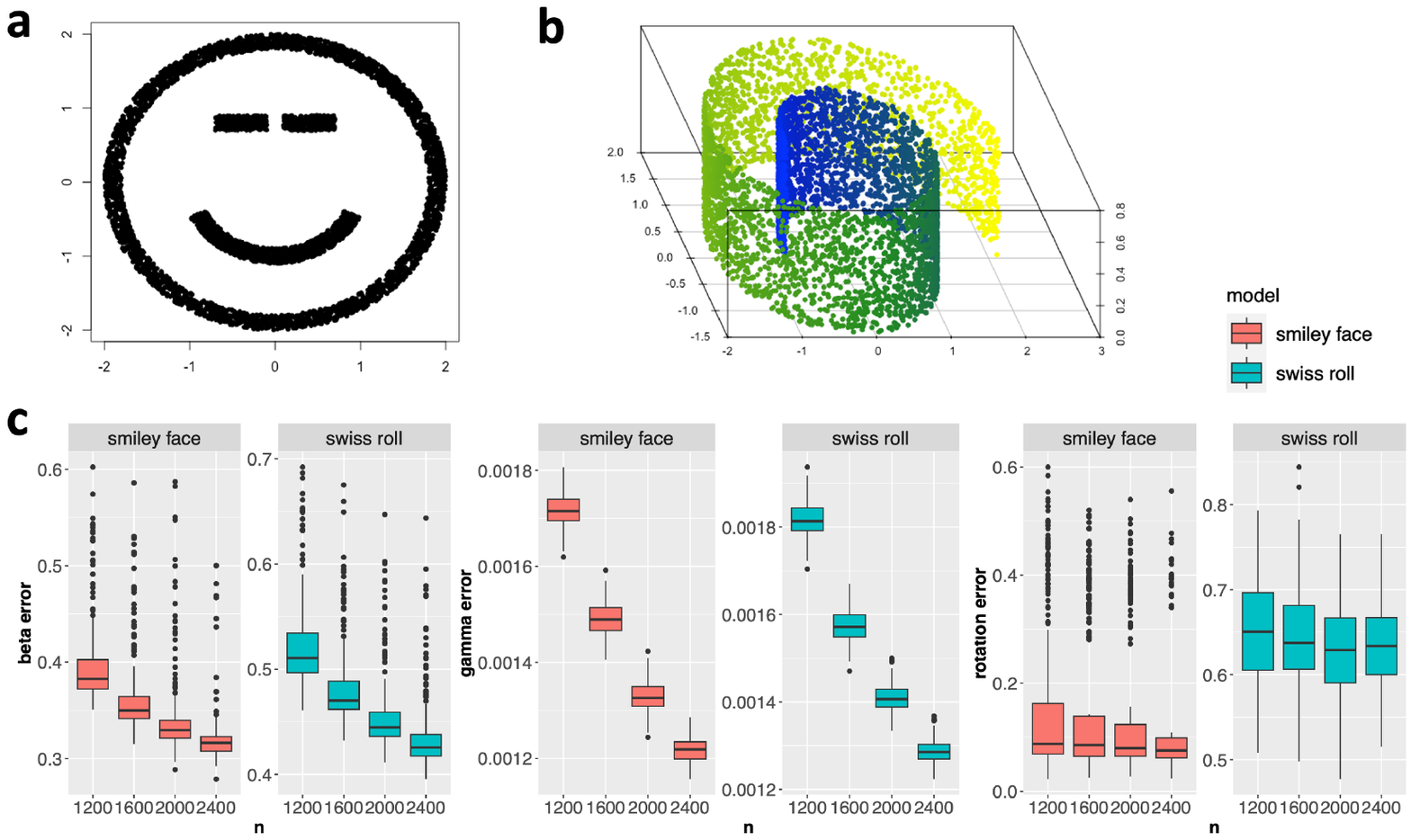}
			\caption{{Results from simulation studies that demonstrate consistency of SMAI-align. (a) Illustration of $n=5000$ samples from the two-dimensional smiley face structure. (b) Illustration of $n=5000$ samples from the three-dimensional Swiss roll manifold. (c) Boxplots (center line, median; box limits, upper and lower quartiles; points, outliers) of estimation errors for $\beta^*$ (left), $\bgamma^*$ (middle), and $\bR^*$ (right) by SMAI-align algorithm, under the simulated data with two different underlying structures and various sample sizes $n$. Each boxplot is based on 300 rounds of simulations. The estimation errors decrease as $n$ increases, suggesting the consistency of SMAI-align in recovering the true alignment function.} } 
	\label{sup.fig2}
		\end{figure}
		
		\begin{figure}
		\centering
	\includegraphics[angle=0,width=15cm]{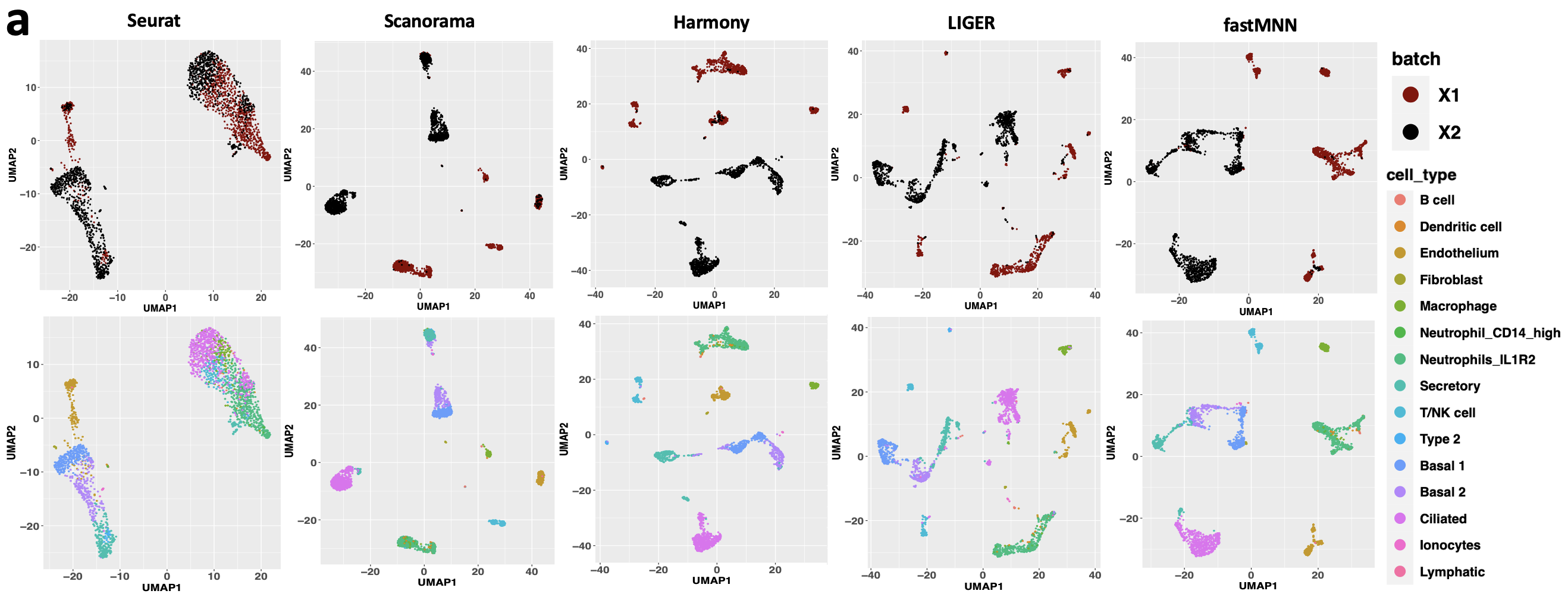}\\\includegraphics[angle=0,width=15cm]{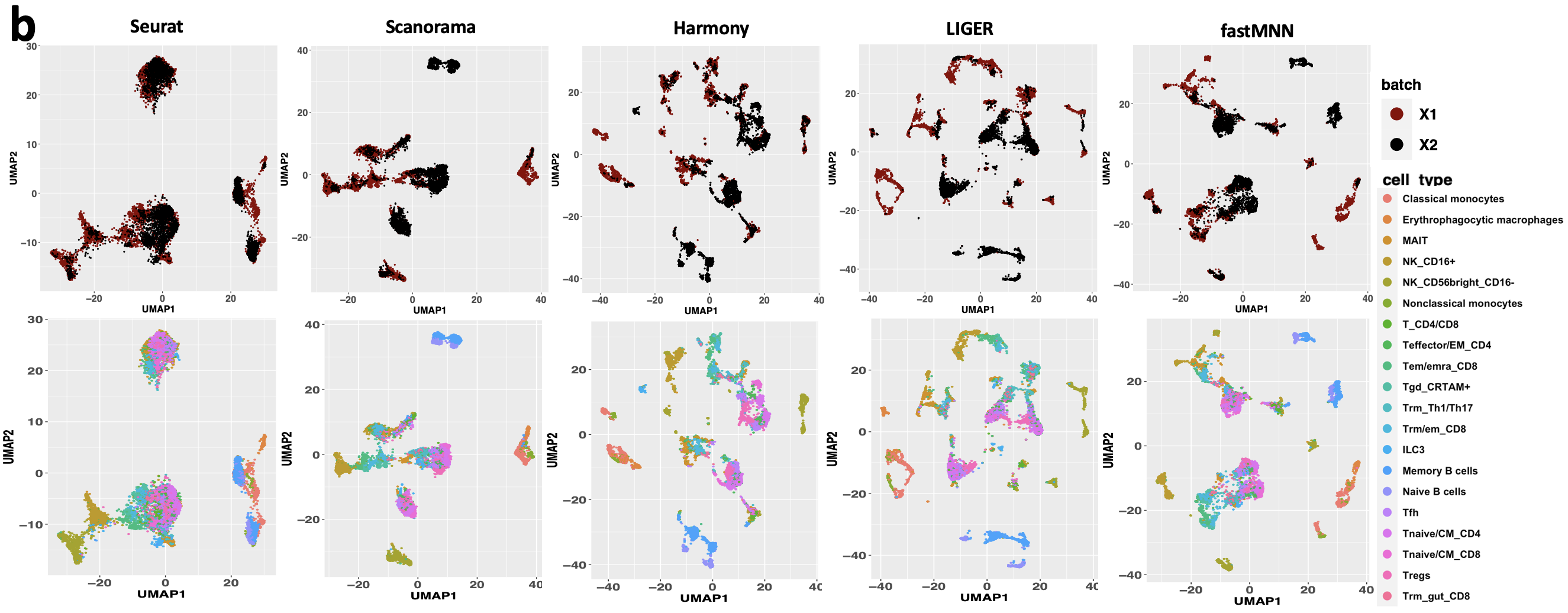}
	\includegraphics[angle=0,width=13cm]{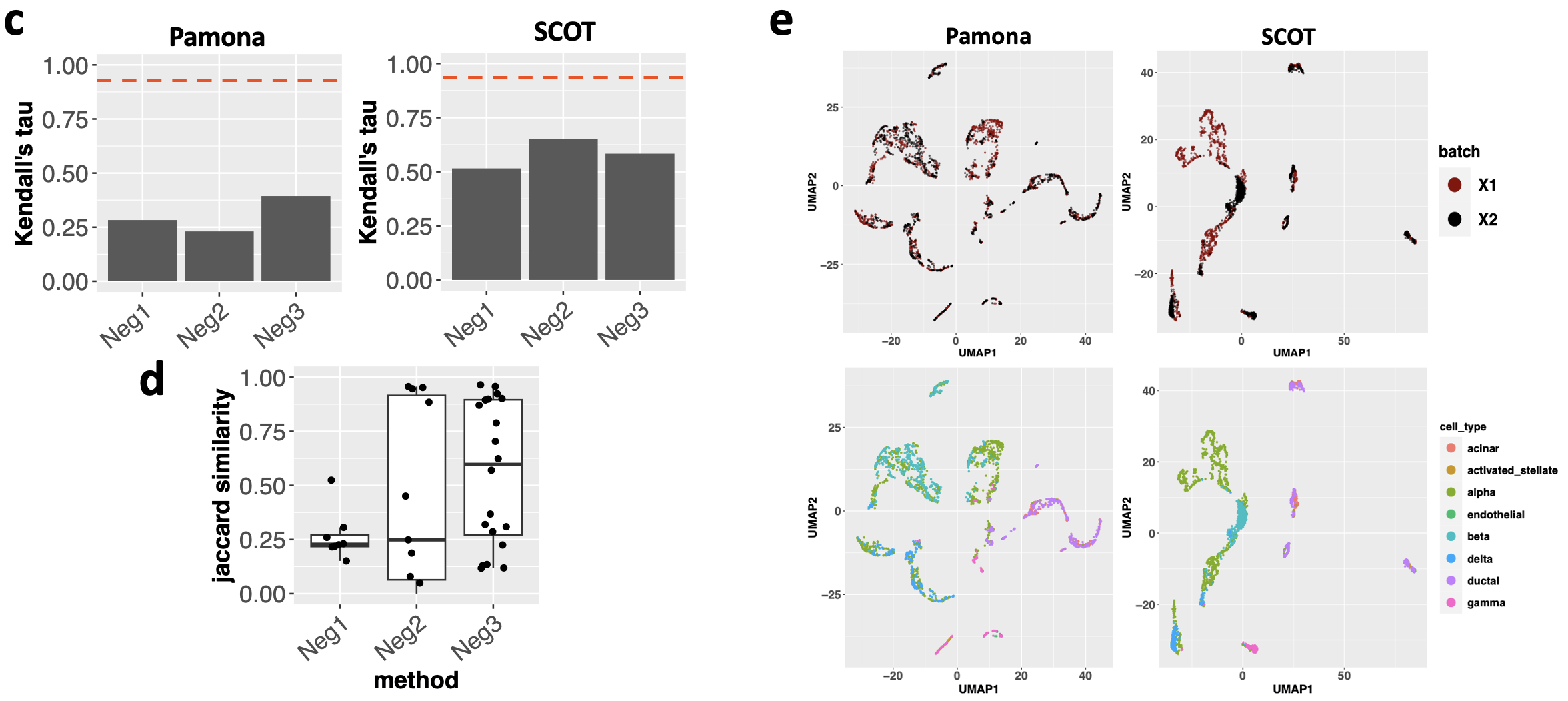}
			\caption{More results on the distortion and misleading alignment due to enforced integration in negative control tasks. (a) and (b): UMAP visualizations of the integrated data under Task Neg2 (a) and Task Neg3 (b), as obtained by five popular methods. For each method, the top figure is colored to indicate the distinct datasets being aligned, whereas the bottom figure colored to indicate different cell types. {(c) Extension of Figure \ref{fig2}b in the main text, demonstrating limited within-data structure-preservation for SCOT and Pamona. The red dashed line benchmarks the average Kendall's tau correlation of 0.9 achieved by SMAI-align over the positive control tasks Pos1-Pos7. (d) Extension of Figure \ref{fig2}c in the main text, showing limited performance in DE analysis with SCOT-integrated data. (e) UMAP visualizations of the Pamona- and SCOT-integrated data under Task Neg1.}} 
	\label{sup.fig2a}
		\end{figure}

	\begin{figure}
		\centering
	\includegraphics[angle=0,width=12cm]{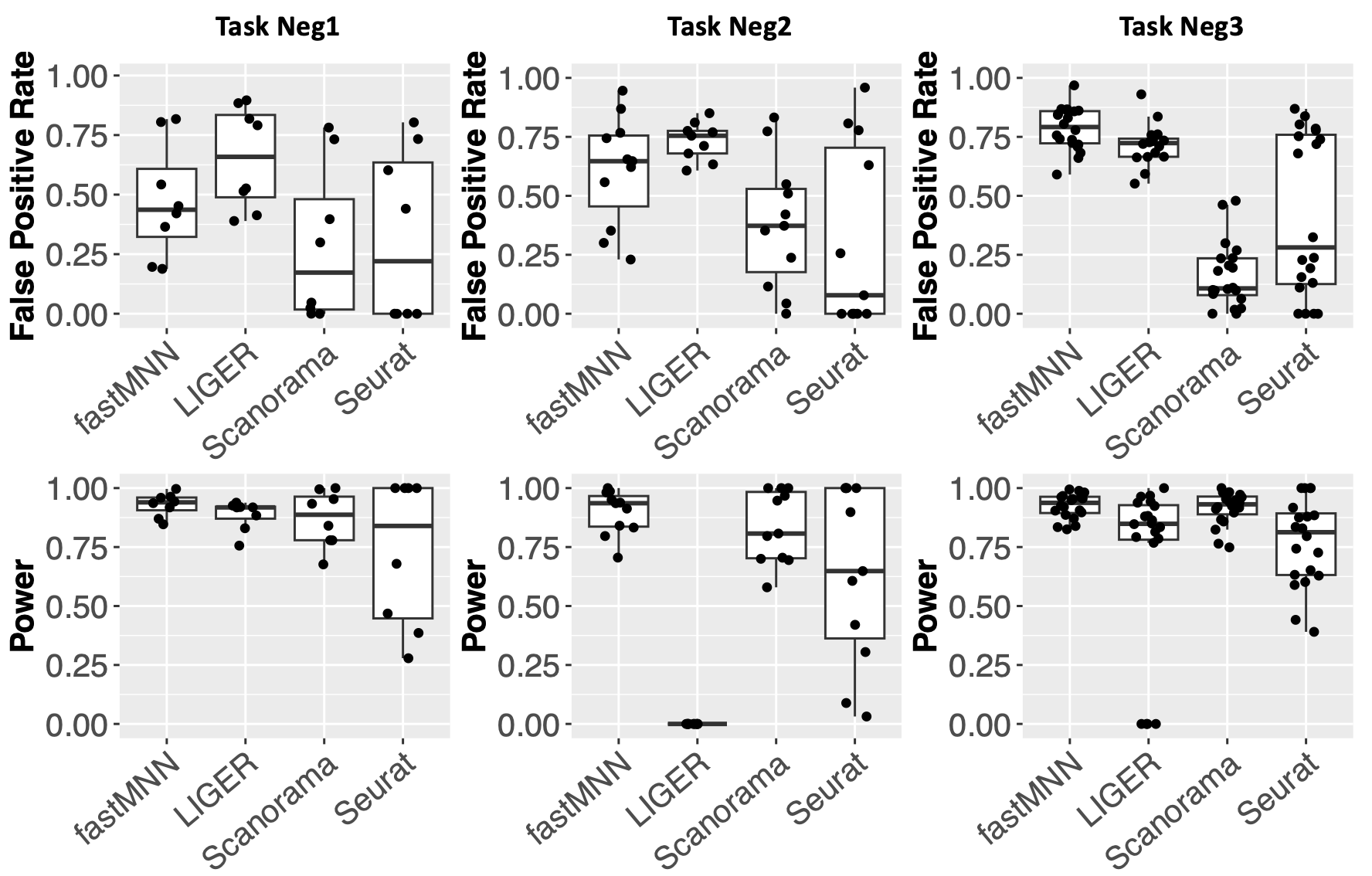}
			\caption{Forcing integration in negative control tasks leads to unreliable DE analysis. Boxplots of false positive rate and power of detecting DE genes based on the integrated data, as compared with the DE genes detected based on the original data. The distortion introduced by enforced alignment of two unalignable datasets may cause false discoveries and reduced power in detecting DE genes.} 
	\label{sup.fig2b}
		\end{figure}

	\begin{figure}
		\centering
		\includegraphics[angle=0,width=8cm]{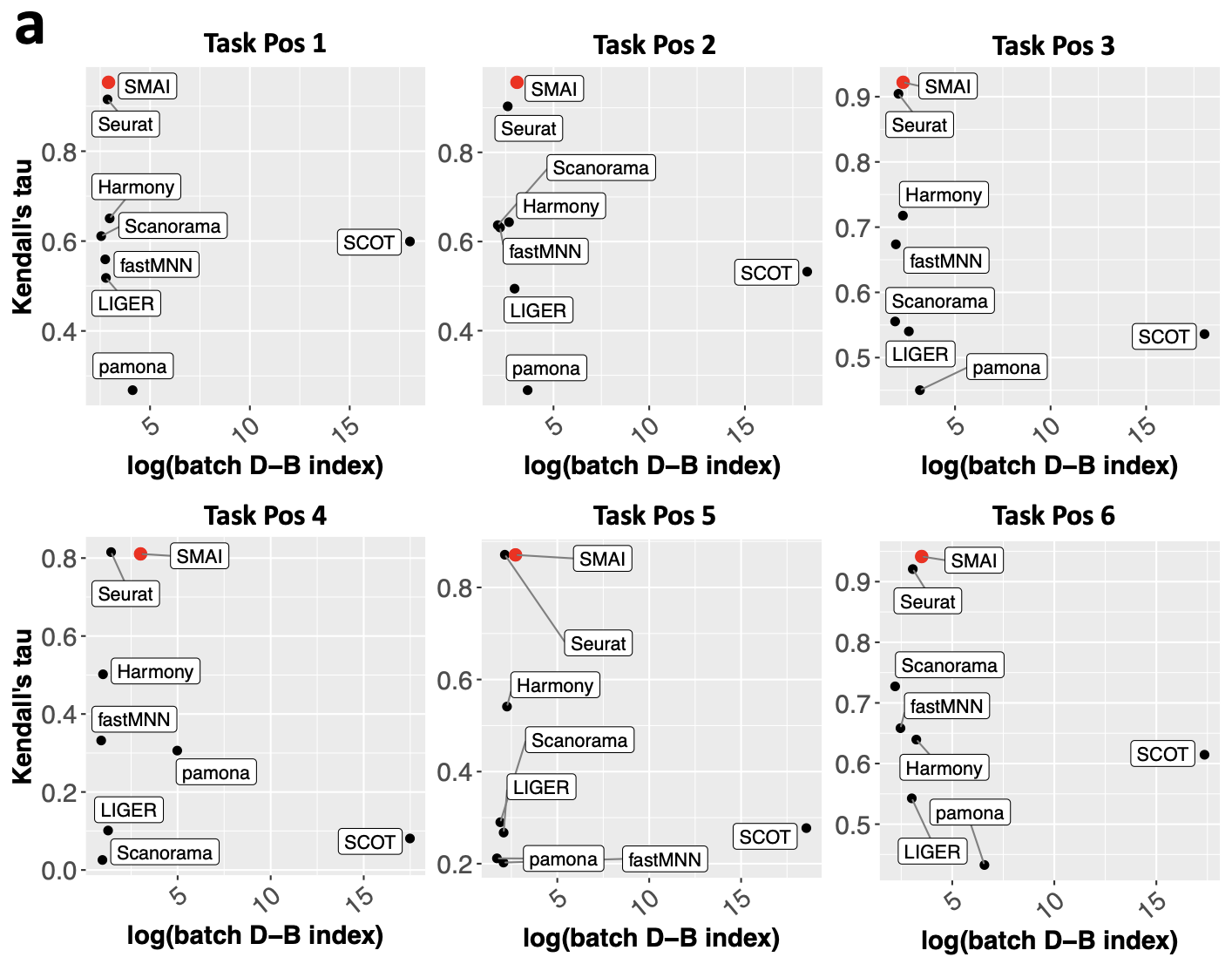}
		\includegraphics[angle=0,width=8cm]{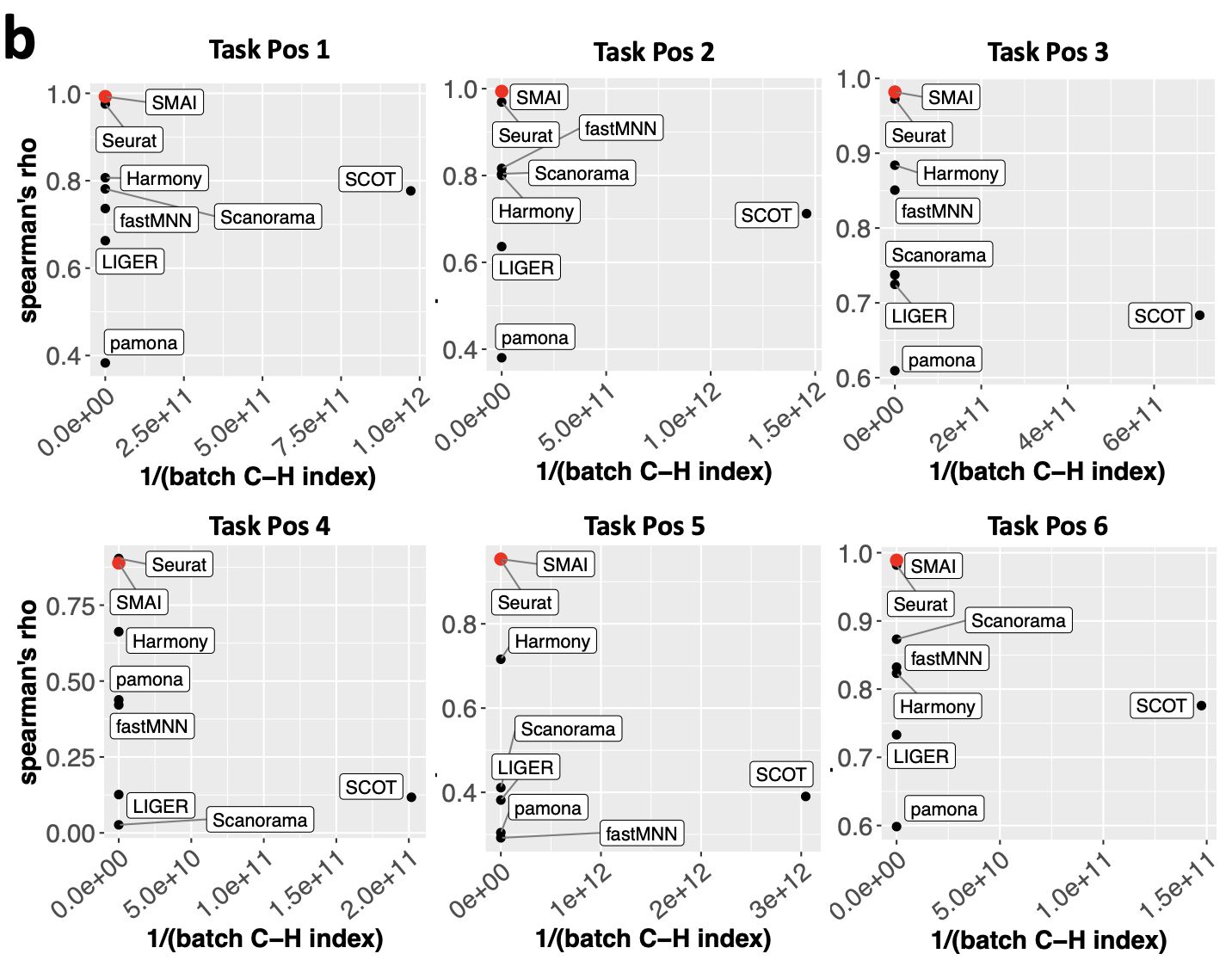}
		\includegraphics[angle=0,width=11.6cm]{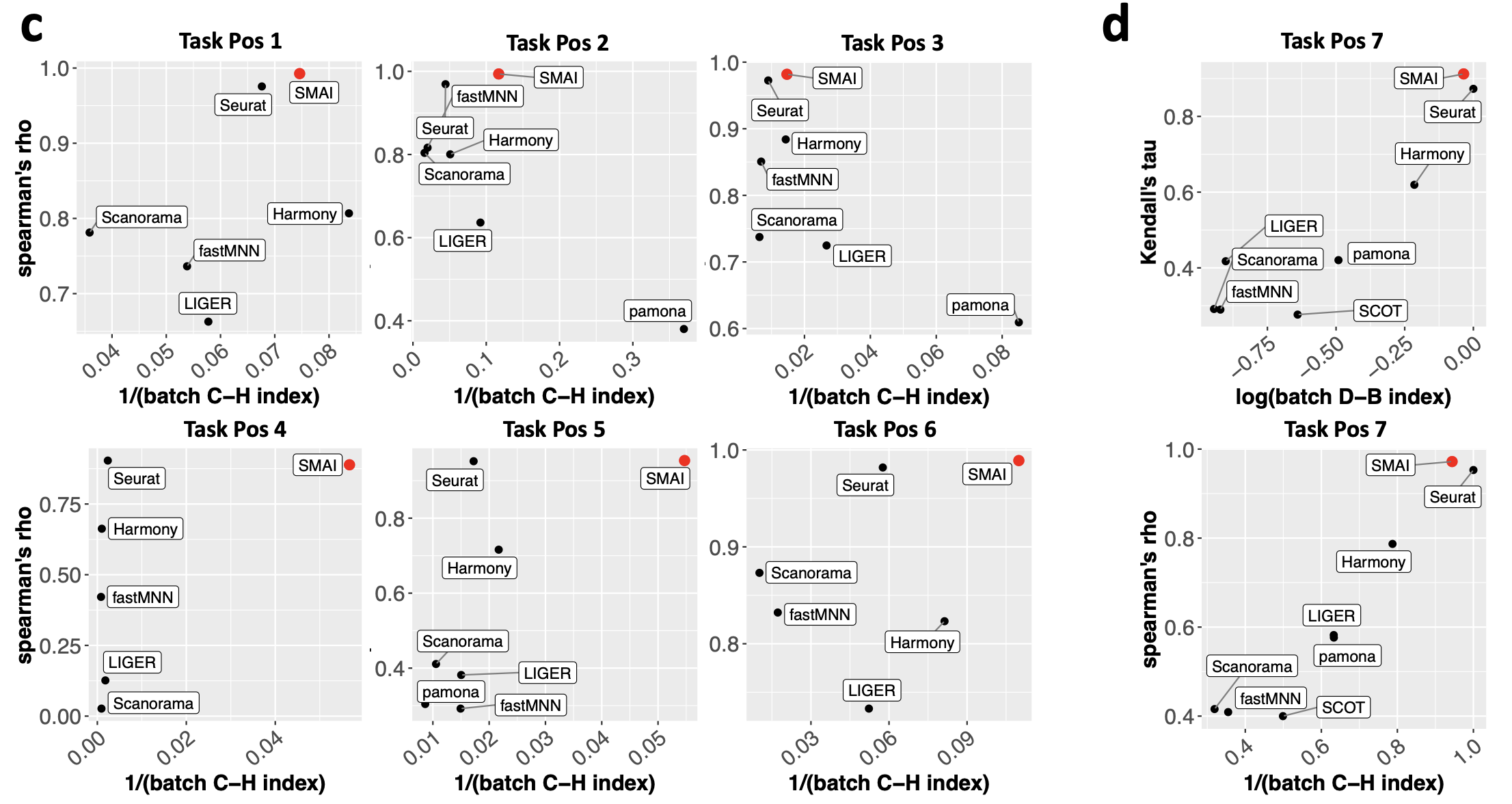}
			\caption{{Evidences of improved structure-preserving integration by SMAI. (a) Extension of Figure \ref{fig3}a in the main text, with SCOT included in the comparison. (b) Comparing seven existing algorithms (black) with SMAI-align (red), in preserving the within-data structures after integration and removing the unwanted variations. The structure-preserving quality is measured by Spearman's rho correlations (y-axis), and the alignment quality is measured by inverse batch-associated C-H index (x-axis). For both metrics, a higher value suggests better performance. (c) Same as panel (b) with SCOT and/or Pamona excluded to better visualize the difference in other methods. (d) Results for Task Pos 7.}  }
	\label{sup.fig3-1}
		\end{figure}

	\begin{figure}
	\centering
	\includegraphics[angle=0,width=16cm]{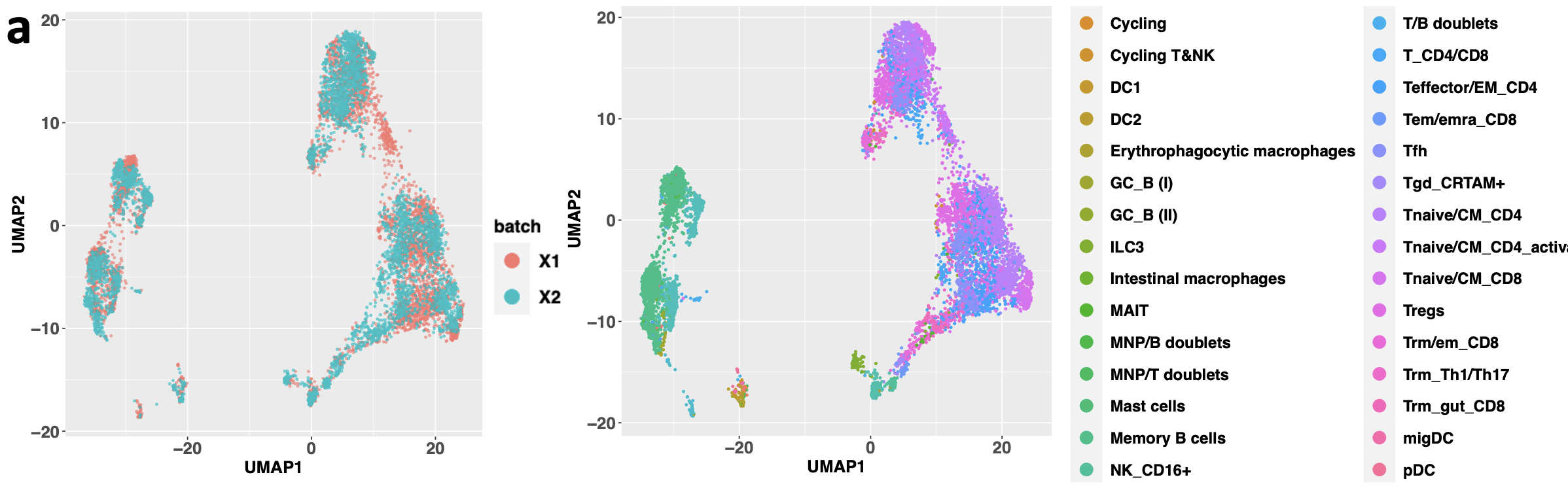}
		\includegraphics[angle=0,width=13cm]{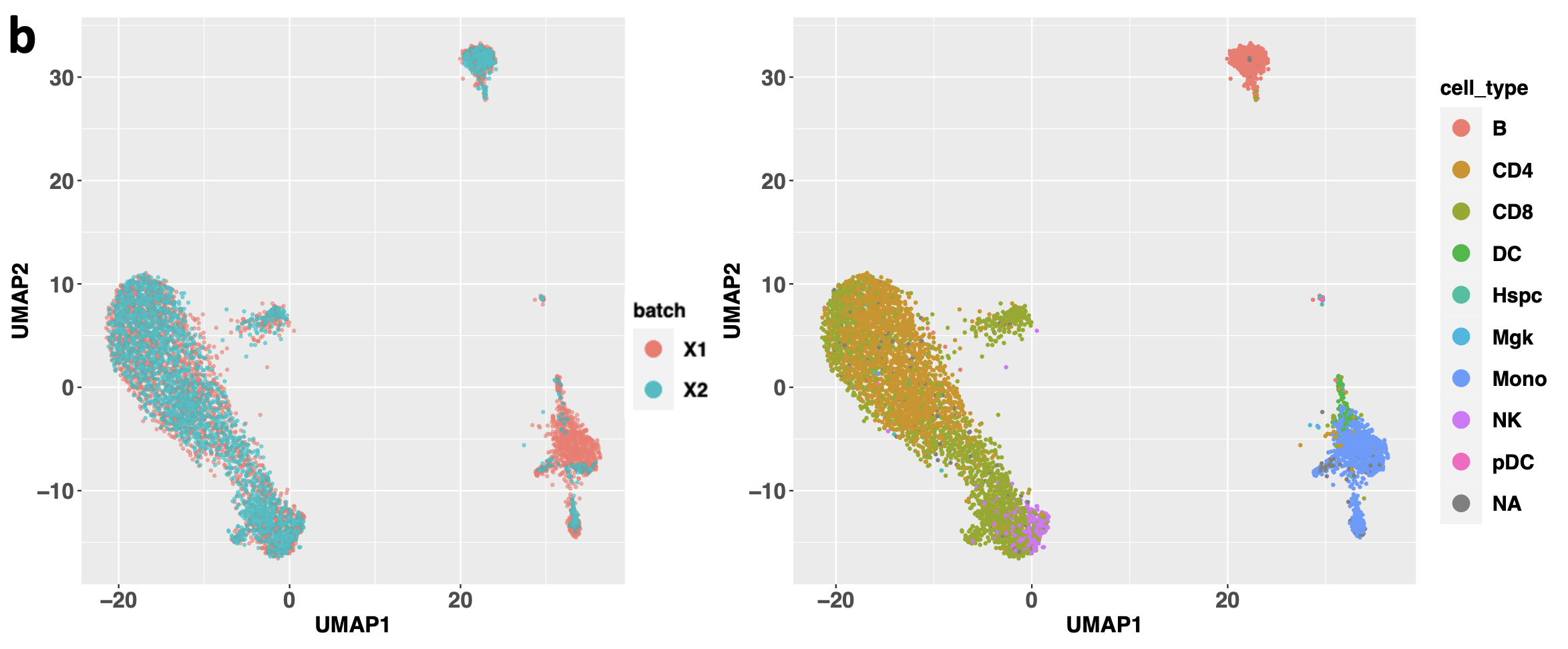}
	\caption{{Visualization of SMAI-aligned datasets. (a) Alignment of the two immune cell datasets in Task Pos6. (b) Alignment of the two PBMCs datasets in Task Pos7. In particular, after SMAI alignment, there is a small portion of cells (``Mono" cells) that are not aligned for two conditions, likely corresponding to the condition-specific mechanisms. } }
	\label{sup.fig3-1-2}
\end{figure}

	\begin{figure}
		\centering
		\includegraphics[angle=0,width=16cm]{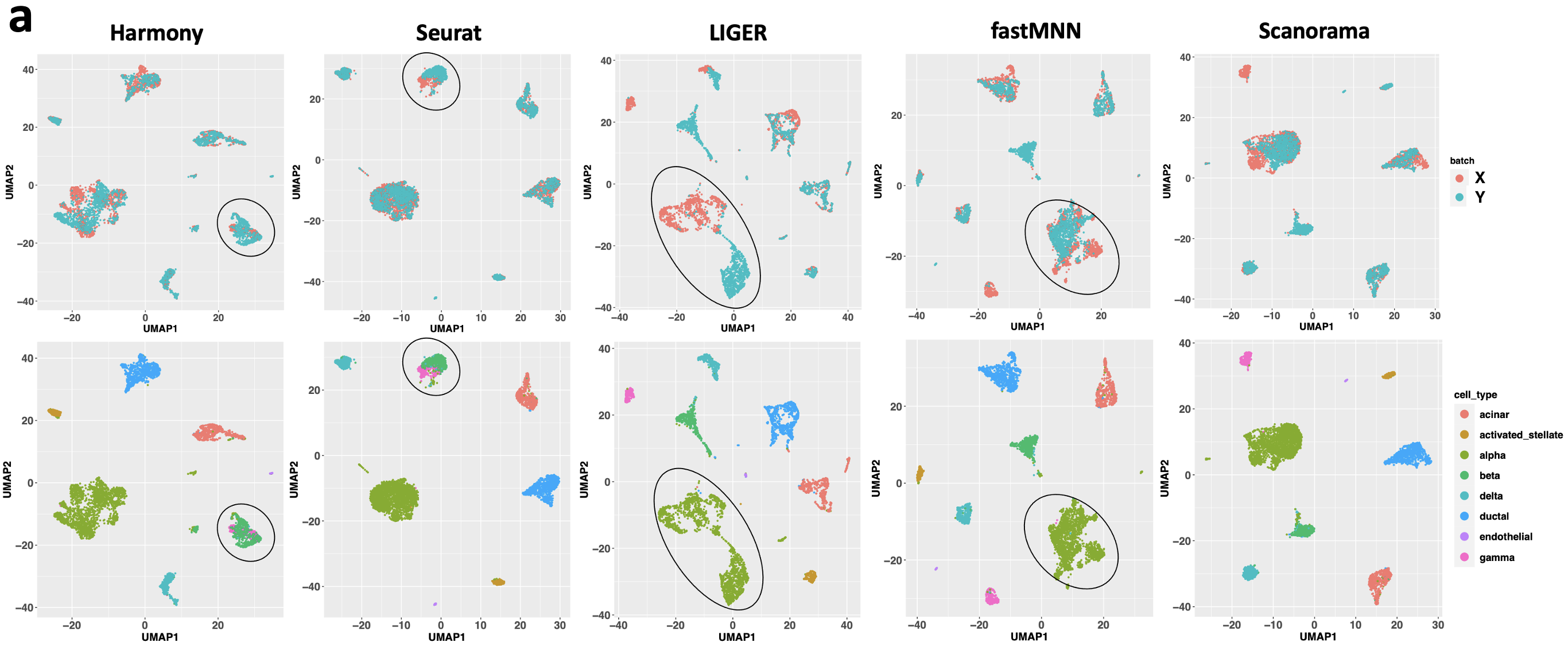}\\
		\includegraphics[angle=0,width=16cm]{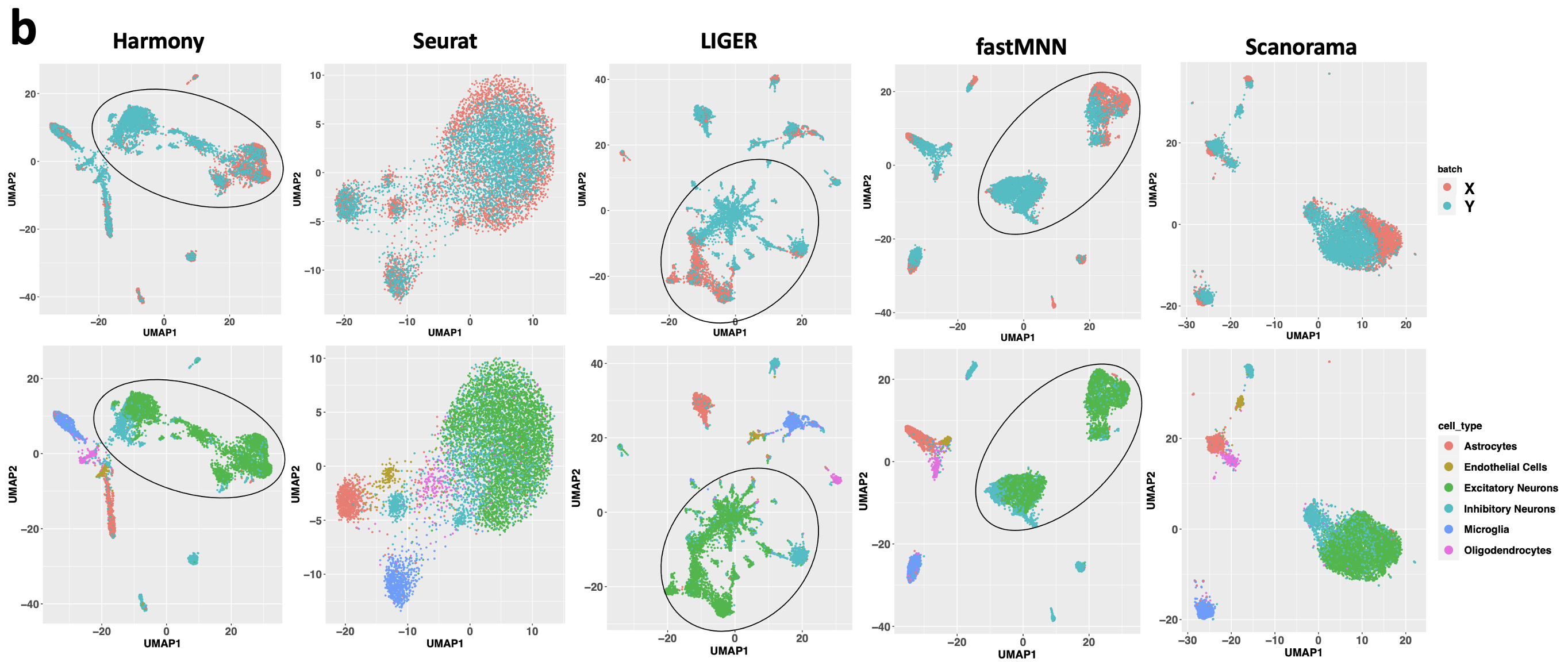}
			\caption{Distortion and misalignment by existing methods over positive control integration tasks. 
			 (a) In Task Pos2, there is false integration of gamma cells and beta cells by Harmony and Seurat, and  significant distortion, that is, stretching and creation of multiple artificial subclusters, of the alpha cell cluster by LIGER and fastMNN. (b) In Task Pos4, there is strong distortion and artificial clustering of the excitatory neurons and the inhibitory neurons, in the data output from Harmony, LIGER and fastMNN.} 
	\label{sup.fig3-2}
		\end{figure}
		
	\begin{figure}
	\centering
	\includegraphics[angle=0,width=12cm]{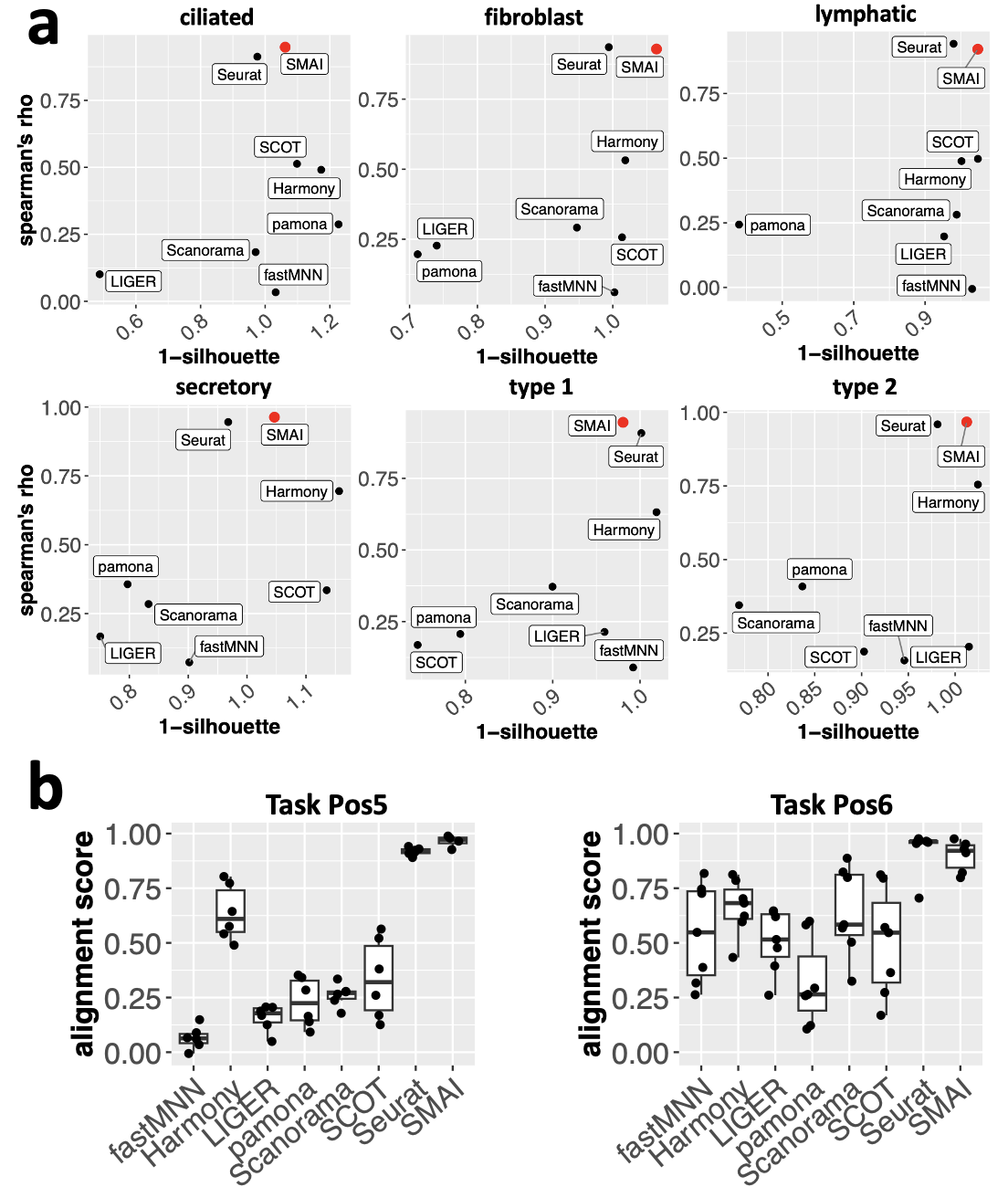}
	\caption{{Evidence of structure-preserving alignment of rare cell types by SMAI. (a) Scatter plot of SMAI and seven existing methods in terms of structure preservation and batch-effect removal for each rare cell type (defined as those containing less than 5\% of the total cells) in Task Pos5. As in Figures \ref{sup.fig2a}, for structure preservation, we used Spearman's rho between the distance matrices of each dataset before and after integration, but restricted to the rows corresponding to a rare cell type. For alignment quality, we used (1-median Silhouette index) to measure how well the same rare cell type is mixed after integration. (b) Boxplots of the overall alignment scores in rare cell type preservation and alignment for each method, across all the rare cell types in Tasks Pos5 ($n=6$ rare cell types) and Pos6 ($n=7$ rare cell types). The alignment score is define as the product of the two metrics (x and y coordinates) in panel (a). The results indicate superior performance of SMAI in dealing with rare cell types.} } 
	\label{sup.fig3-3}
\end{figure}

	\begin{figure}
		\centering
		\includegraphics[angle=0,width=16cm]{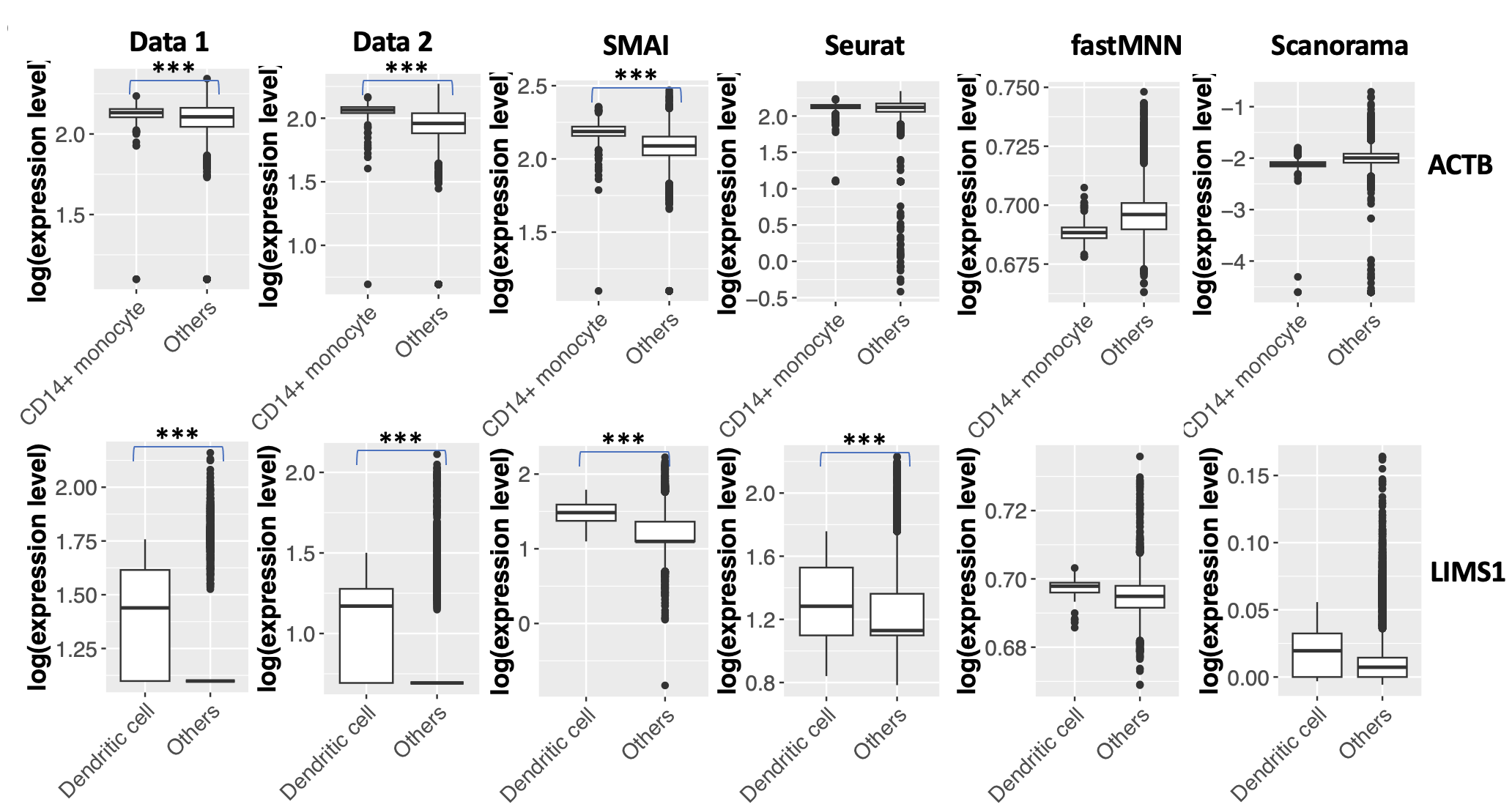}
			\caption{SMAI performs better in preserving DE genes. Top: Boxplots of log-expression levels of ACTB as grouped by cell types in the two datasets (Data 1 with 436 CD14+ monocytes and 2926 other cells, and Data 2 with 354 CD14+ monocytes and 2868 other cells) about human PBMCs (Task Pos3), and in the integrated datasets (790 CD14+ monocytes and 5794 others) as produced by SMAI-align, Seurat, fastMNN, and Scanorama. Bottom: Boxplots of log-expression levels of LIMS1 as grouped by cell types in the two datasets (Data 1 with 76 dendritic cells and 3286 other cells, and Data 2 with 38 dendritic cells and 3184 other cells) about human PBMCs (Task Pos3), and in the integrated datasets (114 dendritic cells and 6470 other cells) as produced by SMAI-align, Seurat, fastMNN, and Scanorama. In both examples, the DE patterns were not sufficiently preserved by fastMNN, Scanorama, and/or Seurat, but were well preserved by SMAI-align, after integration. The stars above the boxplots indicate statistical significance of the p-values. Specifically, * means adjusted p-value $<0.05$;  ** means adjusted p-value $<0.01$; *** means adjusted p-value $<0.001$.} 
	\label{sup.fig3-3-1}
		\end{figure}

	\begin{figure}
		\centering
		\includegraphics[angle=0,width=16cm]{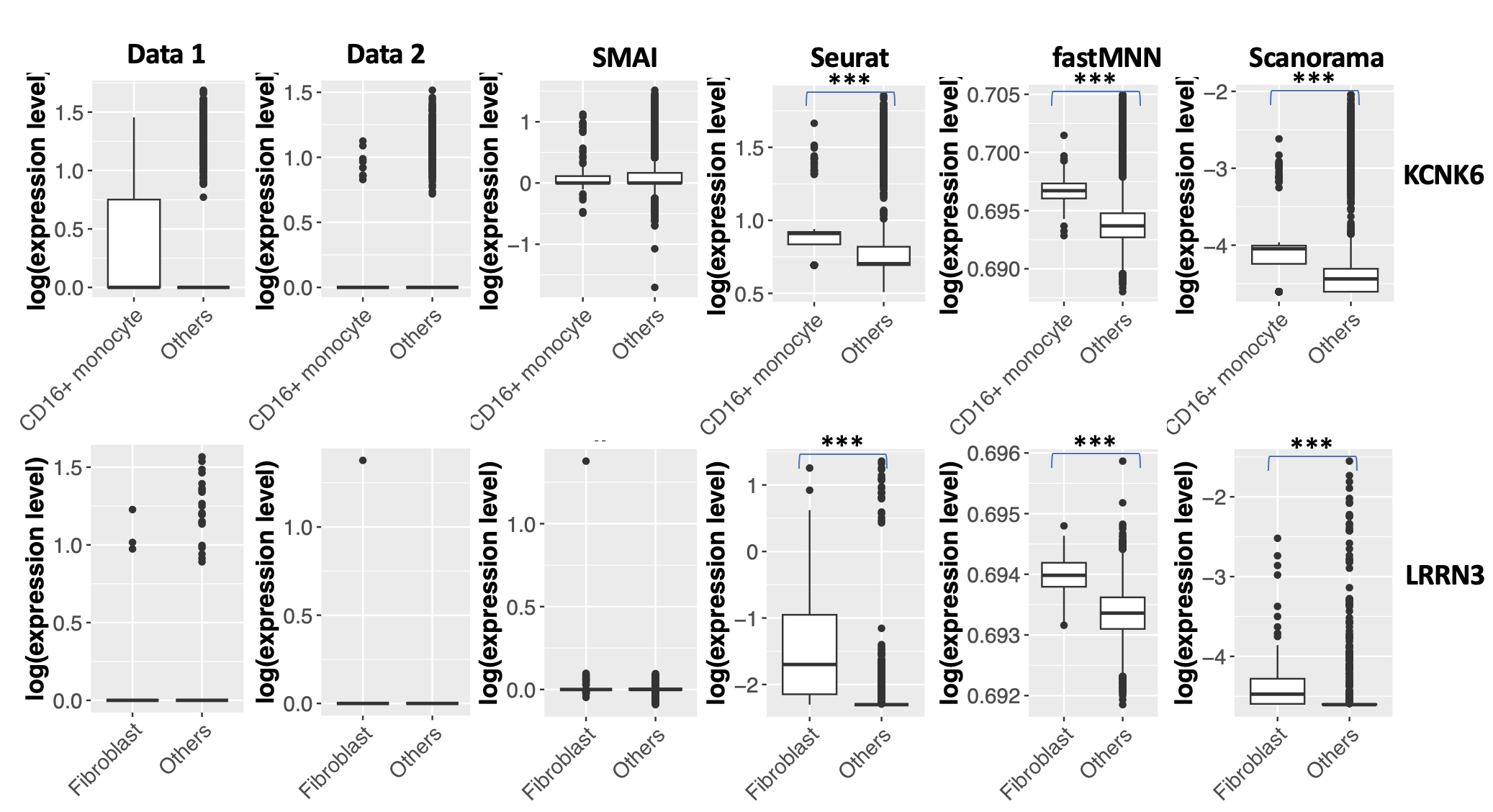}
			\caption{SMAI improves reliability of DE analysis. Top: Boxplots of log-expression levels of KCNK6 as grouped by cell types in the two datasets (Data 1 with 50 CD16+ monocytes and 3312 other cells, and Data 2 with 98 CD16+ monocytes and 3124 other cells) about human PBMCs (Task Pos3), and in the integrated datasets (148 CD16+ monocytes and 6436 others) as produced by SMAI-align, Seurat, fastMNN, and Scanorama. Bottom: Boxplots of log-expression levels of LRRN3 as grouped by cell types in the two datasets (Data 1 with 23 fibroblast cells and 2330 other cells, and Data 2 with 124 fibroblast cells and 1787 other cells) about human lung tissues (Task Pos5), and in the integrated datasets (147 fibroblast cells and 4117 other cells) as produced by SMAI-align, Seurat, fastMNN, and Scanorama. Artificial DE patterns were created by Seurat, fastMNN, and Scanorama after integration, but not by SMAI-align. The stars above the boxplots indicate statistical significance of the p-values. Specifically, * means adjusted p-value $<0.05$;  ** means adjusted p-value $<0.001$; *** means adjusted p-value $<0.0001$.} 
	\label{sup.fig3-3-2}
		\end{figure}
		
	\begin{figure}
		\centering
		\includegraphics[angle=0,width=15cm]{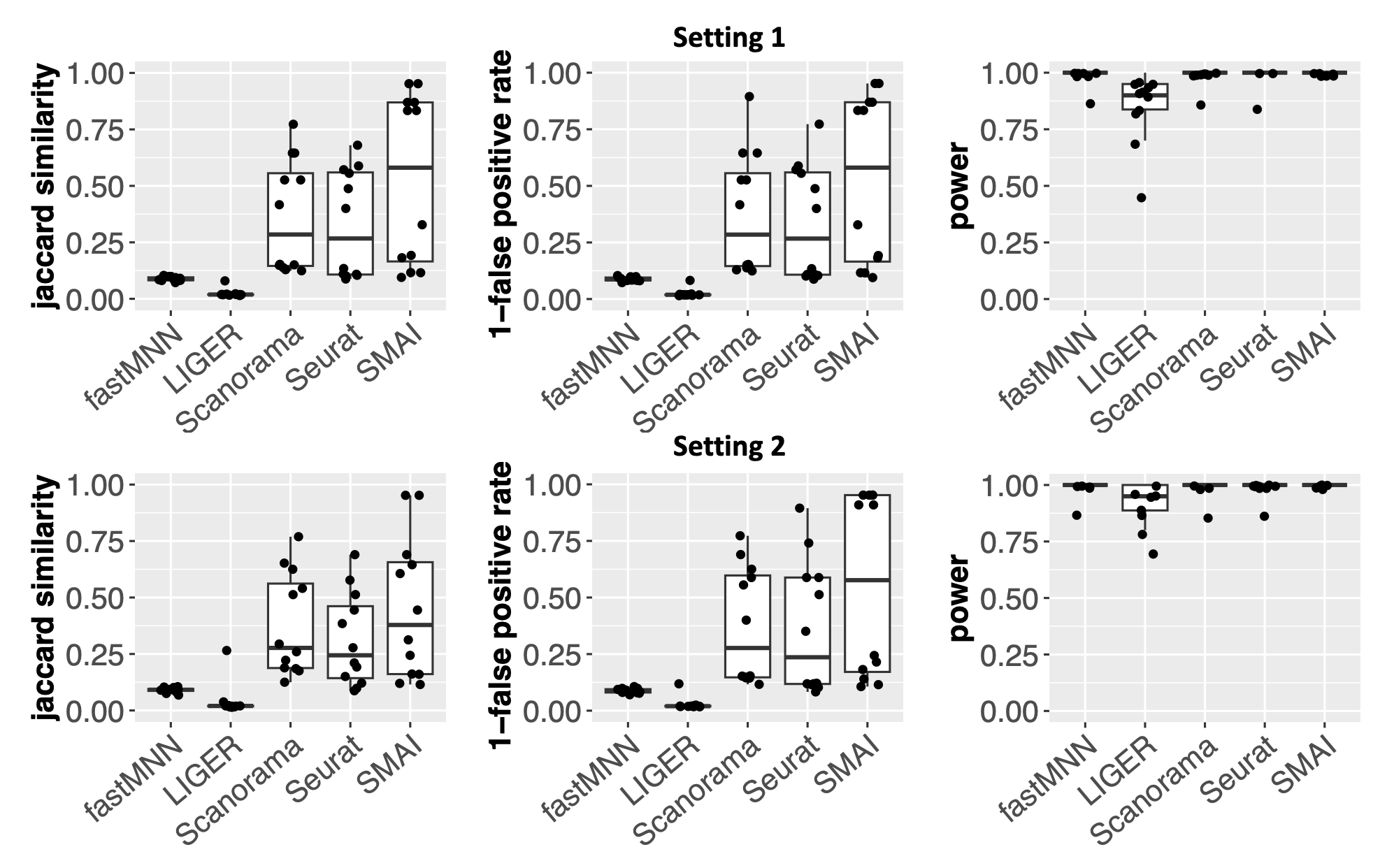}
			\caption{Simulation study reveals the advantage of SMAI in improving DE analysis. We carry out simulation studies by generating pairs of datasets, each containing about 2000 cells of 12 different cell types (clusters) and the expression levels of 1000 genes, allowing for some batch effects and slight differences in cell type proportions. For two different settings of signal-to-noise ratios, we show boxplots of the Jaccard similarity, the ($1-$ false discovery rate), and the power, of the subsequent DE analysis with respect to the true marker genes, based on each integrated dataset as obtained by one of the five methods. Our simulation shows better performance of SMAI as measured by all  three metrics. In particular, our results suggest the tendency of including more false positives by the existing methods than SMAI.} 
	\label{sup.fig4}
		\end{figure}
		
		\begin{figure}
		\centering
		\includegraphics[angle=0,width=15cm]{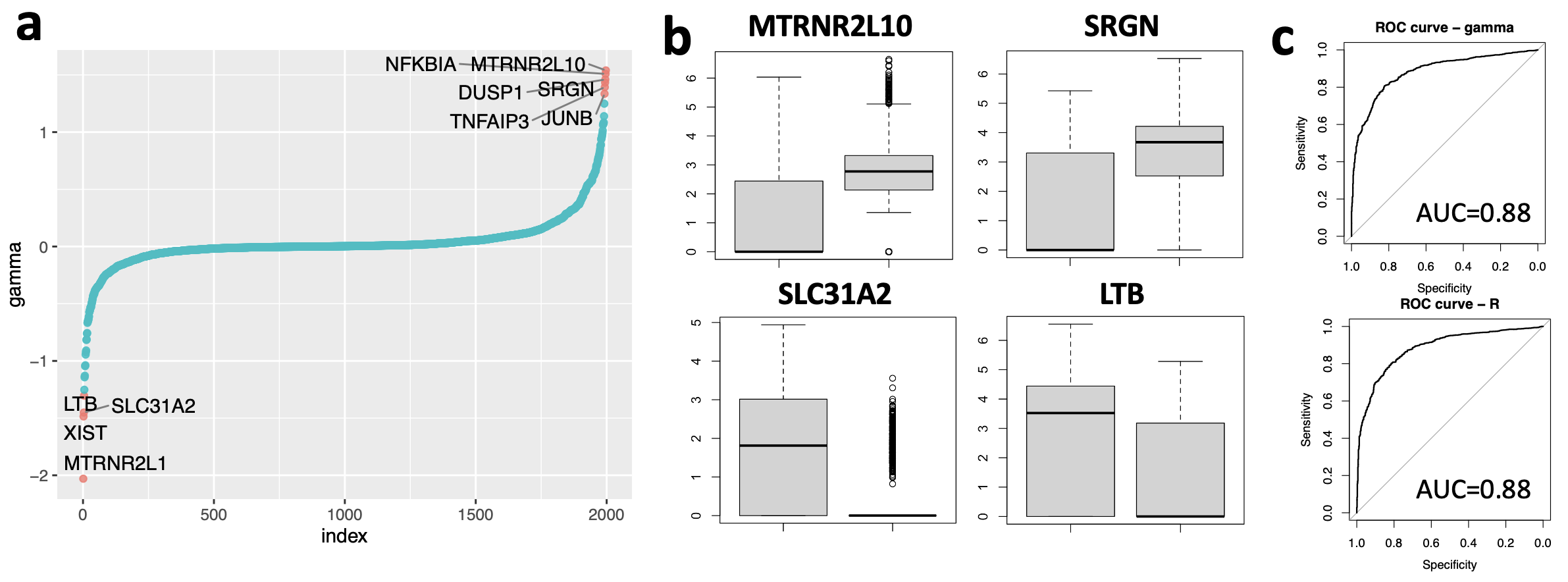}\\
		\includegraphics[angle=0,width=15cm]{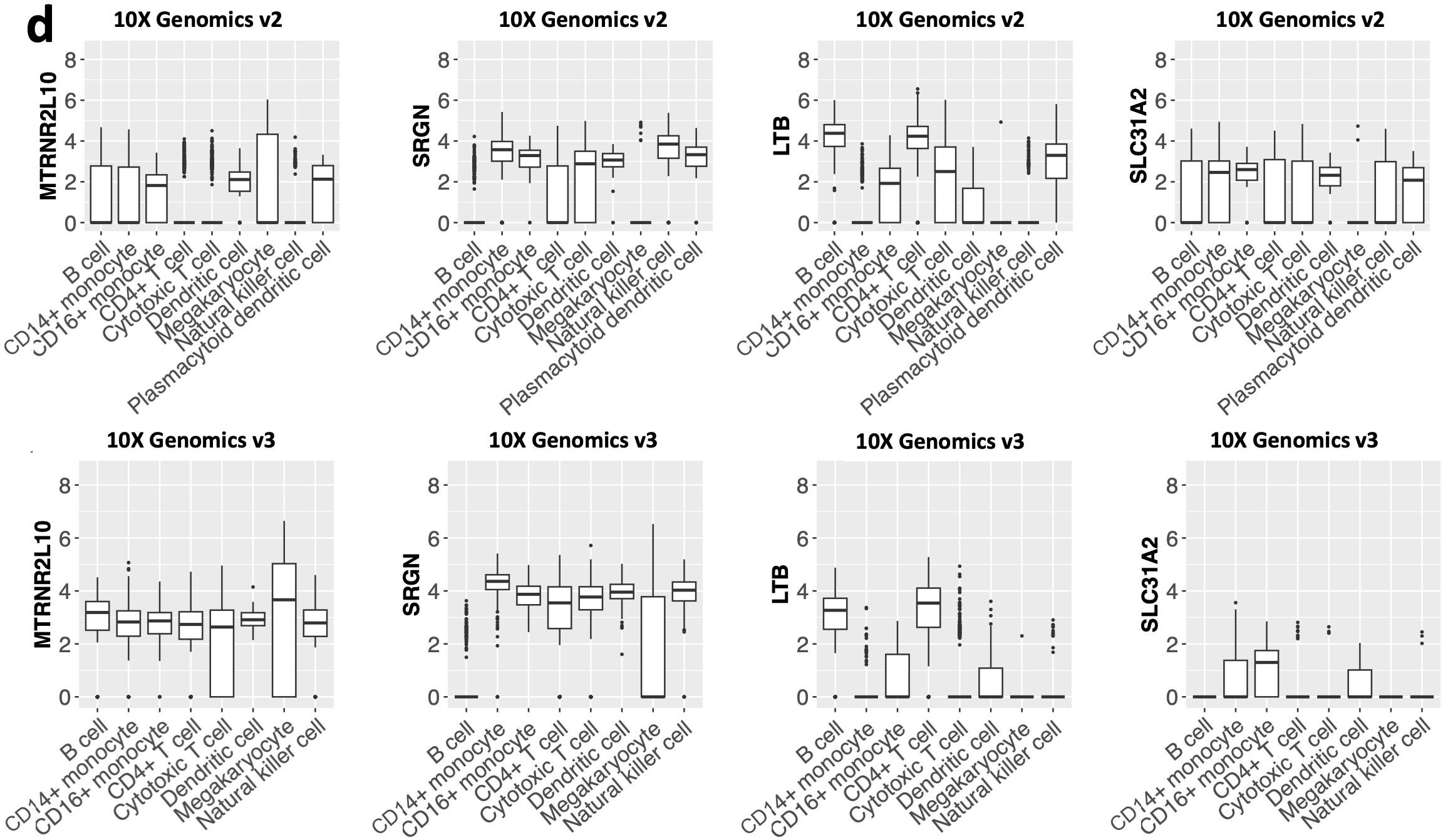}
			\caption{Additional results on SMAI's interpretability, and insights on the batch effects. (a) Visualization of the estimated mean-shift vector $\widehat\bgamma$ from integrating the human PBMCs data (Task Pos3), whose components are ordered from the smallest to the largest. (b) Boxplots of a few genes with largest absolute values in $\widehat\bgamma$ shown in (a). Left: 10X Genomics v2 dataset with $n=3362$. Right: 10X Genomics v3 dataset with $n=3222$. (c) ROC curves between the genes differentially expressed in at least one cell type in the PMBCs data (Task Pos3), and the batch-effect-related genes as captured by $\widehat\bgamma$ (Top), or by $\widehat\bR$ (Bottom). The high AUCs suggest most of the genes contributing to the batch effects are simultaneously DE genes associated with certain cell types. (d) Boxplots of the genes from (b),  grouped by different cell types. The main discrepancy between the measured expression profiles across datasets is likely caused by the differences in the total counts of these transcripts under the respective sequencing technologies. } 
	\label{sup.fig5}
		\end{figure}

			\begin{figure}
		\centering
		\includegraphics[angle=0,width=16cm]{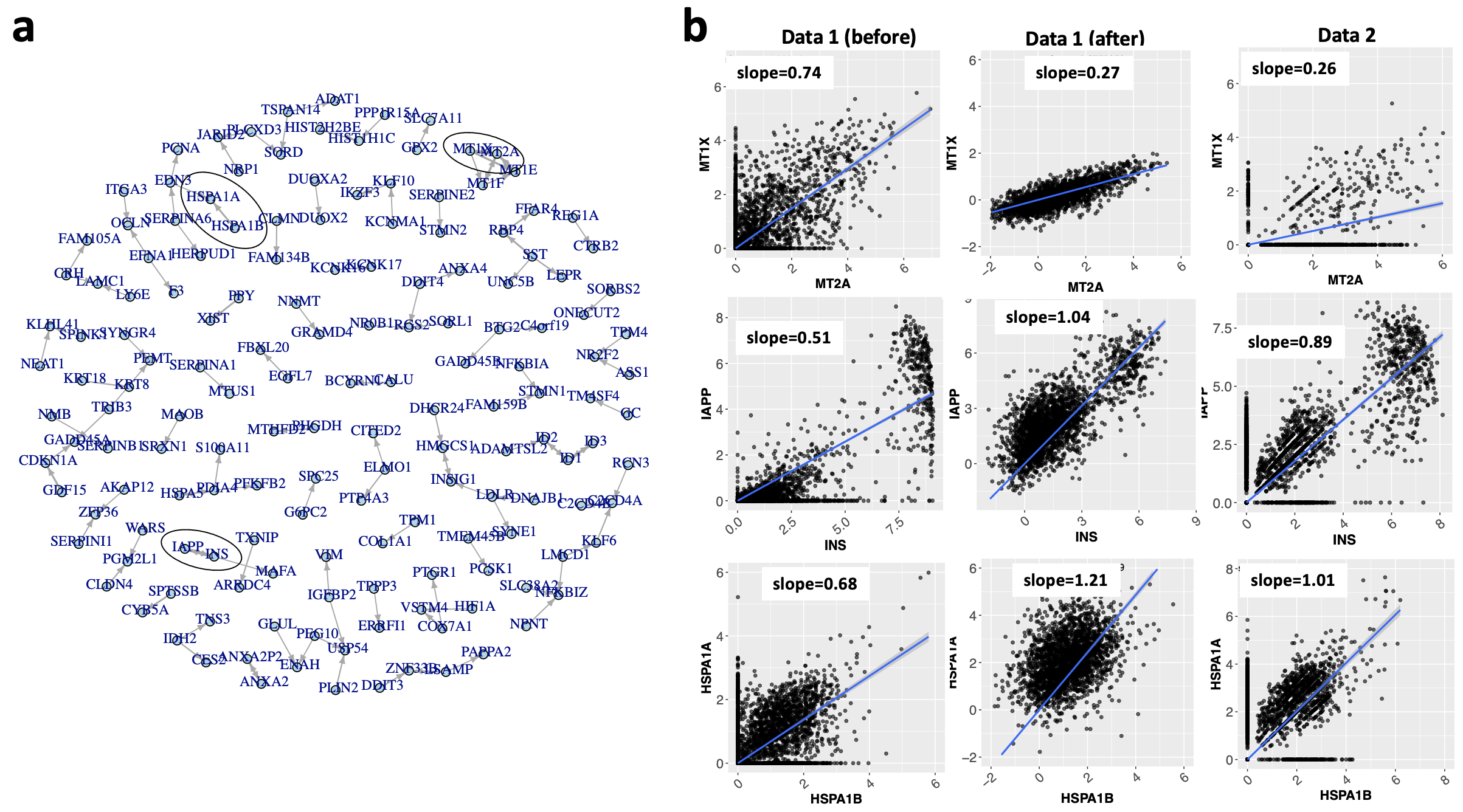}
			\caption{Insights on batch effects revealed by rotation $\widehat\bR$ in Task Pos1. (a) A directed weighted graph representation of the rotation matrix $\widehat\bR$ after removing the edges with weight $<0.1$. The genes linked by thicker edges are more significantly altered during alignment. (b) Scatter plots and the slope of fitted lines between the expression profiles of gene pairs highlighted in (a). Left: measurements in dataset 1 (Smart-Sea2) before alignment to dataset 2 (CEL-Seq2). Middle: measurements in dataset 1 after alignment to dataset 2 by SMAI. Right: measurements in dataset 2. SMAI-align leads to more similar co-occurrence patterns between the genes, as quantified by the slopes, in the two datasets after data alignment.} 
	\label{sup.fig5b}
		\end{figure}
		
			\begin{figure}
		\centering
		\includegraphics[angle=0,width=15cm]{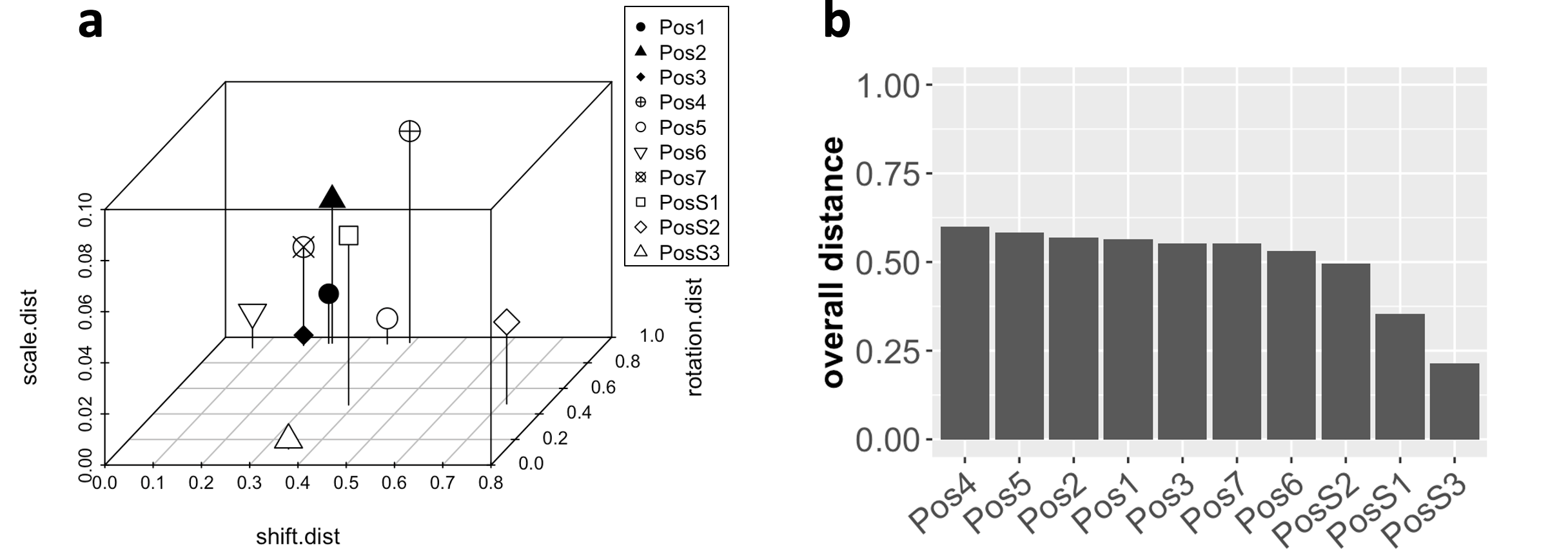}
			\includegraphics[angle=0,width=15cm]{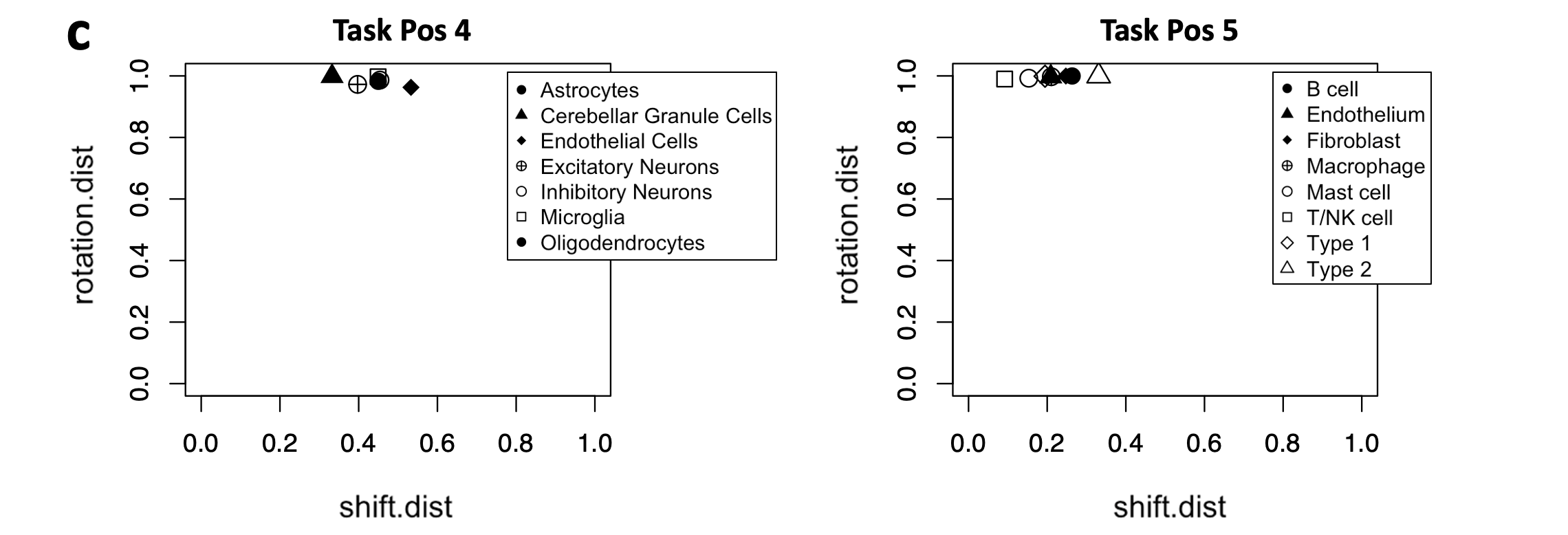}
			\caption{SMAI enables geometric quantification of batch effects. (a) 3-D scatter plot comparing the batch effects across all the positive control tasks, as characterized by the  SMAI-align parameters $(\widehat\beta, \widehat\bgamma, \widehat\bR)$, converted into three distance metrics between 0 and 1, with larger values indicating relatively greater amount of rescaling (z axis, or scale.dist), translation (x axis, or shift.dist), or rotation (y axis, or rotation.dist) needed to achieve alignment. (b) Barplot showing the overall distance between the aligned datasets, as characterized by an average of the three geometric distances shown in (a). These metrics quantify the magnitude and the  geometric composition of the batch effects between two datasets. {(c) 2-D scatter plot of the cell-type specific batch effects obtained by applying SMAI to each cell type within the Tasks Pos4 and Pos5.} The scale.dist is omitted as it depends on the size of each cell type. For each task, there was a remarkable similarity in the obtained cell-type specific translations (x-axis) and rotations (y-axis),  suggesting the suitability of conferring a global alignment function as we did in the main text.} 
	\label{sup.fig5d}
		\end{figure}
		
		\begin{figure}
		\centering
		\includegraphics[angle=0,width=15cm]{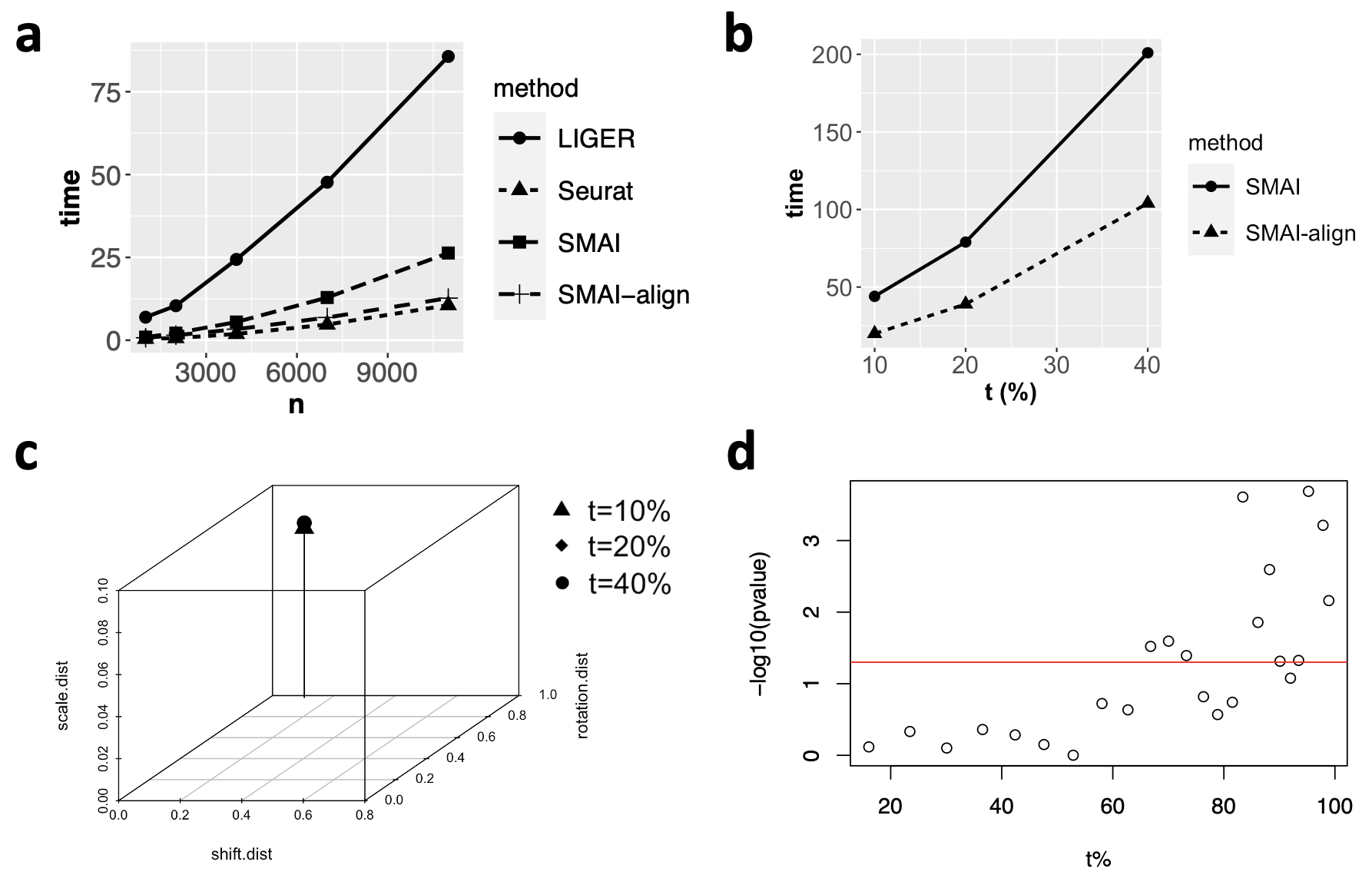}
			\caption{Evaluations of computing time, and the role of the thresholding parameter $t$. (a) By evaluating computing time (minutes) over a series of datasets with improving sample sizes, we find SMAI-align has a similar running time as Seurat, and is in general much faster than LIGER. The complete SMAI algorithm including SMAI-test and SMAI-align still has a reasonable computational time, that is, about twice of the time for running SMAI-align alone. {(b) Evaluation of SMAI and SMAI-align's computing time (minutes) under different subsampling rates on a large dataset containing over 134K cells. (c) 3-D scatter plot of SMAI-align parameters suggests  similarity of the alignment functions learned under different subsampling rates.} (d) Scatter plot between the threshold $t$ and the log-transformed p-values from SMAI-test concerning task Pos5. The red line corresponds to p-value $=0.05$. When $t$ is below certain value (e.g., 60\%), the p-values are not significant as the null hypothesis about partial alignability is likely true; when $t$ gets larger, the null hypothesis are mostly rejected. The transition point is likely the true proportion of alignable samples.} 
	\label{sup.fig6}
		\end{figure}

%
%\end{thebibliography}

\label{lastpage}
\end{document}